\documentclass[aps,prd,showpacs,superscriptaddress,twocolumn,nofootinbib,floatfix]{revtex4}

\usepackage{graphicx}
\usepackage{dcolumn}
\usepackage{bm}
\usepackage{amssymb}
\usepackage{hyperref}

\renewcommand{\case}[2]{{\textstyle \frac{#1}{#2}}}

\newcommand{\quarter}{\ensuremath{\case{1}{4}}}

\renewcommand{\Re}{\ensuremath{\mathop{\mathrm{Re}}}}
\renewcommand{\Im}{\ensuremath{\mathop{\mathrm{Im}}}}

\newcommand{\Dirac}{\ensuremath{D\kern-0.65em/\kern0.2em}}
\newcommand{\dirac}{\ensuremath{D\kern-0.525em/\kern0.2em}}
\newcommand{\Dhisq}{\ensuremath{D_{\mathrm{HISQ}}}}
\newcommand{\Dstag}{\ensuremath{D_{\mathrm{stag}}}}
\newcommand{\Hstag}{\ensuremath{H_{\mathrm{stag}}}}
\newcommand{\Sstag}{\ensuremath{S_{\mathrm{stag}}}}
\newcommand{\Det}{\ensuremath{\mathop{\mathrm{Det}}}}
\newcommand{\tr}{\ensuremath{\mathop{\mathrm{tr}}}}
\newcommand{\Tr}{\ensuremath{\mathop{\mathrm{Tr}}}}
\newcommand{\sign}{\ensuremath{\mathop{\mathrm{sign}}}}

\begin{document}
\preprint{FERMILAB-PUB-11-305-T}

\title{
Staggered fermions, zero modes, and flavor-singlet mesons}

\author{Gordon C. Donald} 
\email[Email: ]{g.donald@physics.gla.ac.uk}
\affiliation{SUPA, School of Physics and Astronomy, 
    University of Glasgow, Glasgow~~G12 8QQ, United Kingdom}

\author{Christine T. H. Davies} 
\email[Email: ]{c.davies@physics.gla.ac.uk}
\affiliation{SUPA, School of Physics and Astronomy, 
    University of Glasgow, Glasgow~~G12 8QQ, United Kingdom}

\author{Eduardo Follana} 
\email[Email: ]{efollana@unizar.es}
\affiliation{Departamento de F\'{i}sica Te\'{o}rica, Universidad de Zaragoza,
    Cl.\ Pedro Cerbuna 12, E-50009 Zaragoza, Spain}

\author{Andreas S. Kronfeld} 
\email[Email: ]{ask@fnal.gov}
\affiliation{Theoretical Physics Department, 
    Fermi National Accelerator Laboratory, 
    Batavia, Illinois 60510-5011, USA}

\collaboration{HPQCD and Fermilab Lattice Collaborations}
\noaffiliation

\date{\today}
    
\begin{abstract}
    We examine the taste structure of eigenvectors of the 
    staggered-fermion Dirac operator.
    We derive a set of conditions on the eigenvectors of modes with 
    small eigenvalues (near-zero modes), such that staggered
    fermions reproduce the 't~Hooft vertex in the continuum limit.
    We also show that, assuming these conditions, the correlators of 
    flavor-singlet mesons are free of contributions singular in $1/m$, 
    where $m$ is the quark mass.
    This conclusion holds also when a single flavor of sea quark is 
    represented by the fourth root of the staggered-fermion determinant.
    We then test numerically, using the highly improved staggered-quark action, whether these 
    conditions hold on realistic lattice gauge fields.
    We find that the needed structure does indeed emerge.
\end{abstract}
   

\maketitle

\section{Introduction}
\label{sec:intro}

Lattice QCD has made several notable strides over the past decade.
A wide variety of calculations with 2+1 flavors of sea quarks 
(corresponding to up, down, and strange) have been found to agree with 
experimental measurements within $\sim2\%$~\cite{ourlattexpt}.
Charmed-meson decay constants~\cite{Aubin:2005ar}, 
semileptonic form factors~\cite{Aubin:2004ej}, 
and the masses of the $B_c$~\cite{Allison:2004be} and
$\eta_b$~\cite{Gray:2005ur} mesons have been 
computed before being confirmed by measurements from experiments.
Calculations at nonzero temperature have shown that QCD possesses not a 
first-order phase transition but a smooth crossover~\cite{Aoki:2006we},
with implications for heavy-ion collisions and a cooling universe.
Some of the most precise determinations of 
the strong coupling~$\alpha_s$~\cite{Davies:2008sw}, 
quark masses~\cite{Davies:2009ih}, and 
flavor-changing couplings~\cite{newfds,Na:2010uf} come from lattice QCD.
It is impractical to cite every development here, but recent 
reviews~\cite{milcreview,Kronfeld:2010aw} cover the breadth of 
progress well.

The results listed above~\cite{ourlattexpt,Aubin:2005ar,Aubin:2004ej,%
Allison:2004be,Gray:2005ur,Aoki:2006we,Davies:2008sw,Davies:2009ih,%
newfds,Na:2010uf} 
have been obtained using staggered 
fermions~\cite{Susskind:1976jm,Sharatchandra:1981si} 
for the sea quarks, because this approach is numerically the 
fastest~\cite{Jansen:2008vs}.
In the continuum limit, one staggered-fermion field yields four 
species with a quantum number nowadays called ``taste.''
In numerical lattice gauge theory, sea quarks are represented by a 
determinant, for staggered fermions,
\begin{equation}
    \Det_4\left(\Dstag + m\right),
    \label{eq:staggered-Det}
\end{equation}
where $\Dstag$ denotes the lattice Dirac operator (see below), $m$ is 
the quark mass, and the subscript 4 is a reminder that the natural 
outcome is 4~tastes. 
To simulate a single species of given mass with staggered fermions, 
the (4-taste) determinant representing the sea is replaced 
with~\cite{Hamber:1983kx}
\begin{equation}
    \left[ \Det_4\left(\Dstag + m\right)\right]^{1/4}.
    \label{eq:fourth-root}
\end{equation}
Below we shall refer to the systems using (\ref{eq:staggered-Det}) and 
(\ref{eq:fourth-root}) as ``unrooted'' and ``rooted'' staggered 
fermions, respectively.
As far as we know, there is no controversy that lattice gauge theory 
with unrooted staggered fermions~(\ref{eq:staggered-Det}) defines a 
four-species continuum gauge theory.

The fourth root is controversial, however, because it is not standard 
quantum field theory.
The arguments supporting its validity hinge on structural 
properties of unrooted staggered fermions, which suggest that the 
continuum limit of $\Det_4(\Dstag+m)$ in Eq.~(\ref{eq:fourth-root}) 
factors into four equivalent 
determinants~\cite{Durr:2005ax,sharpelat06,asklat07,Golterman:2008gt}.
This factorization is verified in weak-coupling perturbation theory,
where the $1/4$ from the exponent multiplies each fermion loop.
Weak coupling also suggests how the symmetries of four species emerge in 
the continuum limit.
In simplified but similar systems where one can retain analytical 
control, the rooted determinant is 
valid~\cite{Adams:2003rm,Adams:2004wp,Adams:2004mf}.
Extensive numerical studies elaborate how the procedure works in the 
Schwinger model~\cite{Durr:2004ta,Durr:2006ze}.
Straightforward analysis of the hadron mass spectrum as a function of 
lattice spacing and quark mass, using chiral perturbation theory, 
substantiates this picture in detail~\cite{milcdecay04,milcreview}.
Further nonperturbative evidence comes from studying the eigenvalues 
of the staggered-fermion operator $\Dstag$, demonstrating that 
they appear in nearly degenerate 
quartets~\cite{oureigsshort,durreigs,Wong:2004nk,oureigslong}.
On lattice gauge fields with nonzero topological charge, sets of
quartets with eigenvalues near zero emerge.
The number of quartets and their
chirality satisfy the index theorem~\cite{Wong:2004nk,oureigslong}.

\begin{figure}
    \includegraphics[width=0.48\textwidth]{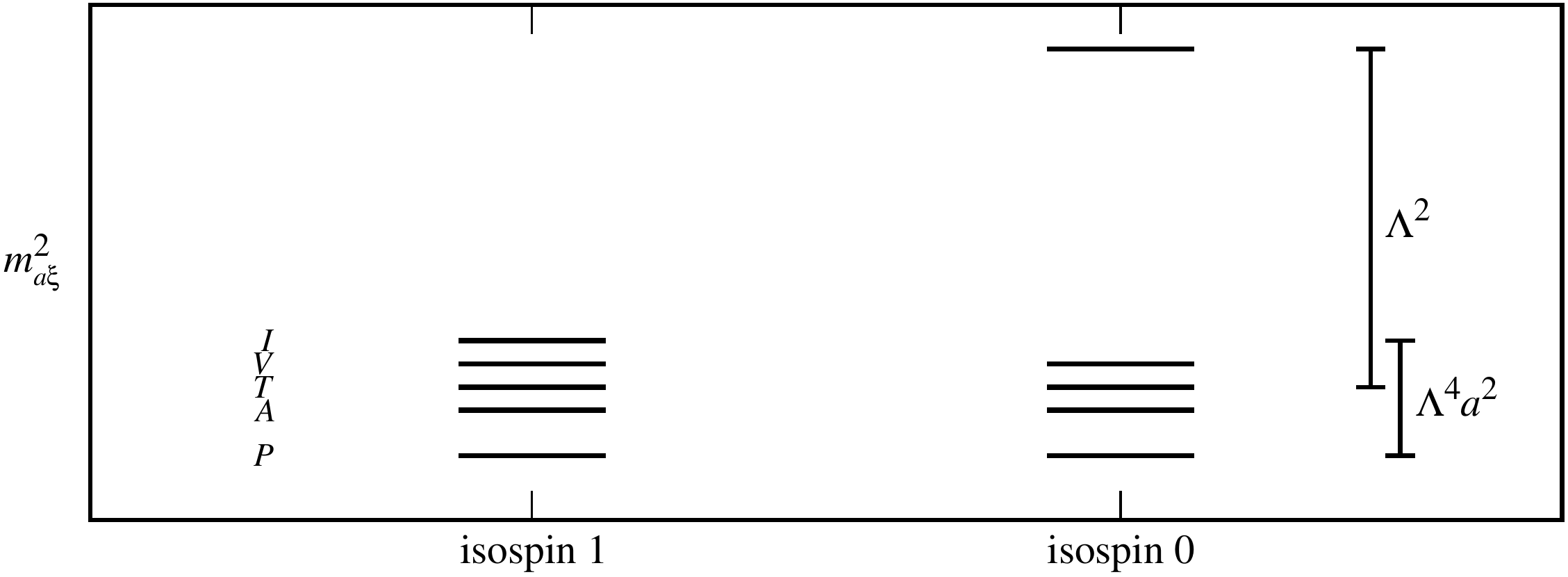}
    \caption[fig:spectrum]{Pattern of flavor and taste quantum numbers 
    in the (pseudoscalar) meson spectrum with two flavors and four tastes.
    The flavor-nonsinglet (isospin~1) mesons are split by small 
    lattice artifacts.
    The flavor-singlet (isospin~0) taste-\emph{nonsinglet} mesons are 
    no different.
    The flavor-taste singlet, however, receives a contribution from 
    mixing with purely gluonic states, an effect studied in 
    Refs.~\cite{Gregory:2007ev,Gregory:2011lat}.}
    \label{fig:spectrum}
\end{figure}

One issue that has not been fully addressed is the behavior of 
flavor-singlet mesons.
Direct calculations of the flavor-singlet meson masses are difficult%
~\cite{Gregory:2007ev,Gregory:2011lat,Jansen:2008wv,Christ:2010dd}, 
because they entail a contribution in which the quark-antiquark of the 
meson annihilates into gluons, and the gluons recreate the 
quark-antiquark pair.
With staggered fermions, it is crucial to bear in mind that only the 
flavor-taste singlet can undergo this process.
Low-energy gluons are taste singlets, so a meson with nontrivial taste 
cannot annihilate into them.
The spectrum with two flavors is sketched in Fig.~\ref{fig:spectrum},
illustrating the roles of the flavor and taste quantum numbers.

Building on the eigenvalue studies, this paper addres\-ses a specific 
concern, namely that flavor-taste-singlet correlators could diverge as 
a power of $m$ as $m\to0$.
Such behavior would be a clear failure of rooted staggered fermions.
We find fault with key steps in an attempted derivation of this 
claim~\cite{Creutz:2007rk,Creutz:2008nk}, which uses the 't~Hooft 
vertex~\cite{tHooft:1976up,tHooft:1976fv} to try to understand the role 
of near-zero modes.
A complementary examination of the same correlators reduces the problem 
to certain properties of the near-zero modes' eigenvectors~\cite{asklat07}.
Then contributions from connected and disconnected correlators cancel 
the divergent behavior; with the correct combinatoric factors~%
\cite{Venkataraman:1997xi,Bernard:2006vv,sharpelat06,asklat07,%
Bernard:2007eh}, the cancellation holds even with the rooted 
determinant of Eq.~(\ref{eq:fourth-root}). 

In this paper, we derive the staggered-fermion 't~Hooft vertex directly 
from the functional integral, both for unrooted and rooted staggered 
fermions.
If unrooted staggered fermions are to obtain a four-species 't~Hooft 
vertex in the continuum limit, we find that the eigenvectors must 
satisfy the same properties derived in Ref.~\cite{asklat07}, namely
Eqs.~(\ref{eq:zetaDef}) and~(\ref{eq:theTest}) below.
References~\cite{Bernard:2006vv,Bernard:2007eh} tacitly assumed these
properties, but we examine the eigenvectors numerically, plotting the 
quantities that enter the 't~Hooft vertex and the flavor-taste-singlet 
correlators.
We find that they behave in precisely the way needed for unrooted and 
rooted staggered fermions to yield four or, respectively, one species 
in the continuum limit.

The rest of this paper is organized as follows.
Section~\ref{sec:staggeigs} discusses staggered fermions and
some of the complaints and concerns about Eq.~(\ref{eq:fourth-root}). 
Section~\ref{sec:hooft} reviews the continuum 't~Hooft vertex 
and its symmetries, constructs the staggered-fermion 't~Hooft vertex,
and sets up the problem of flavor-taste singlets.
This discussion also pinpoints where the analysis of
Refs.~\cite{Creutz:2007rk,Creutz:2008nk} goes astray.
Section~\ref{sec:results} explains details of our numerical setup,
gives our lattice results, and discusses their implications.
The data speak for themselves: they clearly show that the needed 
structure emerges dynamically, ever more so for smaller lattice spacing.
Section~\ref{sec:conclusions} gives our conclusions.
It seems to us that the rooted staggered sea has passed another test in 
its usual way of relying on properties of the unrooted theory.
Appendix~\ref{app:formulas} contains some cumbersome notation that 
lends technical completeness to Secs.~\ref{sec:staggeigs},
\ref{sec:hooft}, and~\ref{sec:results}.
Appendix~\ref{app:actions} writes out improved actions explicitly.
Appendix~\ref{app:creutz} remarks on issues of secondary importance, 
raised in Refs.~\cite{Creutz:2007rk,Creutz:2008nk}.

\section{Staggered fermions}
\label{sec:staggeigs}

In this section, we review unrooted staggered fermions, 
because the way that four species emerge is central to any argument 
that Eq.~(\ref{eq:fourth-root}) is a valid regulator for one species.
We are careful to distinguish between flavor and taste;
the former is a label decoupled from the gauge interaction;
the latter is a property of staggered fermions, described below.

Below we use improved actions to check numerically whether the dynamics 
of staggered fermions are as expected.
For the discussion here, it is enough to start with the original, 
unimproved lattice action~\cite{Sharatchandra:1981si}:
\begin{widetext}
\begin{equation}
	\Sstag = 
		\case{1}{2} a^3 \sum_{x,\mu} \eta_\mu(x)\bar{\chi}(x)\left[
		U_\mu(x)\chi(x+\hat{\mu}a) - 
		U^\dagger_\mu(x-\hat{\mu}a)\chi(x-\hat{\mu}a) \right] + 
		ma^4 \sum_x \bar{\chi}(x)\chi(x),
	\label{eq:Sstag}
\end{equation}
\end{widetext}
where $a$ is the lattice spacing, $\chi(x)$ and $\bar{\chi}(x)$ are
gauge-group multiplets of Grassmann numbers for lattice site~$x$,
$U_\mu(x)$ is a lattice gauge field connecting sites $x$ and $x+\hat{\mu}a$
(such that $\Sstag$ is gauge invariant), 
$m$ is the bare mass, 
$\hat{\mu}$ is a unit vector in the $\mu$ direction, and
$\mu\in\{1,2,3,4\}$.
The staggered-fermion fields carry no Dirac index,
and sign factors appear instead of Dirac matrices:
\begin{equation}
	\eta_\mu(x) = (-1)^{\sum_{\rho<\mu}x_\rho/a}.
	\label{eq:eta}
\end{equation}
The staggered Dirac operator \Dstag\ is defined by writing
\begin{equation}
    \Sstag = a^4\sum_{x,y} \bar{\chi}(x)\left(
        \Dstag + m\delta_{xy} \right)\chi(y).
    \label{eq:Dstag}
\end{equation}
The determinant~(\ref{eq:staggered-Det}) follows from 
integrating the functional integral over $(\chi,\bar{\chi})$.

\Sstag\ is invariant under \emph{shifts}:
\begin{equation}
	S_\mu: \left\{ 
	\begin{array}{l}
		\chi(x) \mapsto \zeta_\mu(x) \chi(x+\hat{\mu}a) \\
		\bar{\chi}(x) \mapsto \zeta_\mu(x) \bar{\chi}(x+\hat{\mu}a)\\
		U_\nu(x) \mapsto  U_\nu(x+\hat{\mu}a) \quad \forall \nu 
	\end{array} \right.,
	\label{eq:shifts} 
\end{equation}
where
\begin{equation}
	\zeta_\mu(x) = (-1)^{\sum_{\sigma>\mu}x_\sigma/a}.
	\label{eq:zeta}
\end{equation}
Acting on fermion fields, $S_\nu S_\mu=-S_\mu S_\nu$.
This built-in Clifford group~$\Gamma_4$ is the origin of the four species 
in the continuum limit and their quantum number taste.
Acting on gauge fields, $S_\nu S_\mu=+S_\mu S_\nu$, from which it 
follows that low-momentum gauge fields are taste singlets.
With $n_f$ flavors of $(\chi,\bar{\chi})$---so $4n_f$ species in all---%
there is still only one gauge field and, thus, only one~$\Gamma_4$.

The kinetic term (for $n_f$ flavors) is also invariant under a 
$\mathrm{U}(n_f)$ symmetry group,
\begin{equation}
    \mathrm{U}_\varepsilon: \left\{ 
	\begin{array}{l}
        \chi(x)       \mapsto e^{\varphi^aT^a\varepsilon(x)} \chi(x) \\
        \bar{\chi}(x) \mapsto \bar{\chi}(x)e^{\varphi^aT^a\varepsilon(x)}
	\end{array} \right.,
	\label{eq:Ueps}
\end{equation}
where the $T^a$ are (anti-Hermitian) flavor generators, including 
flavor-singlet $T^0=i\openone_{n_f}/\sqrt{2n_f}$, and
\begin{equation}
    \varepsilon(x) = (-1)^{\sum_{\mu=1}^4x_\mu/a}.
    \label{eq:varepsilon}
\end{equation}
Three crucial properties of these symmetries~(\ref{eq:Ueps}) are that
\begin{enumerate}
    \item they are exact even at nonzero lattice spacing~$a$;
    \item they are nonsinglets with respect to taste;
    \item they imply that the eigenvalue spectrum of \Dstag\ is pure 
        imaginary and symmetric about~0.
        \label{item:0}
\end{enumerate}
The first property means that these symmetries cannot be anomalous, so 
they cannot be germane to the index theorem.
The second property means that the lack of anomaly is good: 
in QCD, species-nonsinglet symmetries do not have anomalies.
A corollary of the third property ensures that if $i\lambda$ is an 
eigenvalue of \Dstag\ with eigenvector $f(x)$, then $-i\lambda$ is 
also an eigenvalue, now with eigenvector $\varepsilon(x)f(x)$.
This corollary plays an important role in Sec.~\ref{sec:results}.
Unfortunately, the connection between property~\ref{item:0} and the 
spectrum sometimes leads, it seems, to the flavor-singlet 
$\mathrm{U}_\varepsilon$ being misidentified as the analog of continuum 
QCD's anomalous~$\mathrm{U}_A(1)$. 
The first two properties mean, however, that even the flavor-singlet 
$\mathrm{U}_\varepsilon$ cannot be related to $\mathrm{U}_A(1)$.

The analog of the $\mathrm{U}_A(1)$ is a flavor and taste singlet.
It is explicitly broken for $a\neq0$ but restored---apart from the 
anomaly and mass terms---as $a\to0$~\cite{Sharatchandra:1981si}. 
This mechanism is familiar in lattice gauge theory~\cite{Eichten:1985ft};
the same happens with Wilson fermions~\cite{Karsten:1980wd}. 
As $a\to 0$, an anomalous Ward identity emerges with axial-vector 
current, $A^\mu_I(x)$, and pseudoscalar density, $P_I(x)$, that are 
taste-flavor singlets~\cite{Sharatchandra:1981si,Smit:1987zh}.
The subscript~$I$ denotes the trivial representation of the shift 
symmetries~(\ref{eq:shifts}), also called the taste-singlet 
representation.

The way flavor-taste symmetries emerge is crucial to the validity of 
staggered fermions.
In particular,
\begin{eqnarray}
	\Gamma_4 \times \mathrm{SU}_V(n_f) & \subset & \mathrm{SU}_V(4n_f),
	\label{eq:embedtaste} \\
	\mathrm{U}_\varepsilon(n_f) \to
	\mathrm{U}(n_f) \otimes \xi^5 & \subset & \mathrm{SU}_A(4n_f),
	\label{eq:embedaxial}
\end{eqnarray}
where the symmetries on the left are exact (or softly broken) for \Sstag,
and those on the right are desired for continuum QCD.
The $\mathrm{SU}_V(n_f)$ on the left-hand side of 
Eq.~(\ref{eq:embedtaste}) is the obvious flavor-number symmetry of 
Eq.~(\ref{eq:Sstag}) for $n_f$ flavors of equal mass.
The $\xi^5$ on the left-hand side of Eq.~(\ref{eq:embedaxial}) denotes
the taste-nonsinglet nature of~$\mathrm{U}_\varepsilon$.

The pattern of symmetry appears most vividly, both for nonzero $a$ and 
as $a\to0$, in the meson-mass spectrum.
Meson operators can be written $\bar{\chi}\Gamma_\xi\chi$, 
where $\Gamma_\xi$ denotes various choices of sign factors $\eta$ and 
parallel transport within a hypercube, such that the bilinear 
transforms under the $\xi^{\mathrm{th}}$ bosonic representation of the 
shift symmetry group $\Gamma_4$.
As is customary, we label these $\xi\in\{I,V,T,A,P\}$, with $V$ and $A$ 
each grouping together four of these one-dimensional irreps, and $T$~six.
When focusing on a bilinear that transforms under rotations as a scalar,
vector, tensor, axial vector, or pseudoscalar, we shall write for 
$\Gamma_\xi$ either $1_\xi$, $\gamma^\mu_\xi$, $i\sigma^{\mu\nu}_\xi$, 
$\gamma^{\mu5}_\xi$, or $\gamma^5_\xi$, as the case may be.
For example, in this compact notation the taste-singlet pseudoscalar 
density is $P_I=\bar{\chi}\gamma^5_I\chi$.
Appendix~\ref{app:formulas} contains explicit formulas for 
bilinears in the taste-singlet representation~$I$, for all~$\Gamma$.

These operators create states such 
that~\cite{Golterman:1985dz,Kilcup:1986dg,Lee:1999zxa}
\begin{equation}
    \bar{\chi}\Gamma_\xi T^a\chi \doteq \bar{q}\Gamma\xi T^a q +
        \mathrm{O}(a^2),
    \label{eq:bilinears}
\end{equation}
where $q$ and $\bar{q}$ are continuum $4n_f$-species fermion fields, 
on the right-hand side $\Gamma$ is now a (usual) Dirac matrix, and 
$\xi$ is now a four-by-four matrix generator of~$\mathrm{U}(4)$.
Together the tensor products $\xi\otimes T$ generate U$(4n_f)$.
For nonsinglet $\xi\otimes T$ the pseudoscalar meson masses depend 
sensitively on $m$ and $a$, consistent with chiral perturbation 
theory~\cite{milcreview}.
The flavor-taste singlet, with 
$\xi\otimes T\propto\openone_{4n_f}$, should have a mass 
larger than the rest, cf.\ Fig.~\ref{fig:spectrum}, 
but that has not yet been demonstrated numerically~\cite{Gregory:2007ev}.
In this paper, we address this problem by
studying the eigenvectors of~\Dstag.

These lines of theoretical and numerical results lead to the picture 
that~\cite{gplshortpaper:2007}%

\begin{equation}
    \Dstag+m \doteq (\Dirac+m) \otimes \openone_4 +
        a^2\Delta,
    \label{eq:yigal}
\end{equation}
where \Dirac\ is the continuum Dirac operator,
$\openone_4$ is the ${4\times4}$ unit matrix,
$a^2\Delta$ is a taste-symmetry breaking term,
and taste-singlet $\mathrm{O}(a^2)$ corrections are not written out\footnotemark. 
Then
\begin{equation}
    \Det(\Dstag + m) \doteq
        [\Det_1(\Dirac+m)]^4 e^{\Tr_4\ln a^2\Delta(\dirac+m)^{-1}},
\end{equation}
suggesting that
\begin{eqnarray}
    [\Det(\Dstag + m)]^{1/4} & \doteq & \Det_1(\Dirac+m) \times 
        \nonumber \\ & & \hspace*{2em}
        e^{\frac{1}{4}\Tr_4\ln a^2\Delta(\dirac+m)^{-1}}.
        \hspace*{2em}
\end{eqnarray}
It is difficult to prove rigorously that the second factor becomes 
benign as $a\to0$, although a detailed renormalization-group argument 
makes it plausible~\cite{Shamir:2004zc,Shamir:2006nj}.
At nonzero $a$ this factor leads to nonlocality~\cite{Bernard:2006ee}
(though not the nonlocality discussed in 
Refs.~\cite{Bunk:2004br,Hart:2004sz}) and violations of unitarity.
In this paper, we have nothing to add to the arguments marshalled 
elsewhere~\cite{milcreview,Durr:2005ax,sharpelat06,asklat07,%
Golterman:2008gt} that these problems go away as $a\to0$.

A separate line of criticism~\cite{Creutz:2007rk,Creutz:2008nk} focuses 
not on the ultraviolet taste breaking of
$e^{\frac{1}{4}\Tr_4\ln a^2\Delta(\dirac+m)^{-1}}$ 
but on the interplay of the rooted determinant with correlators built 
from valence propagators.  
These papers assert, without derivation, certain symmetries and 
properties of the 't~Hooft vertex that, if true, would imply an 
unphysical $m$ dependence of multipoint meson correlators.
In the next section, we derive, rather than assert, the form of the 
staggered-fermion 't~Hooft vertex.
Our derivations pinpoint where Refs.~\cite{Creutz:2007rk,Creutz:2008nk} 
go astray.
Our derivation further reveals what is needed for staggered fermions to 
generate the continuum-QCD 't~Hooft vertex.
Whether staggered fermions behave in the needed way depends on dynamics, 
for which a numerical test is needed.
The (favorable) results of this test are presented in 
Sec.~\ref{sec:results}.

\section{Near-zero modes and the 't~Hooft~vertex}
\label{sec:hooft}

In this section, we discuss the properties of the near-zero modes in 
more detail.
We review properties of the 't~Hooft vertex in continuum gauge theory, 
with one and with four species.
Then we derive the 't~Hooft vertex for staggered fermions.
We show that the eigenvectors must exhibit a certain structure if 
unrooted staggered fermions are to tend to the continuum gauge theory.
This structure is precisely the criterion presented in 
Ref.~\cite{asklat07} for the rooted theory to have a sensible $\eta'$ 
correlator.

\subsection{Continuum QCD}
\label{subsec:contQCD}

In continuum gauge theories, the Dirac operator can have genuine zero 
modes.
For a single species, the eigenfunctions and eigenvalues are denoted 
$\Dirac\phi_\sigma=i\lambda_\sigma\phi_\sigma$, where $\lambda$ is 
real, and integer~$\sigma$ labels the modes.
For the modes with nonzero eigenvalue, it is convenient to take
$\sigma>0$ ($\sigma<0$) for modes with $\lambda>0$ ($\lambda<0$).
These modes come in conjugate pairs: 
$\lambda_{-\sigma}=-\lambda_\sigma$, 
$\phi_{-\sigma}=\gamma^5\phi_\sigma$. 
In the subspace~of zero modes, $\lambda=0$,
the eigenfunctions can be chosen such that
$\gamma^5\phi^{(\pm)}_\iota=\pm\phi^{(\pm)}_\iota$,
with the integer label $\iota$ ranging from 1 to $k_\pm$.
For $n$ species, the Dirac operator is $\Dirac\openone_n$, with
eigenfunctions $\phi_\sigma e^{(\tau)}$, where the $e^{(\tau)}$ form an 
orthonormal basis in species space.
The number and chirality of zero modes is related to the topological 
charge $Q$ via the index theorem~\cite{Atiyah:1963nq,Atiyah:1967ih}
\begin{equation}
    n_+-n_-=nQ,
    \label{eq:index}
\end{equation}
where $n$ is the number of species, and $n_\pm=nk_\pm$ accounts for the 
species multiplicity.

The determinant acquires a factor of mass $m$ from each zero mode.
As $m\to0$ it would seem that such gauge fields would drop out of the 
ensemble average.
But if one looks at the eigenvalue-eigenfunction representation of the 
propagator, one finds powers of $1/m$ that cancel the powers of $m$ 
from the determinant.
Focusing on $|Q|=1$ and $n=1$, so that there is one zero mode,
the propagator is 
($\langle\bullet\rangle_{|Q|=1}$ denotes average over $|Q|=1$ gauge fields)
\begin{widetext}
\footnotetext{Reference~\cite{gplshortpaper:2007} shows that 
the $\mathrm{U}_\varepsilon$ symmetry requires the dimensions-5 terms 
of the \emph{off-shell} Lee-Sharpe~\cite{Lee:1999zxa} effective 
Lagrangian to possess coefficients proportional to~$ma$. 
The apparent $\mathrm{O}(a)$ effects in Refs.~\cite{Gliozzi:1982ib,%
KlubergStern:1983dg} are an artifact of the choice of field variables.}
\begin{equation}
    \langle \psi(x)\bar{\psi}(y) \rangle = \left\langle 
        m\prod_{\sigma>0}(\lambda_\sigma^2+m^2)
        \sum_\sigma 
            \frac{\phi_\sigma(x)\phi_\sigma^\dagger(y)}{i\lambda_\sigma+m} 
        \right\rangle_{\hspace{-0.5em}A} = 
        \left\langle \prod_{\sigma>0}(\lambda_\sigma^2+m^2)\,
        \phi_0(x)\phi_0^\dagger(y) \right\rangle_{\hspace{-0.5em}|Q|=1}
        + \mathrm{O}(m),
    \label{eq:hooft}
\end{equation}
where $\phi_0$ is now used for the zero-mode eigenfunction.
One sees that the mode with $\lambda=0$ has a canceling factor of 
$1/m$.
The factor $\phi_0(x)\phi_0^\dagger(y)$ is the 't~Hooft ``vertex''
\cite{tHooft:1976up,tHooft:1976fv}.
(If $\phi_0$ is localized, as it is around instantons, then the ``vertex'' 
has support only for $x,y$ near the center of localization.)
For the four-point function, there are superficially two powers of 
$1/m$, but two contributions identical apart from their opposite sign 
cancel each other.
This is simply the Pauli exclusion principle arising from the
Grassmann nature of the fields.

With $n=4$ fermion species, each mode is replicated four times,
so gauge fields with $|Q|=1$ yield four zero modes, one per species.
The determinant yields a factor~$m^4$, which is not compensated until 
the eight-point function: 
\begin{equation}
    \left\langle\prod_{f=1}^4\psi_f(x_f)\bar{\psi}_f(y_f)\right\rangle =
        \left\langle\prod_{\sigma>0}(\lambda_\sigma^2+m^2)^4
        \prod_{f=1}^4 \phi_0(x_f)\phi_0^\dagger(y_f) 
        \right\rangle_{\hspace{-0.5em}|Q|=1}
        + \mathrm{O}(m),
    \label{eq:4species}
\end{equation}
with four factors like that in Eq.~(\ref{eq:hooft}).
In higher-point functions, Pauli exclusion again ensures that 
contributions singular in $m$ cancel.
Below we are interested in flavor-singlet meson correlators, such as 
(flavor index contracted; $n_f=4$)
\begin{equation}
    \left\langle\prod_{f=1}^4\bar{\psi}\Gamma_f\psi(x_f)\right\rangle =
        24
        \left\langle\prod_{\sigma>0}(\lambda_\sigma^2+m^2)^4
        \prod_{f=1}^4 \phi_0^\dagger\Gamma_f\phi_0(x_f) 
        \right\rangle_{\hspace{-0.5em}|Q|=1}
        + \mathrm{O}(m),
    \label{eq:24}
\end{equation}
\end{widetext}
where the combinatoric factor 24 obtains after cancellations between 
many (dis)connected terms.

Let us now examine a property of the 't~Hooft vertex that is central to 
Creutz's arguments~\cite{Creutz:2007rk,Creutz:2008nk}.
Under the anomalous $\mathrm{U}_A(1)$ transformation
\begin{equation}
    \psi\mapsto e^{i\gamma^5\alpha/2}\psi,\quad\quad
    \bar{\psi}\mapsto \bar{\psi}e^{i\gamma^5\alpha/2},
    \label{eq:UA1}
\end{equation}
the $n$-species 't~Hooft vertex transforms as
\begin{equation}
    \prod_{f=1}^n \phi_0(x_f)\phi_0^\dagger(y_f) \mapsto
        e^{\pm in\alpha} \prod_{f=1}^n \phi_0(x_f)\phi_0^\dagger(y_f),
\end{equation}
where the sign is the chirality of the zero mode, 
$\gamma^5\phi_0=\pm\phi_0$.
If $\alpha$ is a multiple of $2\pi/n$, the prefactor is unity; thus, 
the 't~Hooft vertex remains invariant under a $\mathbb{Z}_n$ 
subgroup of $\mathrm{U}_A(1)$~\cite{Creutz:2007yr}.

This invariance holds for the full determinant~\cite{Creutz:2009zq,Creutz:2011hy}.
Under the transformation (\ref{eq:UA1}) with $\alpha=2\pi/n$, one has
\begin{eqnarray}
    \hspace*{-1em}
    m^{n(n_++n_-)}e^{i(n_+-n_-)2\pi/n} 
        {\Det_n}'[\Dirac+me^{i\gamma^52\pi/n}] = \quad\nonumber \\
    m^{n(n_++n_-)}
        {\Det_n}'[\Dirac+me^{i\gamma^52\pi/n}] ,
    \label{eq:rot}\hspace*{0.5em}
\end{eqnarray}
where $\Det_n'$ denotes the $n$-species determinant with zero modes 
projected out.
The right-hand side follows because, by Eq.~(\ref{eq:index}), 
the phase on the left-hand side is trivial.
Because $e^{i2\pi/n}=e^{-i2\pi(n-1)/n}$, 
the twisted mass $me^{i\gamma^52\pi/n}$ can be removed with nonsinglet 
$\mathrm{SU}_A(n)$ transformations, namely,
\begin{equation}
    \psi \mapsto e^{-i\gamma^5\Xi\pi/n}\psi,
        \quad \quad
    \bar{\psi} \mapsto \bar{\psi}e^{-i\gamma^5\Xi\pi/n},
    \label{eq:notUA1}
\end{equation}
where $\Xi = \mathop{\rm diag}(1,\,\ldots\,1,\;-(n-1))$,
or any permutation thereof.
The composition of transformations~(\ref{eq:notUA1})
and (\ref{eq:UA1}) with $\alpha=2\pi/n$ returns the original 
determinant, $m^{n(n_++n_-)}\Det_n'(\Dirac+m)$.
We have shown here that the $\mathbb{Z}_n$ in question is not only a 
subset of the anomalous~$\mathrm{U}_A(1)$, but also the center of the 
exact $\mathrm{SU}_A(n)$.
In fact, $\mathbb{Z}_n$ is the intersection of the $\mathrm{SU}_A(n)$ 
and~$\mathrm{U}_A(1)$.

\subsection{Unrooted staggered fermions}

Now we would like to see how staggered fermions reproduce the 
four-species 't~Hooft vertex.
Let us now denote the eigenvectors and eigenvalues
$\Dstag f_s(x)=i\lambda_sf_s(x)$.
We use $f$ for the eigenvectors of $\Dstag$, instead of $\phi$ for 
the eigenfunctions of \Dirac, because our aim is to study whether and 
how a structure like $\phi_\sigma e^{(\tau)}$ arises from the~$f_s$.
As before, it is convenient to choose $s>0$ ($s<0$) for $\lambda_s>0$ 
($\lambda_s<0$).
As mentioned above, the function $f_{-s}(x)=\varepsilon(x)f_s(x)$ has
eigenvalue $\lambda_{-s}=-\lambda_s$, which follows from the 
$\mathrm{U}_\varepsilon$ symmetry.
One must bear in mind that the relation between eigenvectors 
$f_{\pm s}$ originates from a different flavor of symmetry than the 
relation between eigenfunctions $\phi_{\pm\sigma}$.
In the notation introduced above Eq.~(\ref{eq:bilinears}), 
multiplication by $\varepsilon(x)$ corresponds to $\gamma^5_P$, 
a taste nonsinglet that, in a continuum four-species theory, looks like
$\gamma^5\xi^5$, not~$\gamma^5\openone_4$.

The first step is to single out the modes analogous to the zero modes 
in the continuum theory.
With staggered (and most other lattice) fermions, no exact zero modes 
arise, but one expects \Dstag\ to have some exceptionally small 
eigenvalues~\cite{Smit:1987zh}.
A crisp way to identify them is via the spectral flow of the 
operator~\cite{Adams:2009eb}
\begin{equation}
    \Hstag=-i\Dstag+\mu\gamma^5_I,
    \label{eq:Hstag}
\end{equation}
with $\Hstag f_s(x;\mu)=\lambda_s(\mu)f_s(x;\mu)$;
the eigenvalues of \Dstag\ are $i\lambda_s(0)$.
From the $\mathrm{U}_\varepsilon$ symmetry,
$f_{-s}(x;\mu)=\varepsilon(x)f_s(x;-\mu)$,
$\lambda_{-s}(\mu)=-\lambda_s(-\mu)$.
Near-zero modes are those with a nearby zero crossing, 
${\lambda(\mu_0)=0}$ for $\mu_0\ll\Lambda$.
The (taste-singlet) chirality is then
\begin{equation}
    \hat{\mathcal{X}}_s = \sign\lambda'_s(\mu_0),
    \label{eq:Adams}
\vspace*{0.7em}
\end{equation}
\noindent
where the prime denotes differentiation with respect to~$\mu$.
Taking the $\mathrm{U}_\varepsilon$ symmetry into account, we can 
label the positive-chirality modes $f_i^{(+)}$ with 
$i>0$ ranging from $1,\ldots,\ell_+$ ($\lambda_i$ slightly positive) and
$i<0$ ranging from $-1,\ldots,-\ell_+$ ($\lambda_i$ slightly negative).
A similar labeling scheme can be adopted for the $2\ell_-$  
negative-chirality modes~$f_i^{(-)}$.
Note that~\cite{Adams:2009eb}
\begin{equation}
    \lambda'_s(\mu_0)\approx \lambda'_s(0) =
        \sum_{x}f_s^\dagger(x)\gamma^5_If_s(x) \equiv \mathcal{X}_s,
    \label{eq:chirality}
\end{equation}
where $\mathcal{X}_s$ is a more common way to identify 
chirality~\cite{Vink:1988ss}.
Modes $s$ and $-s$ have the same value of taste-singlet chirality
(whether defined by $\mathcal{X}_s$ or $\hat{\mathcal{X}_s}$),
because $\gamma^5_I$ implies transport over an even number of links 
and, consequently, the $\varepsilon$ sign factors at the two ends of 
$\gamma^5_I$ are the same.

The spectral flow is elegant but computationally demanding.
It is also possible to identify the near-zero modes by looking for 
modes with $\lambda$ sufficiently small and $\mathcal{X}$ sufficiently 
close to $\pm1$.
Although the spectral flow is (presumably) more decisive in 
borderline cases, in practice, especially for the scope of this paper, 
the computational demand seems prohibitive.
In Sec.~\ref{sec:results}, we shall therefore rely on our experience in 
Refs.~\cite{oureigsshort,oureigslong} of using $(\lambda,\mathcal{X})$ 
to identify the near-zero modes.

If staggered fermions generate four species in the continuum limit, then 
the eigenvalues should arrange themselves into closely spaced quartets.
For nonzero modes, four modes should cluster around some distinctly 
nonzero value.
For near-zero modes, on the other hand, such quartets lie slightly 
above and below the real axis. 
$\mathrm{U}_\varepsilon$~symmetry dictates that a mode and its 
$\varepsilon$ partner have the same chirality and, thus, may be 
assigned to the same quartet. 
If the gauge-field dynamics yield even $\ell_\pm=2k_\pm$, then 
one has quartets.
The resulting index theorem is then ($n_\pm=4k_\pm$)
\begin{equation}
    n_+-n_-=4Q,
    \label{eq:staggered-index}
\end{equation}
where $Q$ is a pure-gauge definition of topological charge.
For smooth enough fields and for extensions of Eq.~(\ref{eq:Sstag}) 
that smooth out the interaction, both kinds of quartets 
emerge~\cite{oureigsshort,durreigs}, as does the connection between 
gauge-field topology and the 
index~\cite{oureigsshort,durreigs,Wong:2004nk,oureigslong}.

With one fermion field but sets of four near-zero modes, the 
combinatorics underlying the 't~Hooft mechanism are less 
straightforward than in four-species continuum theories.
Let us focus on $|Q|=1$.
Two pairs of near-zero modes appear with eigenvalues 
$\pm i\lambda_i$, $i=1,2$.
``Small'' means $|\lambda_i|\sim (a\Lambda)^{p_\lambda}\Lambda$;
a~power law with $p_\lambda=1$ or $2$ suffices, 
and one expects $p_\lambda=2$ \cite{gplshortpaper:2007}.
Moreover, $\mathcal{X}_1$ and $\mathcal{X}_2$ have the same sign 
(with several actions~\cite{oureigsshort,oureigslong}), and we shall 
see in Sec.~\ref{sec:results} that these features also hold for the 
highly-improved staggered-quark (HISQ) action~\cite{hisq}.

\begin{widetext}
To derive the 't~Hooft vertex explicitly, let us examine the (fermion) 
eight-point function, which for staggered fermions~is
\begin{equation}
    \left\langle\prod_{f=1}^4\chi(x_f)\bar{\chi}(y_f)\right\rangle =
        \left\langle \prod_{i=1}^2(\lambda_i^2+m^2)\,
            \prod_{s>0}(\lambda_s^2+m^2)\det_{(f,g)} G(x_f,y_g)
        \right\rangle_{\hspace{-0.4em}|Q|=1},
    \label{eq:4staggered}
\end{equation}
\end{widetext}
where the propagator
\begin{equation}
    G(x,y) = \langle\chi(x)\bar{\chi}(y)\rangle_{\chi,\bar{\chi}} =
        \sum_{\mathrm{all}~s} 
        \frac{f_s(x)f_s^\dagger(y)}{i\lambda_s+m}
\end{equation}
with the sum running over near-zero and nonzero modes.
Neglecting in Eq.~(\ref{eq:4staggered}) the near-zero $\lambda_i$ 
relative to $m$, the near-zero-mode terms contribute 
to Eq.~(\ref{eq:4staggered}) as
\begin{equation}
    m^4\det_{(f,g)}G(x_f,y_g) = \det_{(i,f)}f_i(x_f) 
        \det_{(j,g)}f_j^\dagger(y_g) + \mathrm{O}(m),
    \label{eq:4modes}
\end{equation}
where $i,j\in\{-2,-1,1,2\}$.
In higher-point functions, the Pauli exclusion again ensures that 
contributions singular in $m$ cancel.

The product of determinants on the right-hand side of 
Eq.~(\ref{eq:4modes}) is the 't~Hooft vertex for (unrooted) staggered 
fermions. 
To reproduce the product of four factors of $\phi_0\phi_0^\dagger$ in 
Eq.~(\ref{eq:4species}), the four staggered eigenfunctions $f_i$, 
$i\in\{-2,-1,1,2\}$, must 
have structure similar to $\phi_0(x)e^{(i)}$.
One could seek such structure in a basis where a taste index looks 
obvious, but because taste is, fundamentally, a quantum number of the 
shifts, i.e., single-link translations, gauge-dependent roughness of 
the gauge field would obscure it.

The way forward is to contract the $\chi$ and $\bar{\chi}$ fields 
into color singlets.
The contractions must also be taste singlets, because a nonsinglet 
corresponds to ${e^{(i)}}^\dagger\xi e^{(j)}$, $\xi\neq\openone_4$,
which need not vanish when $j\neq i$.
In Eq.~(\ref{eq:4staggered}) we thus replace $\chi(x_f)\bar{\chi}(y_f)$ 
with a taste singlet $\bar{\chi}\Gamma^f_I\chi(x_f)$.
Contracting Eq.~(\ref{eq:4modes}) in this way, one is led to consider
\begin{equation}
    \zeta^\Gamma_{ij}(x) = f_{i}^\dagger\Gamma_If_{j}(x)
    \label{eq:zetaDef}
\end{equation}
with, recall, some parallel transport implied by $\Gamma_I$.
The 't~Hooft vertex simplifies in the desired way if
\begin{equation}
    \zeta^\Gamma_{ij}(x) \propto \delta_{ij}\left[1+
        \mathrm{O}(a^{p_{\zeta^\Gamma}}) \right],
    \label{eq:theTest}
\end{equation}
for $\Gamma_I=1_I$, $\gamma^5_I$, $i\sigma^{\mu\nu}_I$.
If, further, the proportionality fulfilled by an (approximately) 
$i$-independent diagonal $\zeta^\Gamma_{ii}(x)$,
$\zeta^\Gamma_{ij}$ would then mimic 
$\phi^\dagger_0\Gamma\phi_0{e^{(i)}}^\dagger e^{(j)}\propto\delta_{ij}$.
Approaching this limit as a power law with $p_{\zeta^\Gamma}=1$ or~2 
suffices, and one expects $p_{\zeta^\Gamma}=2$ \cite{gplshortpaper:2007}.
Section~\ref{sec:results} presents numerical results for these 
local overlaps, including $a$ dependence.

For $\Gamma_I=\gamma^\mu_I$, $\gamma^{\mu5}_I$, the local overlaps 
$\zeta^\Gamma_{ij}$ behave somewhat differently. 
In continuum gauge theory, the zero modes satisfy 
${\phi_\iota^{(\pm)}}^\dagger\gamma^\mu\phi_\iota^{(\pm)}=%
 {\phi_\iota^{(\pm)}}^\dagger\gamma^{\mu5}\phi_\iota^{(\pm)}=0$, 
because $\gamma^5$ anticommutes with $\gamma^\mu$ and 
$\gamma^5\phi_\iota^{(\pm)}=\pm\phi_\iota^{(\pm)}$.
The spin and taste degrees of freedom emerge from staggered 
fermions via the same dynamical mechanism, so the diagonal
$\zeta^{\gamma^\mu}_{ii}$ and $\zeta^{\gamma^{\mu5}}_{ii}$ 
should vanish commensurately with the off-diagonal~$\zeta^\Gamma_{ij}$, 
$\Gamma_I=1_I$, $\gamma^5_I$, $i\sigma^{\mu\nu}_I$.

The local overlaps of continuum nonzero modes of different species also 
vanish (trivially, because 
${e^{(\tau_1)}}^\dagger e^{(\tau_2)}=\delta_{\tau_1\tau_2}$).
Therefore, within a quartet of staggered-fermion nonzero modes, 
continuum QCD is reproduced if $\zeta^\Gamma_{rs}$, $r\neq s$, also 
vanish as $a\to0$.

\begin{widetext}
Assuming Eq.~(\ref{eq:theTest}) holds, it is easy to see that 
    \begin{equation}
    \left\langle\prod_{f=1}^4\bar{\chi}\Gamma_{If}\chi(x_f)\right\rangle =
        \left\langle\prod_{\sigma>0}(\lambda_\sigma^2+m^2)^4
        \sum_{(ijkl)}
        \zeta^{\Gamma_1}_{ii}(x_1) 
        \zeta^{\Gamma_2}_{jj}(x_2) 
        \zeta^{\Gamma_3}_{kk}(x_3) 
        \zeta^{\Gamma_4}_{ll}(x_4) 
        \right\rangle_{\hspace{-0.5em}|Q|=1}
        + \mathrm{O}(m),
    \label{eq:24perms}
\end{equation}
\end{widetext}
where the sum runs over the $4!=24$ ways of choosing distinct
$(ijkl)$ from $\{-2,-1,1,2\}$.

Let us now discuss the $\mathbb{Z}_n$ (now $\mathbb{Z}_{4n_f}$) symmetry 
mentioned at the end of Sec.~\ref{subsec:contQCD}.
The anomalous~$\mathrm{U}_A(1)$ and most of the softly broken, 
nonanomalous $\mathrm{SU}_A(4n_f)$ emerge only in the continuum limit.
Some passages in Refs.~\cite{Creutz:2007rk,Creutz:2008nk} seem to 
assign a pertinent role to the $\mathrm{U}_\varepsilon(n_f)$ 
symmetries, which are exact even at nonzero~$a$.
These symmetries are a distraction at best:
the group $\mathrm{U}_\varepsilon(n_f)$ intersects with the relevant
$\mathbb{Z}_{4n_f}$, which is the center of $\mathrm{SU}(4n_f)$,
only at $-\openone_{4n_f}$.

\subsection{Rooted staggered sea}

With rooted staggered fermions, two changes are carried out.
In addition to using the rooted determinant~(\ref{eq:fourth-root}), 
the simple combinatorics of $\det_{(f,g)}G(x_f,y_g)$ must also 
change~\cite{Venkataraman:1997xi}.
For example, the taste-singlet pseudoscalar meson propagator is 
replaced with
\begin{equation}
    \langle\bar{\chi}\gamma^5_I\chi(x)
           \bar{\chi}\gamma^5_I\chi(y) \rangle_U \to -
        \case{1}{4} C(x,y) + \case{1}{16}D(x,y),
    \label{eq:etacd}
\end{equation}
where the connected and disconnected contributions are
\begin{eqnarray}
    C(x,y) &=& \left\langle \mathbb{D} \tr \left[ 
         \gamma^5_IG(x,y)\gamma^5_I G(y,x) \right] \right\rangle_U,
    \label{eq:Cdef} \\
    D(x,y) &=& \left\langle \mathbb{D} 
        \tr\left[\gamma^5_IG(x,x)\right] 
        \tr\left[\gamma^5_IG(y,y)\right] \right\rangle_U,
    \label{eq:Ddef}
\end{eqnarray}
where $\mathbb{D}$ is the rooted determinant~(\ref{eq:fourth-root}),
the trace is over color, and the translations implied by $\gamma^5_I$ 
act to the right (left) on the first (second) argument of~$G$.
The correlator in Eq.~(\ref{eq:etacd}) couples to the analog of the 
flavor-singlet $\eta'$ meson in QCD, 
and similar constructions hold for other taste-singlet bilinears.

The combinatoric factors in Eq.~(\ref{eq:etacd}) follow immediately 
from considering~\cite{Bernard:2006vv,sharpelat06,asklat07,%
Bernard:2007eh}
\begin{equation}
    \left\{ \Det_{n_f}\left[
        (\Dirac+m)\otimes\openone_4\right]\right\}^{n/4},
    \label{eq:RCT}
\end{equation}
where---inside the braces---one has four copies of $n_f$ 
noncontroversial fermions. 
Equation~(\ref{eq:RCT}) together with a source for a single species 
provide an engine to generate the combinatorics of rooting (in general):
to obtain $n$ species from 4, a term with $t$ traces over color 
receives a factor~\cite{Venkataraman:1997xi} \hspace*{-0.7em}
\begin{equation}
    \left(-\frac{n}{4}\right)^t. \label{eq:n/4}
\end{equation} \hspace*{1em}

For Eq.~(\ref{eq:RCT}) to be relevant to staggered fermions, the 
dynamics must ensure Eq.~(\ref{eq:yigal}) and, in particular, 
Eq.~(\ref{eq:theTest}), as we now show.
The single-flavor determinant becomes (for $|Q|=1$)
\begin{equation}
    \prod_{i=1}^2(\lambda_i^2+m^2)^{1/4}\,
        \prod_{s>0}(\lambda_s^2+m^2)^{1/4}.
    \label{eq:rootDet}
\end{equation}
Neglecting $\lambda_i$ compared to $m$ again, 
the first product collapses to $|m|$.
The near-zero-mode contributions are then
\begin{eqnarray}
    C(x,y) & = & \sum_{i,j} \left\langle \frac{|m|\mathbb{D}'}{m^2}
        \zeta^{\gamma^5}_{ij}(x) \, \zeta^{\gamma^5}_{ji}(y)
        \right\rangle_{\hspace{-0.4em}|Q|=1} ,
    \label{eq:Czero} \\[1.0em]
    D(x,y) & = & \sum_{i,j} \left\langle \frac{|m|\mathbb{D}'}{m^2}
        \zeta^{\gamma^5}_{ii}(x) \, \zeta^{\gamma^5}_{jj}(y)
        \right\rangle_{\hspace{-0.4em}|Q|=1},
    \label{eq:Dzero}
\end{eqnarray}
where $\mathbb{D}'$ is the $s>0$ product in Eq.~(\ref{eq:rootDet}),
and $i,j\in\{-2,-1,1,2\}$.
If Eq.~(\ref{eq:theTest}) holds, then the sum in Eq.~(\ref{eq:Czero}) 
collapses to terms with $i=j$, apart from lattice artifacts.
Thus, $C$ has 4 contributions singular in~$1/|m|$, whereas $D$ has 16.
With the correct combinatoric factors, they cancel.

It is, perhaps, instructive to exhibit the three-point correlator.
Assuming Eq.~(\ref{eq:theTest}) and homing in on the zero-mode 
contributions,
\begin{widetext}
\begin{eqnarray}
    \bar{\chi}\Gamma_{I1}\chi(x_1)
    \bar{\chi}\Gamma_{I2}\chi(x_2) 
    \bar{\chi}\Gamma_{I3}\chi(x_3) & \to & {  } -
        \case{1}{4}\{\tr[\Gamma_{I1}G(x_1,x_2)\Gamma_{I2}G(x_2,x_3)
            \Gamma_{I3}G(x_3,x_1)] + {\rm 1~perm}\} \nonumber \\ & &  {  } +
        \case{1}{4^2}\{\tr[\Gamma_{I1}G(x_1,x_1)]\tr[\Gamma_{I2}G(x_2,x_3)
            \Gamma_{I3}G(x_3,x_2)] + {\rm 2~perms}\} \nonumber \\ & & {  } -
       \case{1}{4^3}\,\tr[\Gamma_{I1}G(x_1,x_1)]\tr[\Gamma_{I2}G(x_2,x_2)
            \tr[\Gamma_{I3}G(x_3,x_3)] \label{eq:3mess} \\ & \to & 
            \frac{|m|}{m^3} \left(-2+3-1\right)
            \zeta^{\Gamma_1}(x_1)
            \zeta^{\Gamma_2}(x_2)
            \zeta^{\Gamma_3}(x_3)
            \zeta^{\Gamma_4}(x_4),
\end{eqnarray}
\end{widetext}
where $|m|$ comes from the rooted determinant.
Here sums over the four staggered-fermion near-zero modes cancel the 
explicit factors of $\quarter$.
The $|m|/m^2$ contributions cancel in a similar way.
Earlier work~\cite{Bernard:2006vv,sharpelat06,Bernard:2007eh}, tacitly 
assumed Eq.~(\ref{eq:theTest}); in particular, Ref.~\cite{Bernard:2007eh}
shows how the combinatorics work for higher-point 't~Hooft-vertex effects.

In Refs.~\cite{Creutz:2007rk,Creutz:2008nk,Creutz:2008kb}, 
Creutz disregards the cancellations stemming from the correct weighting 
of different contributions to flavor-taste-singlet correlators.
He considers more primitive combinations, like any individual line in 
Eq.~(\ref{eq:3mess}), which clearly are singular as $m\to0$.
He then draws two incorrect inferences.
First, he claims that the normal cancellations connected with Pauli 
statistics cannot arise.
Combining the correct weights with the assumption (tested below) 
Eq.~(\ref{eq:theTest}), one sees that this is not the case.
The outcome is not too mysterious: as taste emerges into a species-like 
quantum number, the correct set of correlators averages over them.

The other misstep is to assert that the $\mathbb{Z}_{4n_f}$ symmetry of 
the unrooted 't~Hooft vertex cannot be reduced to $\mathbb{Z}_{n_f}$.
This is incorrect, because, while the rooted determinant clearly 
retains the symmetries of the unrooted determinant, the 't~Hooft vertex 
stems from the combined behavior of determinant and valence propagators.
The replacement of the combinatoric factors of traces 
with~(\ref{eq:n/4}) effectively projects the symmetry emerging in the 
chiral limit from $\mathrm{SU}(4n_f)$ [taking $n=1$ in~(\ref{eq:n/4})]
to $\mathrm{SU}(n_f)$.
Since the relevant symmetry is the center of the emergent flavor symmetry, 
one has $\mathbb{Z}_{n_f}$.

Many of these points have been made before~\cite{Bernard:2006vv,%
sharpelat06,Bernard:2007eh,Bernard:2008gr}, but until now it  
has always been assumed that the tastes decouple as posited in 
Eq.~(\ref{eq:theTest}).
(Reference~\cite{asklat07} noted the necessity of this assumption.)
Our approach can easily be extended to taste-nonsinglet flavor-singlet 
correlators, and the properties of the local overlaps with nonsinglet 
$\Gamma$s will not enjoy the cancellation.
We shall now compute the $\zeta^\Gamma_{ij}$ nonperturbatively, to find 
out whether the tastes couple to each other at the strong scale 
$\Lambda_{\rm QCD}$ or at the cutoff scale~$a^{-1}$.

\section{Numerical Results}
\label{sec:results}

In this section we present our numerical methods and results.
First we explain the motivation for studying improved discretizations 
and why it suffices to compute their eigenvalue spectrum on quenched
gauge fields.
We present results for eigenvalues and chirality with the HISQ 
action.
These results are qualitatively similar to those obtained 
with the Asqtad and Fat7$\times$Asqtad actions in 
Refs.~\cite{oureigsshort,oureigslong}, so we focus here on $|Q|=1$.
Then we show results for the overlaps, $\zeta_{ij}^\Gamma$, defined 
in Eq.~(\ref{eq:zetaDef}), and test their behavior as a function 
of lattice spacing against Eq.~(\ref{eq:theTest}). 
Finally we discuss correlators for 
mesons of different $J^{P}$ in turn, starting 
with pseudoscalars where the issues are particularly important. 
Taken together, our results demonstrate how the behavior of the 
different contributions from near-zero and nonzero 
modes matches that expected in the continuum. 

\subsection{Methods}
\label{subsec:methods}

In this paper, we use the same ensembles of SU(3) gauge fields as in 
earlier studies of eigenvalues and chirality~\cite{oureigsshort,oureigslong}. 
They are quenched configurations, omitting the effects of sea quarks. 
They are generated with a Symanzik-improved gauge action, 
so that the tree-level $a^2$ errors are removed~\cite{Weisz:1982zw},
and tadpole-improved couplings in this action, so that loop corrections 
are reduced~\cite{Lepage:1992xa}.
Three different values of the gauge coupling are used, giving three 
widely separated values of the lattice spacing, covering the range of 
typical unquenched lattice-QCD calculations~\cite{milcreview},
so our results should pertain directly to them.
At the middle value of the three lattice spacings, we have three 
different-sized lattices in order to check the volume dependence. 
The parameters for the configurations are given in Table~\ref{tab:configs}. 
\begin{table}[tbp]
\centering
    \caption{Details of the gauge configurations used: 
        $\beta$ is the bare gauge coupling, 
        $a$ the lattice spacing~\cite{australians},
        $V$ the spacetime volume in lattice units, and
        $L$ the linear size in physical units. 
        The final column gives the number of configurations in each 
        ensemble with~$|Q|=1$. We refer to set~1 as having a ``coarse'' 
        lattice spacing, sets~2, 3, and 4 as ``intermediate'',
        and set~5 as ``fine''.}
    \label{tab:configs}
    \begin{tabular}{cccccc}
    \hline\hline
    Ensemble & $\beta$ & $a$ (fm) &   $V$  & $L$ (fm) & $\#\{|Q|=1\}$ \\ \hline
        1    &   4.6   &  0.125   & $12^4$ &   1.50   &    294        \\
        2    &   4.8   &  0.093   & $12^4$ &   1.12   &    806        \\
        3    &   4.8   &  0.093   & $16^4$ &   1.49   &    424        \\
        4    &   4.8   &  0.093   & $20^4$ &   1.86   &    288        \\
        5    &   5.0   &  0.077   & $20^4$ &   1.54   &    430        \\
    \hline\hline
    \end{tabular}
\end{table}

It is sufficient to study these issues in the quenched approximation, 
because we aim to test a structural property of staggered fermions in 
fixed-$Q$ sectors.
In particular, omitting the determinant decouples Creutz's infrared 
concerns from others' ultraviolet concern that taste breaking remains 
in the continuum limit.
If the eigenvectors satisfy Eq.~(\ref{eq:theTest}) strongly enough, then
the 't~Hooft vertex and the consequent cancellation of mass-singular
contributions to the connected and disconnected flavor-singlet meson 
correlators should work out in general.
We shall see that this is the case.

With the original staggered-fermion action, Eq.~(\ref{eq:Sstag}), 
the interaction connects adjacent sites.
Very large discretization errors arise in a wide range of observables, 
washing out the expected quartet structure in the eigenvalue spectrum.
These discretization errors have been traced to taste-changing 
interactions from gluons with one or more components of momentum 
$p_\mu\approx\pi/a$~\cite{Lepage:1996jw}.
Because of the gluon exchange, these effects are formally of order 
$\alpha_sa^2$, i.e., $\alpha_s$ times smaller than normal 
discretization effects~\cite{Naik:1986bn}.
In order to reduce these taste-changing effects, it is necessary to 
smear the gauge field, replacing $U_\mu$ and $U^\dagger_\mu$ in 
Eq.~(\ref{eq:Sstag}) with sums of products of link matrices tracing out 
more complicated paths between $x$ and $x\pm\hat{\mu}a$%
~\cite{Lepage:1996jw,Lagae:1998pe,Orginos:1999cr}.

Several staggered-fermion actions have been developed along these lines.
The Asqtad~\cite{gplasqtad} and Fat7$\times$Asqtad~\cite{oureigslong} 
actions exhibit a reduction, relative to the nearest-neighbor action in
Eq.~(\ref{eq:Sstag}), in splittings between pseudoscalar mesons 
of different taste~\cite{milcdecay04,oureigslong}. 
Similarly, with these actions the quartet structure of the eigenvalue 
spectrum more clearly emerges~\cite{oureigsshort,oureigslong}.  

Here we have calculated low-lying eigenvalues and eigenvectors for the 
highly improved staggered-quark (HISQ) action~\cite{hisq}, reusing the 
same gauge-field configurations.
The HISQ action supersedes the Fat7$\times$Asqtad 
action; it is essentially the same but corrects 
the smearing at the second stage to remove fully the 
discretization errors that the smearing introduces. 
As we shall see in Sec.~\ref{subsec:evals} this 
change makes only a small effect. The  
eigenvalue quartet structure is very clear with the HISQ action, which
is reflected in other properties that,
by now, have been thoroughly tested: small pseudoscalar 
mass splittings and small discretization errors, even 
for heavy 
quarks~\cite{hisq,Follana:2007uv,newfds,Bazavov:2010ru}. 

Appendix~\ref{app:actions} provides explicit equations for the smeared 
actions.

We use the Lanczos algorithm to calculate the low-lying 
eigenvalues, $i\lambda$, of the anti-Hermitian massless HISQ Dirac 
operator, \Dhisq, defined implicitly in Eq.~(\ref{eq:SHISQ}).
Owing to its red-black checkerboard structure, the calculations can be 
simplified by using the Hermitian positive semi-definite operator 
$-\Dhisq^2$, projected onto either the red (even) or black (odd) sites 
of the lattice.
This yields $\lambda^2$, from the smallest values upwards, and 
eigenvector~$f$, on the chosen half of the lattice.
The eigenvalues of $\Dhisq$ are then $\pm i\lambda$, and the
corresponding eigenvector on the other half of the lattice is 
$\pm\Dhisq f/i\lambda$.
This construction automatically implements the requirement that the 
eigenvectors corresponding to eigenvalues $i\lambda$ and $-i\lambda$ 
are simply related by multiplication with $\varepsilon(x)$.
Thus, on the odd (even) sites, the $-s$th eigenvector is opposite 
(same) in sign as the $+s$th eigenvector.

\subsection{Eigenvalues and chirality with HISQ}
\label{subsec:evals}

\begin{figure}[bp]
    \includegraphics[width=0.48\textwidth]{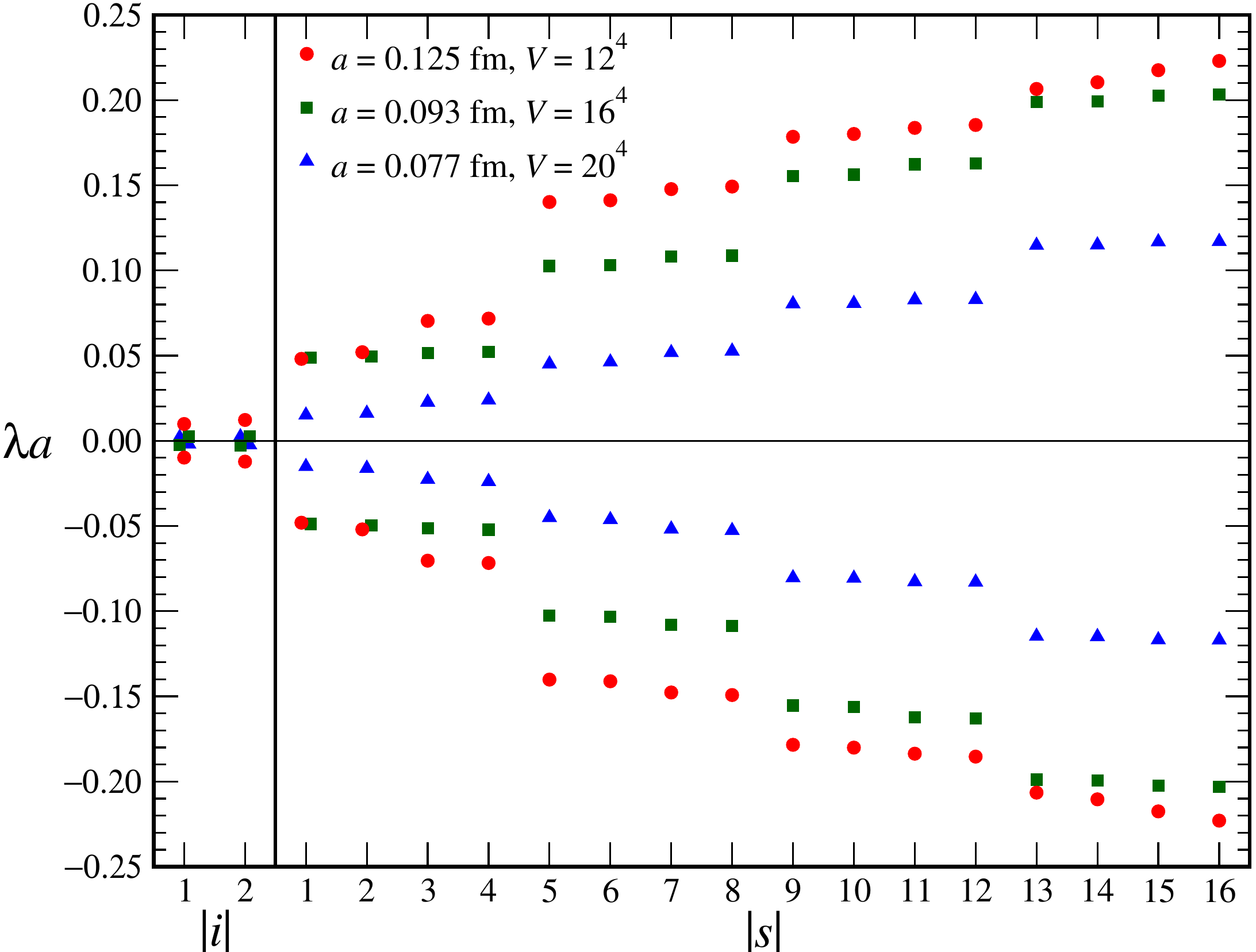}
    \caption{The four near-zero eigenvalues (left panel) and the
        16 lowest-lying nonzero pairs of \Dhisq\ eigenvalues on a typical 
        $|Q|$=1 configuration from sets 1 (red circles), 
        3 (green squares), and 5 (blue triangles).
        For clarity, some modes are offset horizontally.}
    \label{fig:eigs}
\end{figure}

Figure~\ref{fig:eigs} shows the four near-zero eigenvalues as well as the
16~pairs of nonzero eigenvalues of \Dhisq\ with smallest $|\lambda|$, 
obtained on typical $|Q|=1$ configurations from ensembles labeled 
1 (coarse), 3 (intermediate), and 5 (fine) in Table~\ref{tab:configs}.
These lattices have similar physical volume but lattice spacing varying 
from 0.125 to 0.077~fm.
The anticipated picture is unmistakable: four (and only four) very 
small eigenvalues appear, followed by distinct quartets.
As the lattice spacing decreases, eigenvalues within a quartet come 
closer and closer to being degenerate, typically by forming two 
close-by almost degenerate pairs.
The near-zero modes are typically, on these lattices, at least an order of 
magnitude smaller than the low-lying nonzero modes. 

The Lanczos algorithm also gives the eigenvectors corresponding to 
these eigenvalues.
Normalizing them to have modulus~1, we compute the chirality $\mathcal{X}$ 
in Eq.~(\ref{eq:chirality}), using the smeared $W_\mu$ matrices 
[Eq.~(\ref{eq:W4HISQ})] instead of $U_\mu$.
Reference~\cite{oureigslong} showed that it makes little qualitative 
difference to the results whether the original $U_\mu$, 
Asqtad $V_\mu$ [Eq.~(\ref{eq:V4Fat7})], 
or Fat7$\times$Asqtad $\check{W}_\mu$ [Eq.~(\ref{eq:W4Fat7Asqtad})] are used.
The numerical values of 
the chirality may change, but the picture remains qualitatively the same.

Because lattice artifacts break the taste-singlet symmetry, the 
chirality defined in Eq.~(\ref{eq:chirality}) takes values that 
are not simply 1 and~0~\cite{Vink:1988ss}. 
References~\cite{oureigsshort,oureigslong} found, however, that it is 
easy, especially with improved gauge and staggered-fermion actions, to 
separate the near-zero modes with relatively large chirality, 
close to 1, from the other modes with chirality close to~0. 
The number of near-zero modes defined this way agrees with the index 
theorem, Eq.~(\ref{eq:staggered-index}), and pure-gauge definitions of 
the topological charge. 
The agreement between the index and the gauge-field topological charge 
improves as the lattice spacing gets smaller. 
On the $a=0.077$~fm ensemble, the disagreement for Asqtad and 
Fat7$\times$Asqtad is just 2\%~\cite{oureigslong}, which is no worse 
than the ambiguity between different gluonic definitions. 
For this paper, we therefore simply take the index to classify the 
topology. 

\begin{figure}[bp]
    $|\mathcal{X}|$ \hfill { } \\[-3.6em] 
    { } \hfill \includegraphics[width=0.45\textwidth]{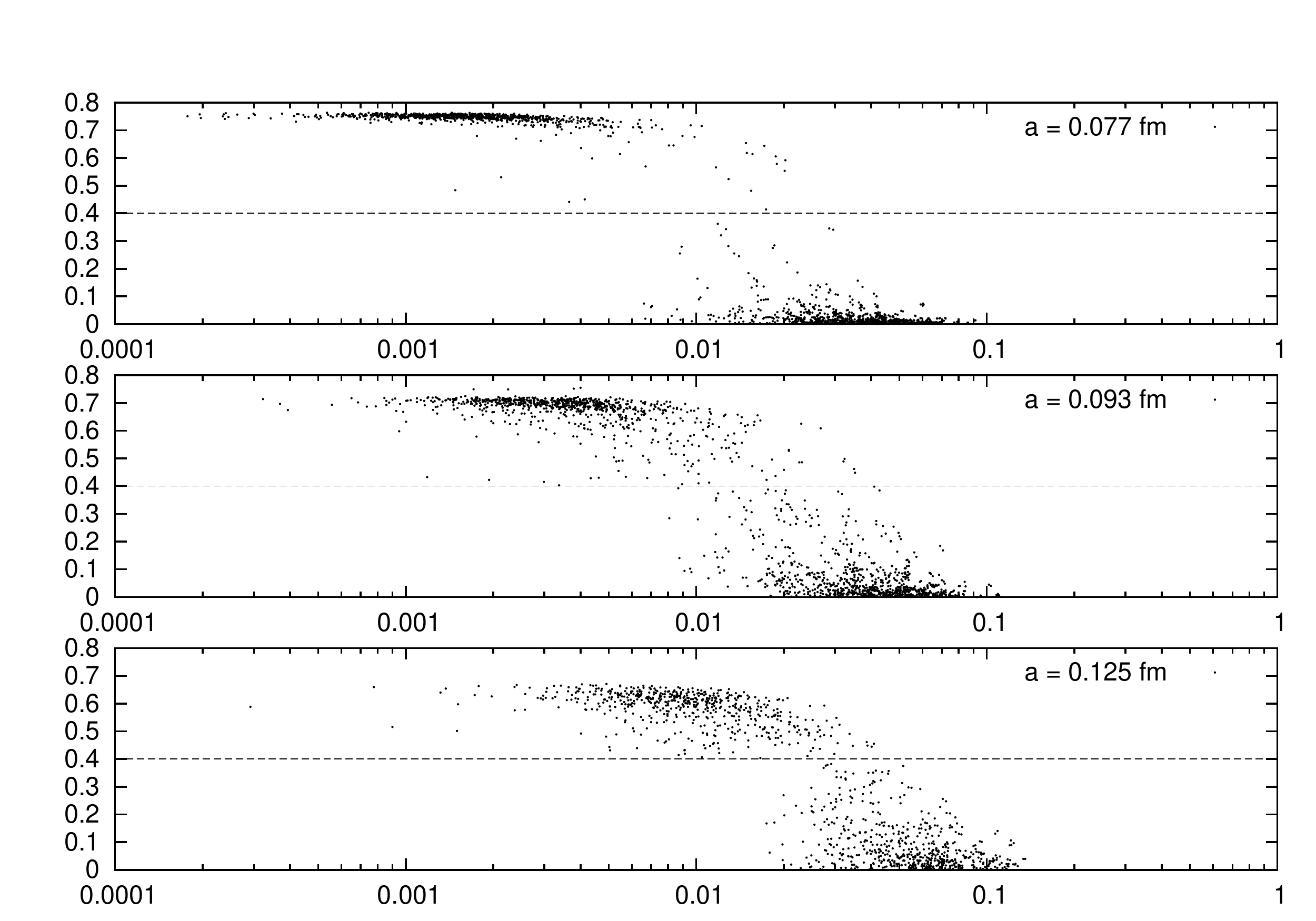} \\
    { } \hfill $\lambda a$\hspace*{2em}
    \caption{The absolute value of the chirality $|\mathcal{X}|$ plotted 
        against eigenvalue, $\lambda a$, in lattice units for the four lowest 
        (positive) eigenvalues for the $|Q|$ = 1 configurations in 
        ensembles 1, 3, and 5.
        The dotted line on each graph indicates $|\mathcal{X}|=0.4$,
        which is used to separate large and 
        small chirality in determining the value of $Q$ (see text).}
    \label{fig:evchi}
\end{figure}

Figure~\ref{fig:evchi} shows the chirality values for the 
HISQ action versus eigenvalues on 
all configurations defined to be of topological charge $\pm 1$ via the index.
To reduce clutter, Fig.~\ref{fig:evchi} shows only the two near-zero modes 
and the two lowest-lying nonzero modes.
(Because $\mathcal{X}_{-s}=\mathcal{X}_s$, we count only the 
positive-$\lambda$ modes here.)
One sees a clear separation of large and small chirality values, 
especially so on the finer configurations.
Although the values corresponding to the maximum chirality do
not change very markedly from coarse to finer lattices, the 
spread of results becomes much narrower.
The small chirality values, corresponding to nonzero eigenmodes, fall 
rapidly to zero with lattice spacing. 
We take $\mathcal{X}>0.4$ (drawn on the graphs) to indicate 
large chirality and then count the number of eigenvalues (with 
positive $\lambda$) that have large chirality.
Configurations with two (positive-$\lambda$) large-chirality modes 
are taken to be $|Q| = 1$ configurations. 
Table~\ref{tab:configs} lists the number of such configurations 
for each ensemble. 
More general scatter plots with results at $|Q|>1$ and the Asqtad and 
Fat7$\times$Asqtad actions have been given in Ref.~\cite{oureigslong}, 
and with HISQ look very similar.

\subsection{Results for \boldmath$\zeta^\Gamma_{ij}$}
\label{subsec:zeta-plots}

\begin{figure}[p]
    \hskip -10pt
    \includegraphics[width=0.5\textwidth]{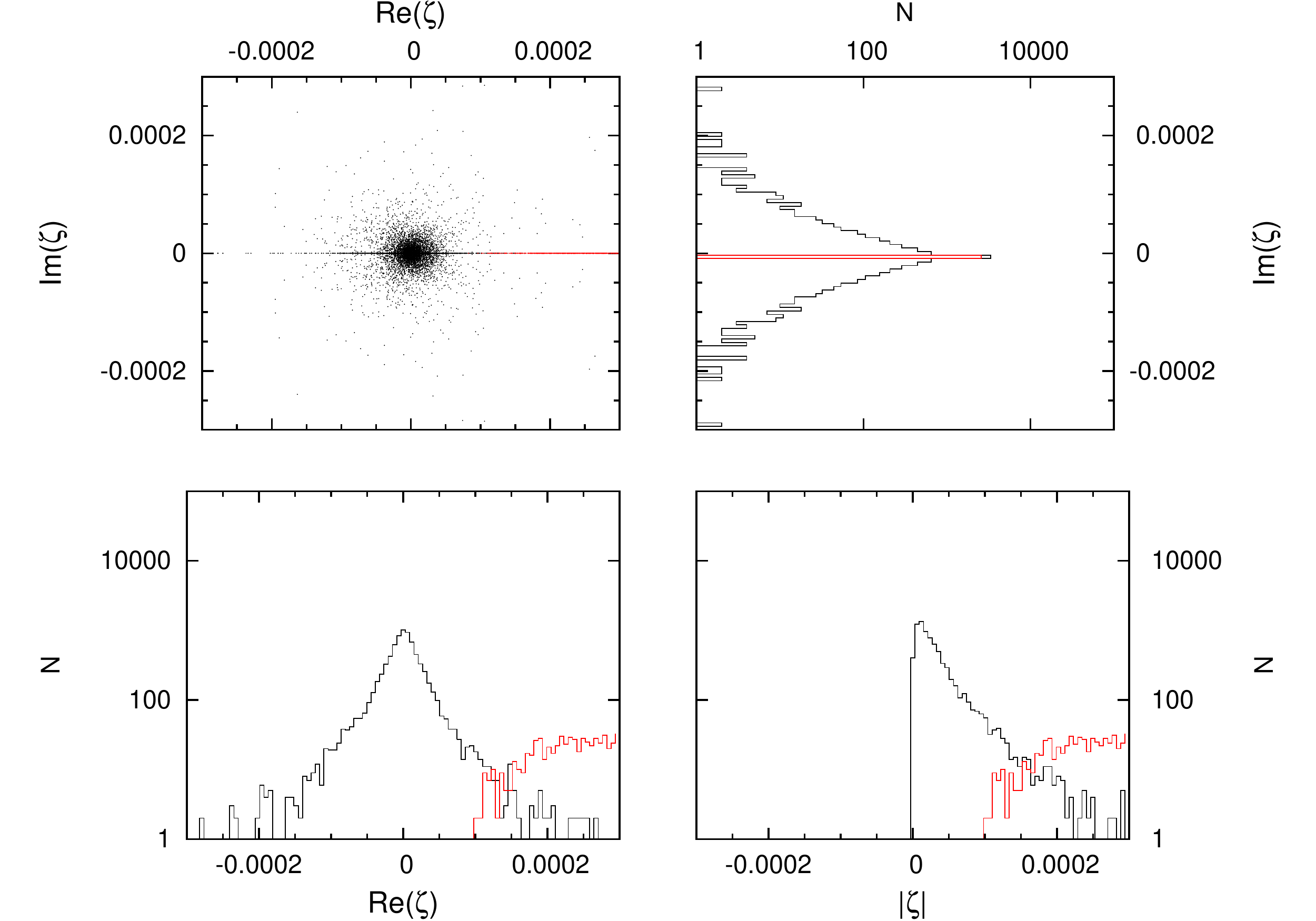} \\[1em]
    \hskip -10pt
    \includegraphics[width=0.5\textwidth]{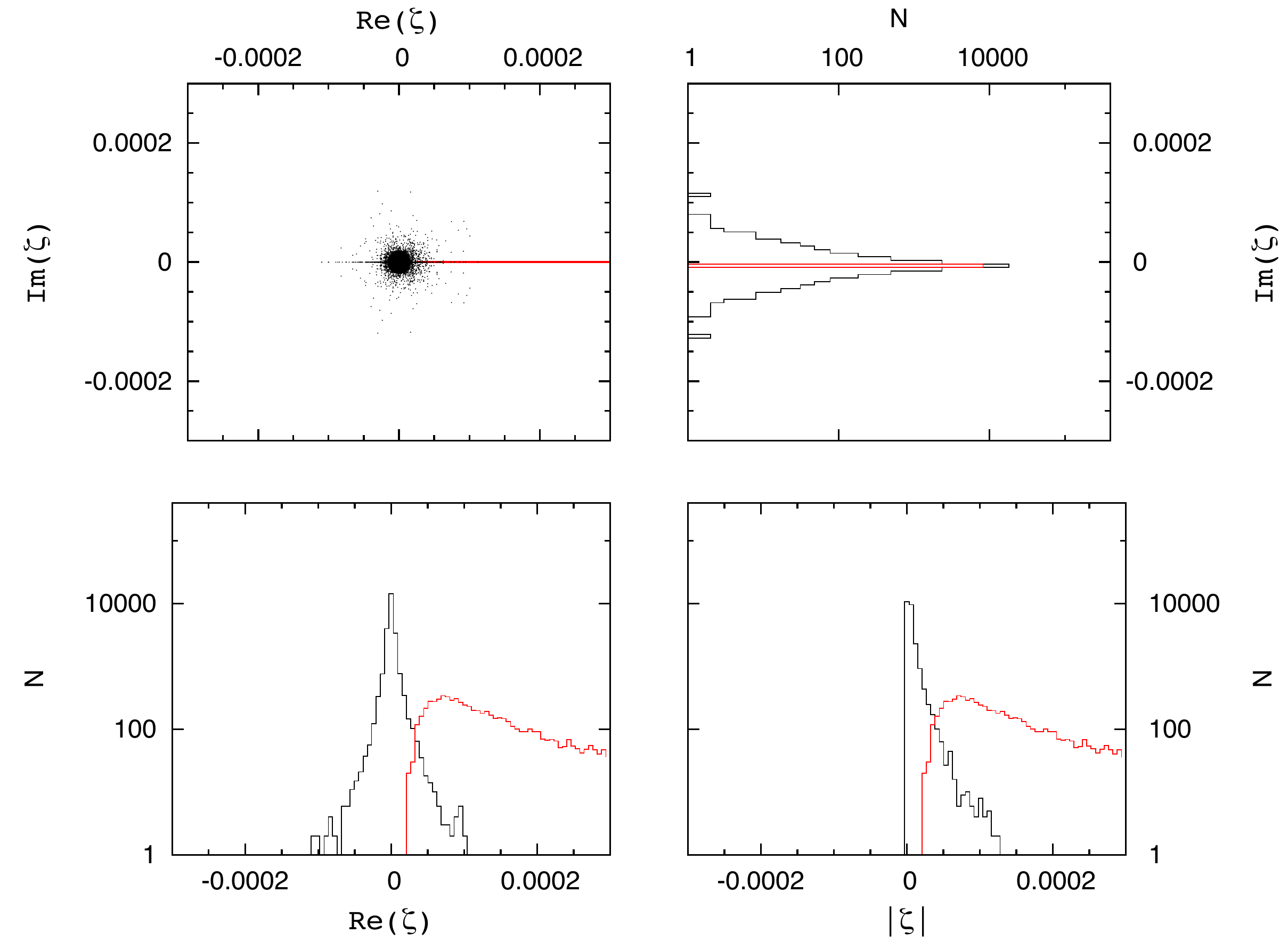} \\[1em]
    \hskip -10pt
    \includegraphics[width=0.5\textwidth]{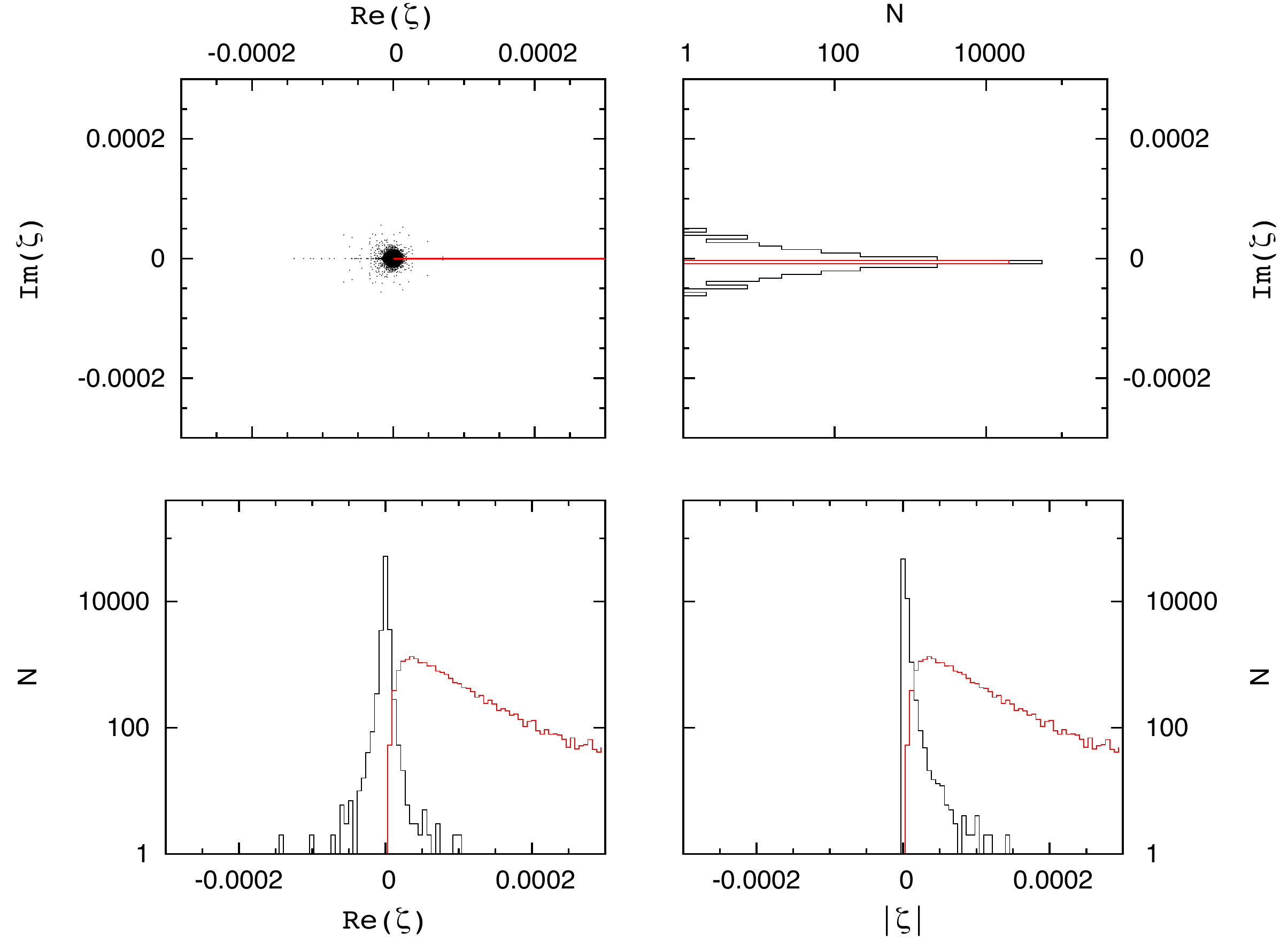} 
    \caption{$\zeta^1_{ij}$ on coarse (top), intermediate
        (middle), and fine (bottom) $|Q|=1$ gluon field configurations, 
        with $j=i$ (red) and $j\neq i$ (black), $i,j = \pm 1, \pm 2$. 
        Note the logarithmic $y$-axis scale for the histograms.}
    \label{fig:zetaS}
\end{figure}

\begin{figure}[p]
    \hskip -10pt
    \includegraphics[width=0.5\textwidth]{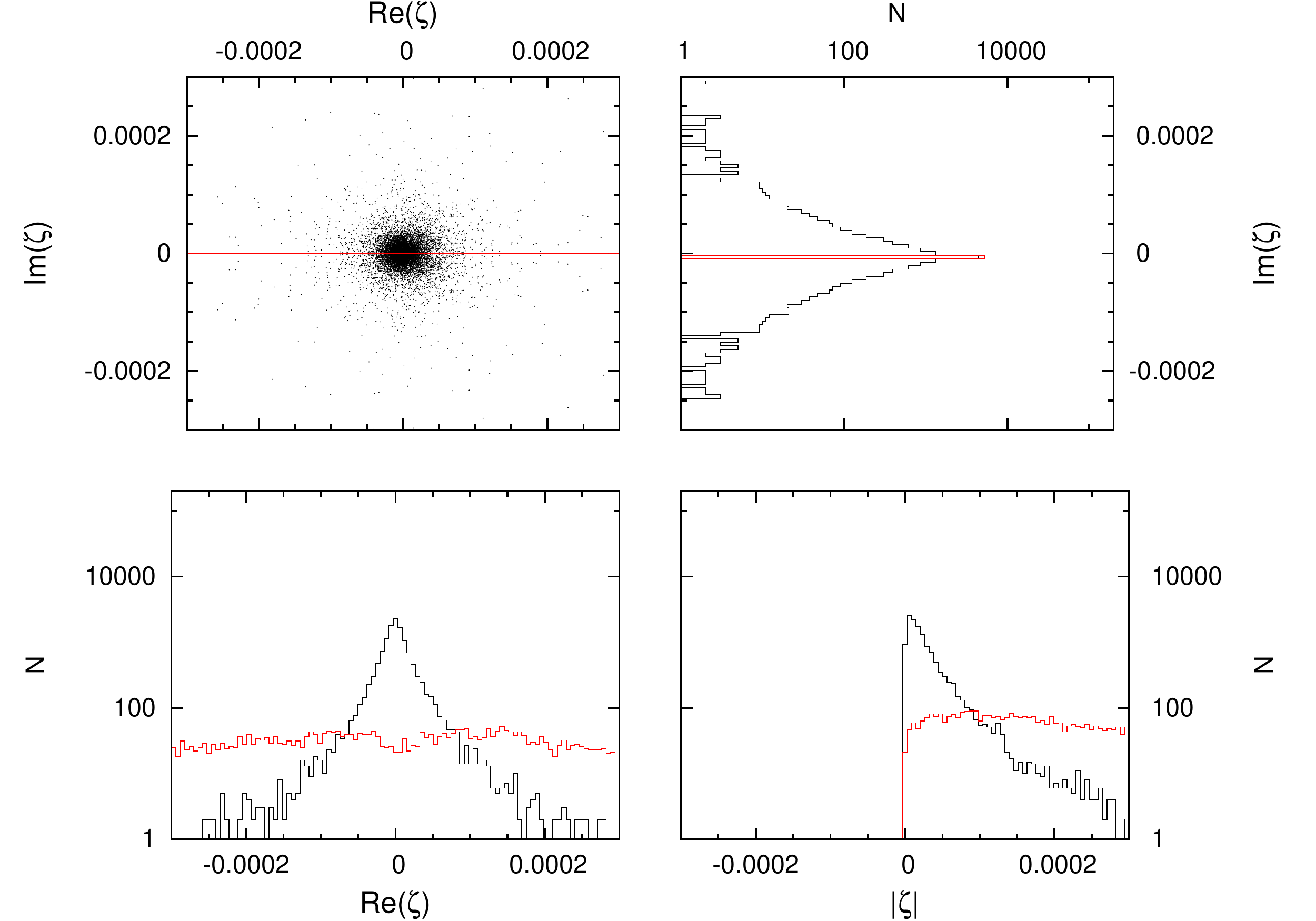} \\[1em]
    \hskip -10pt
    \includegraphics[width=0.5\textwidth]{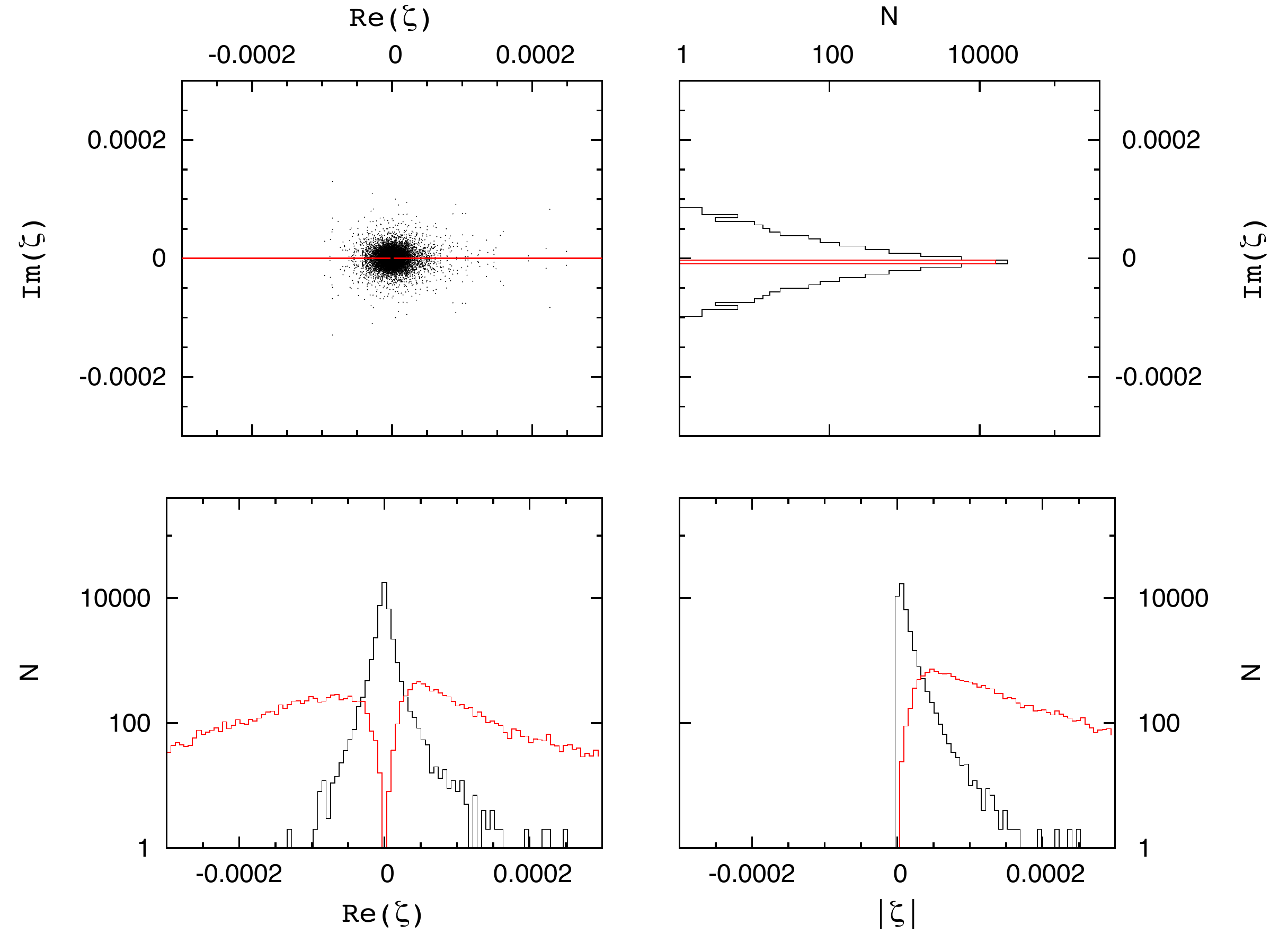} \\[1em]
    \hskip -10pt
    \includegraphics[width=0.5\textwidth]{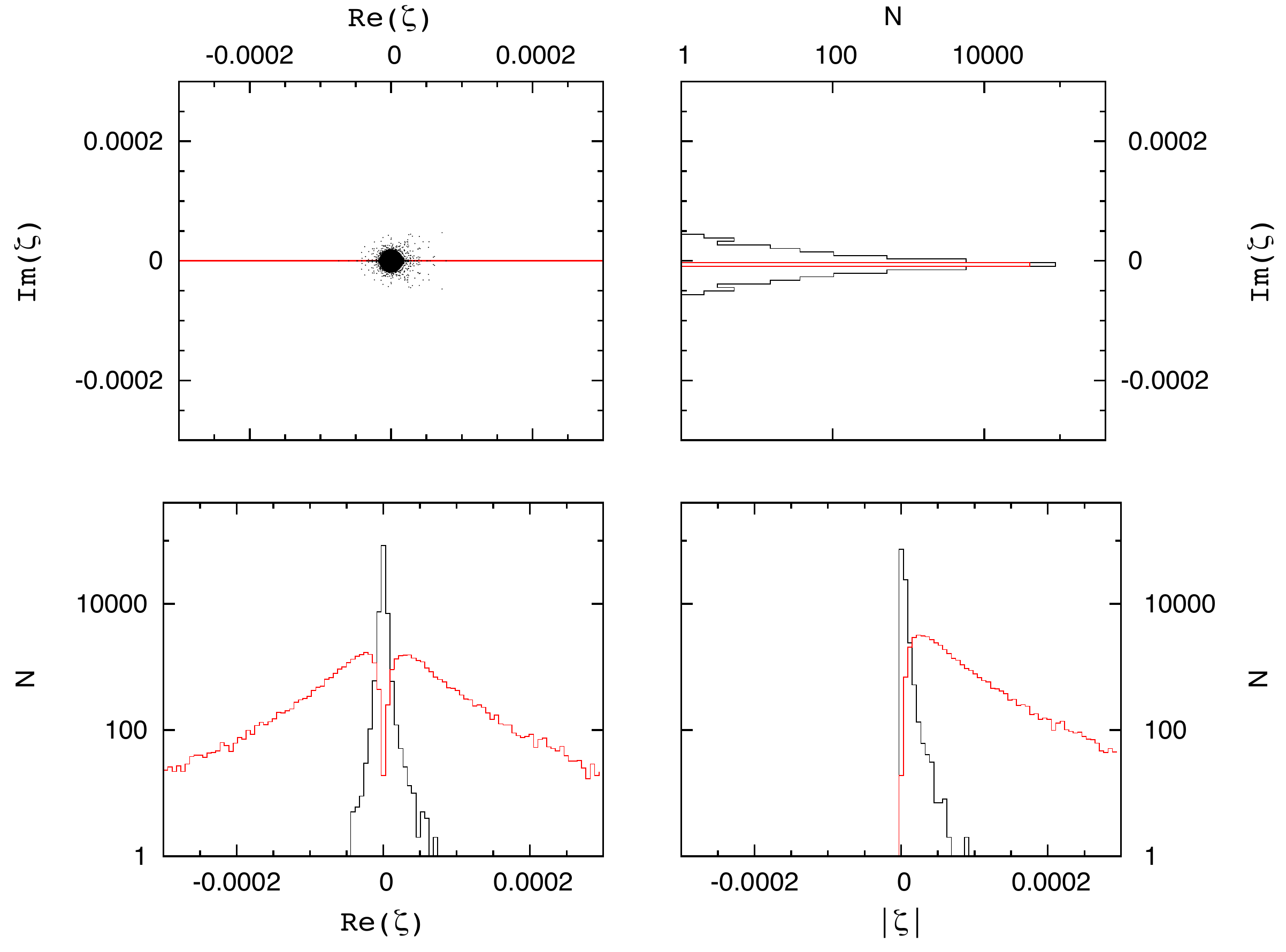} 
    \caption{$\zeta^{\gamma^5}_{ij}$ on coarse (top), intermediate
        (middle), and fine (bottom) $|Q|=1$ gluon field configurations, 
        with $j=i$ (red) and $j\neq i$ (black), $i, j = \pm 1, \pm 2$. 
        Note the logarithmic $y$-axis scale for the histograms.}
    \label{fig:zetaP}
\end{figure}

\begin{figure}[p]
    \hskip -10pt
    \includegraphics[width=0.5\textwidth]{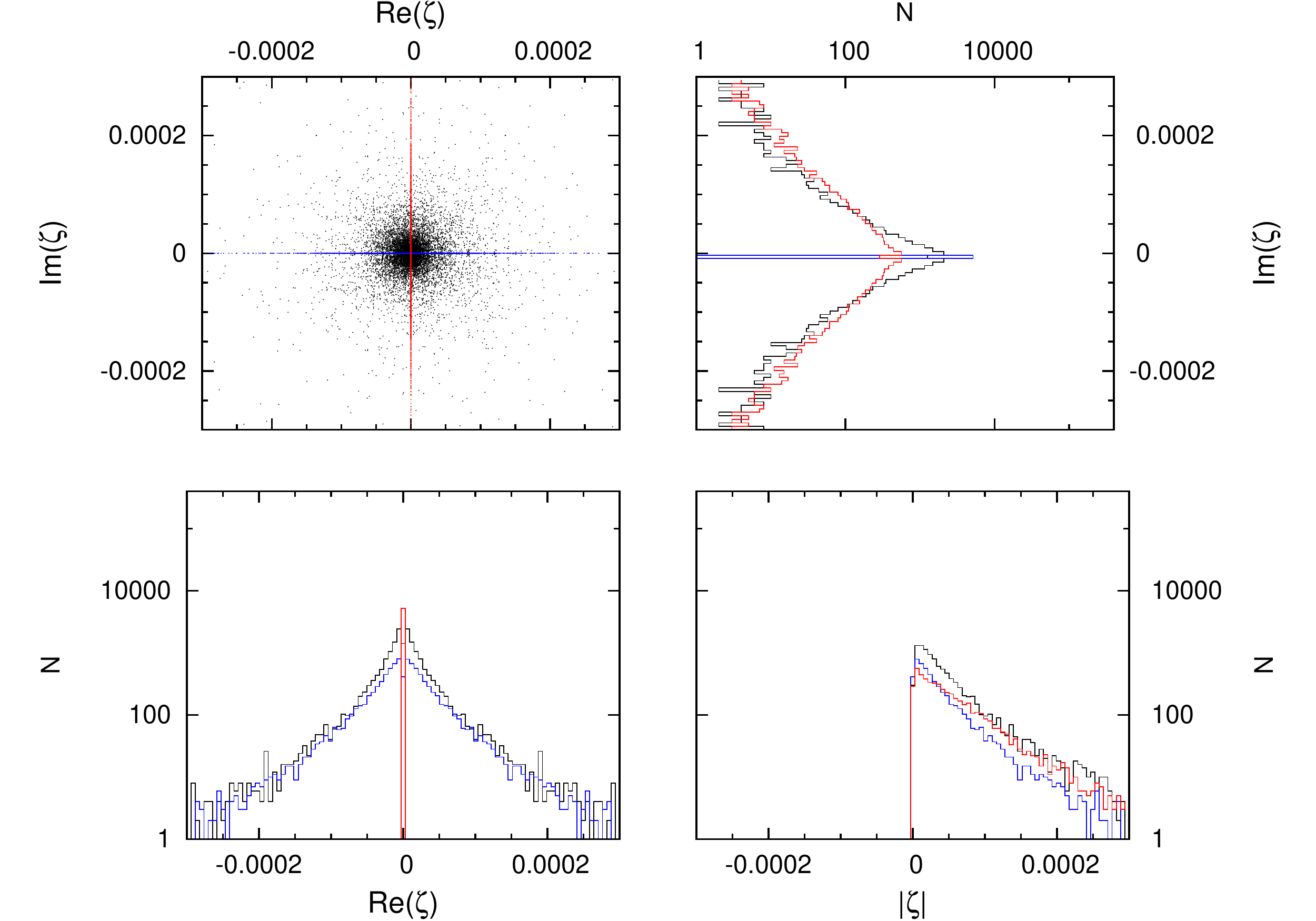} \\[1em]
    \hskip -10pt
    \includegraphics[width=0.5\textwidth]{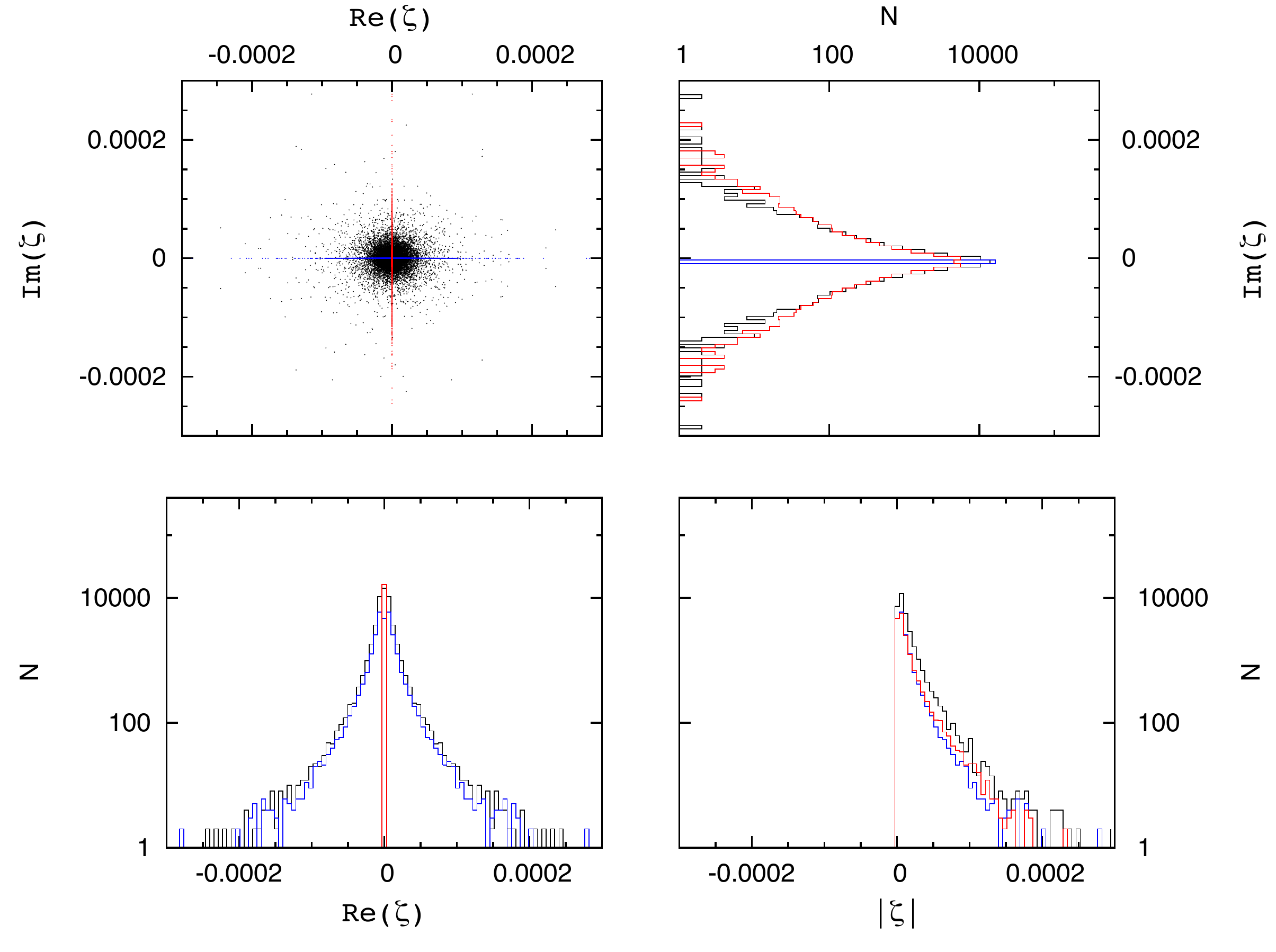} \\[1em]
    \hskip -10pt
    \includegraphics[width=0.5\textwidth]{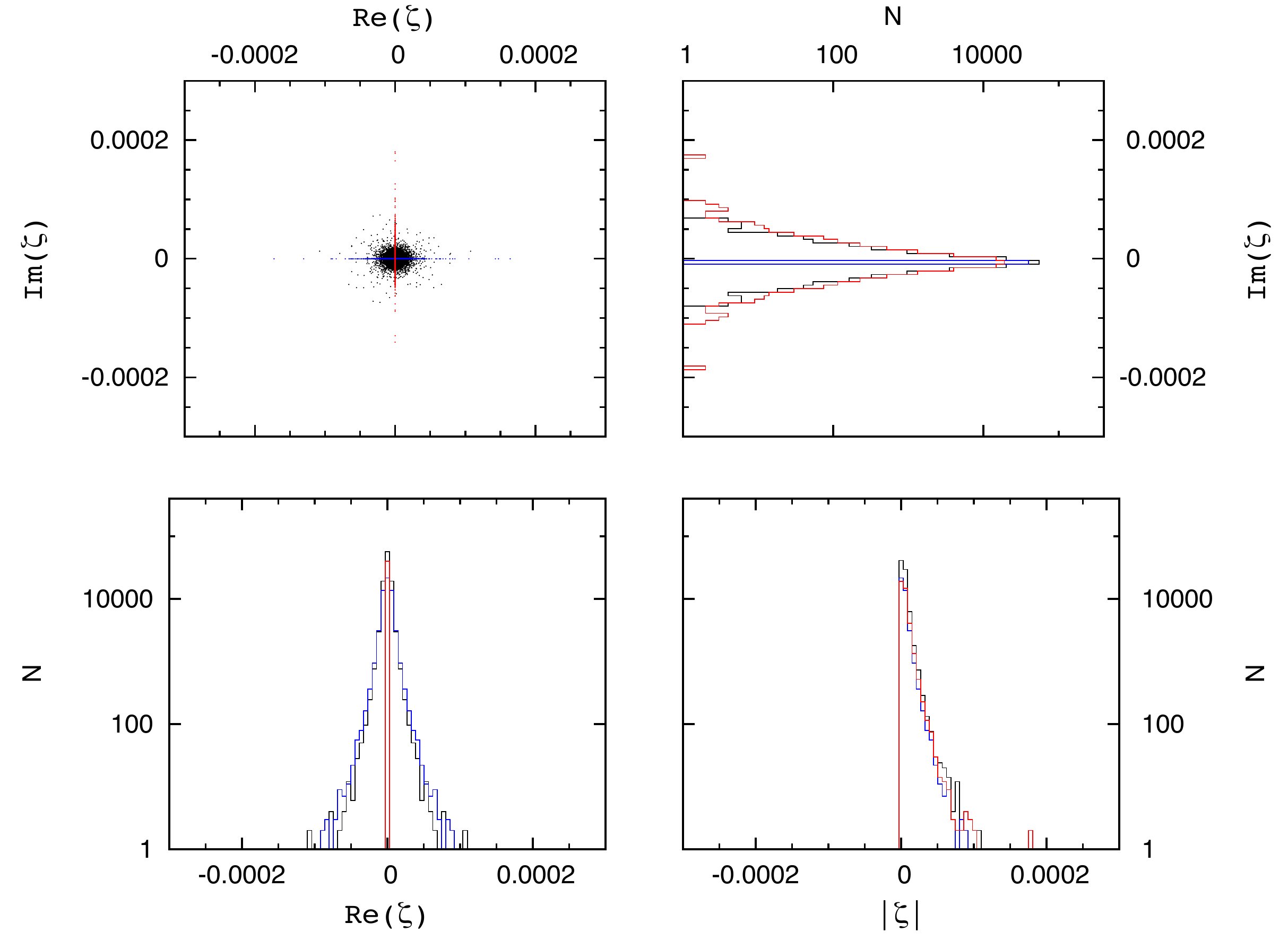} 
    \caption{$\zeta^{\gamma^\mu}_{ij}$ on coarse (top), intermediate
        (middle), and fine (bottom) $|Q|=1$ gluon field configurations, 
        with $j=i$ (red), $j=-i$ (blue), and $|j|\neq|i|$ (black), 
        $i, j = \pm 1, \pm 2$. 
        Note the logarithmic $y$-axis scale for the histograms.}
    \label{fig:zetaV}
\end{figure}

Using the eigenvectors determined in the previous 
section we now go on to look in more detail at the 
overlaps of the near-zero-mode eigenvectors that 
are relevant to the 't~Hooft vertex. 
Figures~\ref{fig:zetaS}--\ref{fig:zetaV} show scatter plots and 
histograms of the $\zeta^\Gamma_{ij}$ distributions for 
$\Gamma=1$, $\gamma^5$, and $\gamma^\mu$, and $i, j = \pm 1, \pm 2$ 
on one or two 
configurations, ranging over all~$x$.
Each figure displays this information, from top to bottom, 
for the coarse ($a=0.125$~fm),
intermediate ($a=0.093$~fm), and
fine ($a=0.077$~fm) lattices,
at (nearly) fixed physical volume (sets 1, 3, and 5).
The four panels in each case show 
the scatter of $\zeta^\Gamma_{ij}$ in the complex plane (upper left), 
the histogram for $\Re\zeta^\Gamma_{ij}$ (lower left),
the histogram for $\Im\zeta^\Gamma_{ij}$ (upper right), and
the histogram for  $|\zeta^\Gamma_{ij}|$ (lower right).
The number of points in the histograms for each set is the 
lattice volume, $V$, of Table~\ref{tab:configs}. 
Note the logarithmic scale on the histogram plots. 
Red points and lines denote diagonal $\zeta^\Gamma_{ii}$,
and black off-diagonal $\zeta^\Gamma_{ij}$ ($j\neq i$).
In the case of the vector overlap, $\Gamma=\gamma^\mu$,
we separate the off-diagonal $\zeta^\Gamma_{ij}$ into two.  
Black is reserved for $|j|\neq|i|$ and
the case of $j=-i$ is shown in blue.

The most striking feature for the scalar (Fig.~\ref{fig:zetaS}) and 
pseudoscalar (Fig.~\ref{fig:zetaP}) is how different the diagonal and 
off-diagonal distributions are.
The diagonal scalar overlap $\zeta^1_{ii}$ is a sum of absolute 
squares, so it is real and positive.
Because $f_{-i}(x) = \varepsilon(x) f_i(x)$
and the taste-singlet scalar operator is local,
$\zeta^1_{i,-i}$ is equal to $\zeta^1_{i,i}$ on even sites but 
real and negative on odd sites. Upon averaging over a 
hypercube in Eq.~(\ref{eq:scalar}), cancellations render 
$\zeta^1_{i,-i}$ relatively small.
It is visible on Fig.~\ref{fig:zetaS} 
as a black line stretching along the negative real 
axis; the positive part being invisible underneath the 
red line for $\zeta^1_{i,i}$. 
The off-diagonal ($|j|\neq |i|$) scalar overlap $\zeta^1_{ij}$ is  a
complex number of random phase.
The width of all the histograms falls going down the column of 
plots as the lattices become finer. What is crucial for 
the taste structure of the eigenvectors, however, is the relative
width of the histograms for $|\zeta^\Gamma_{ij}|$ for $i\ne j$ 
compared to that for $|\zeta^\Gamma_{ii}|$.  
From the plots it can be seen that the width of 
the off-diagonal distribution 
is falling faster with lattice spacing than that of the diagonal.
Figure~\ref{fig:zetaS} shows a single configuration with $|Q|=1$, but 
we have examined others, and they look the same.

Figure~\ref{fig:zetaP} for the pseudoscalar case 
shows two configurations, one each with $Q=+1$ 
and~$-1$.
The plots behave in the same way as the scalar overlaps, except that 
$\zeta^{\gamma^5}_{ii}$ is real and \emph{negative} for $Q=-1$, as 
a consequence of parity. From the same arguments as above, 
since the taste-singlet pseudoscalar operators links odd sites 
to odd sites and even sites to even sites, $\zeta^{\gamma^5}_{i,-i}$ 
is also real and takes the same or opposite sign to 
$\zeta^{\gamma^5}_{i,i}$ on odd sites or even sites. 
$\zeta^{\gamma^5}_{i,-i}$ is therefore not visible beneath 
$\zeta^{\gamma^5}_{i,i}$ on Fig.~\ref{fig:zetaP}. 
Once again, looking down the plots, we see clearly that the width 
of the off-diagonal distribution 
(combining $j=-i$ and $|j| \ne |i|$) 
decreases with lattice spacing, relative to the diagonal 
distribution.

With $\zeta^{\gamma^\mu}_{ij}$ the behavior differs.
Recall that $\zeta^{\gamma^\mu}_{ij}$ should vanish for all $i,j$, 
even~$j=i$.
As seen in Fig.~\ref{fig:zetaV}, we find
$\zeta^{\gamma^\mu}_{ii}$ to be pure imaginary, which follows 
from the definition of the operator~$\gamma^\mu_I$, Eq.~(\ref{eq:vector});
we find $\zeta^{\gamma^\mu}_{i,-i}$ to be pure real, which follows by 
changing the sign of the odd pieces of $f_{-i}(x)$ relative 
to $f_{i}(x)$, because the vector operator couples even to odd sites 
and vice versa; and we find
$\zeta^{\gamma^\mu}_{ij}$, $|j|\neq|i|$, to be complex and of random phase.
In this case, however, the widths of all three distributions not only are
the same (when nonzero) but also decrease with decreasing lattice 
spacing together.
Indeed the widths of all $|\zeta^{\gamma^\mu}_{ij}|$ distributions are 
similar to the widths of the $|\zeta^1_{ij}|$ and
$|\zeta^{\gamma^5}_{ij}|$ distributions, $j\neq i$.

\begin{table*}[tbp]
\centering
    \caption{Widths, in lattice units and multiplied by $10^4$,
        obtained for different $|\zeta^{\Gamma}_{ij}|$ 
        histograms on each set of $|Q|=1$ gauge configurations. 
        The first column gives the set and then subsequent columns 
        list the width, with an estimate of the error, for different 
        $\Gamma$ and $i,j$ combinations, $i, j = \pm 1$, $\pm 2$.}
    \label{tab:widths}
    \begin{tabular}{clllllll}
    \hline\hline
Set & $\gamma^5$,   $i=j$ & $\gamma^5$,   $i\ne j$    
    & $\gamma^\mu$, $i=j$ & $\gamma^\mu$, $i=-j$ 
    & $\gamma^\mu$, $|i|\ne|j|$ & $1$, $i=j$ & $1$, $i\ne j$ \\
\hline

1 & $3.93(31) $ & $0.296(26)$ & $0.242(11) $ & $0.202(21) $ & $0.224(1) $ & $6.84(17) $ & $0.285(23) $ \\ 
2 & $4.08(56) $ & $0.180(11)$ & $0.188(23)$ & $0.146(12)$ & $0.155(16)$ & $6.07(70)$ & $0.158(9)$ \\ 
3 & $1.49(3) $ & $0.0635(38)$ & $0.0567(36)$ & $0.0541(42)$ &  $0.0561(22)$ & $2.15(6)$ & $0.0566(27)$ \\ 
4 & $0.548(26) $ & $0.0390(16)$ & $0.0242(10)$ & $0.0246(17)$ & $0.0258(4)$ & $0.846(29)$ & $0.0406(22)$ \\ 
5 & $0.716(14) $ & $0.0230(7)$ & $0.0211(2)$ & $0.0206(2)$ & $0.0219(5)$ & $0.954(16)$ & $0.0199(5)$ \\ 
    \hline\hline
    \end{tabular}
\end{table*}

\begin{figure}
    \includegraphics[width=80mm]{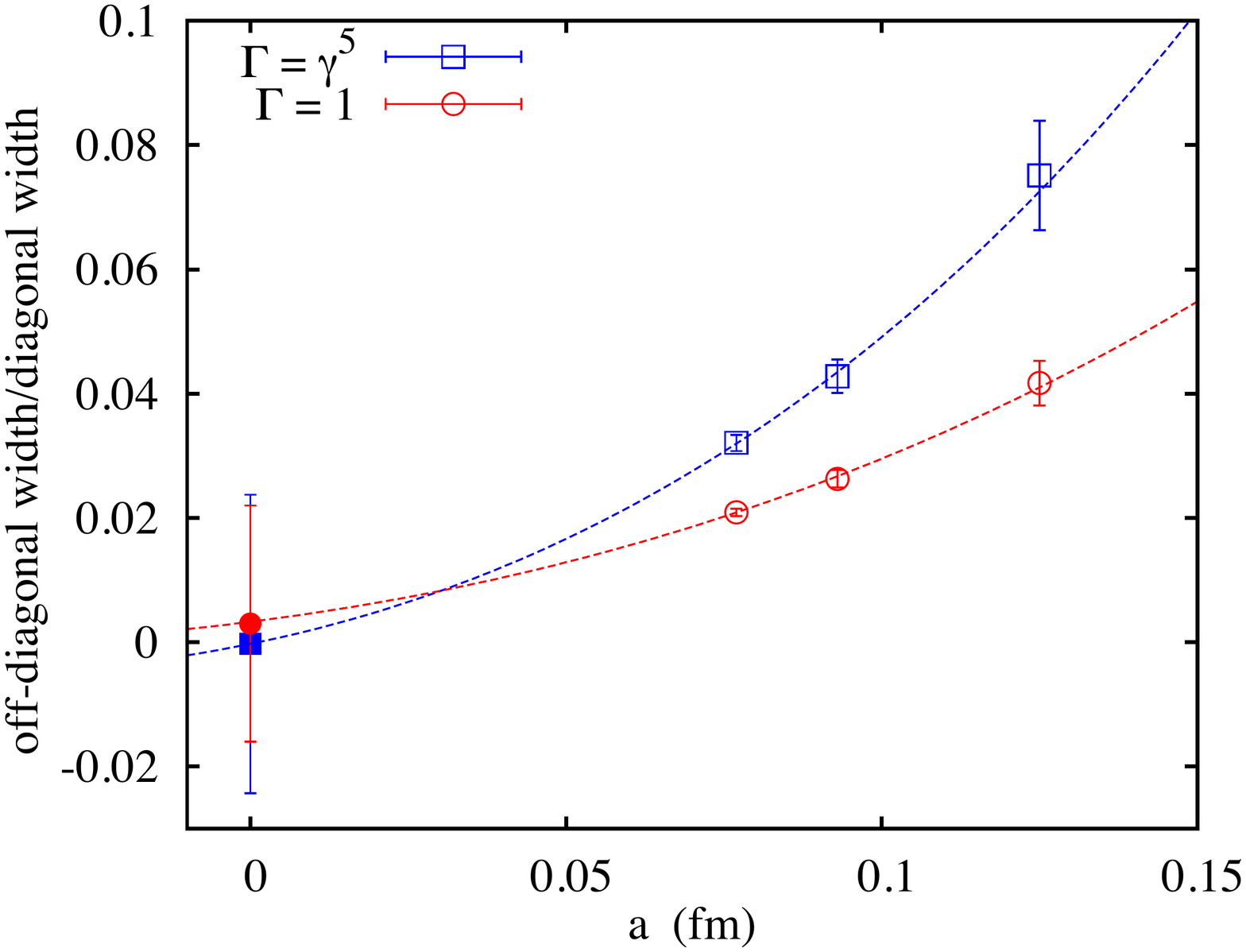} 
    \includegraphics[width=80mm]{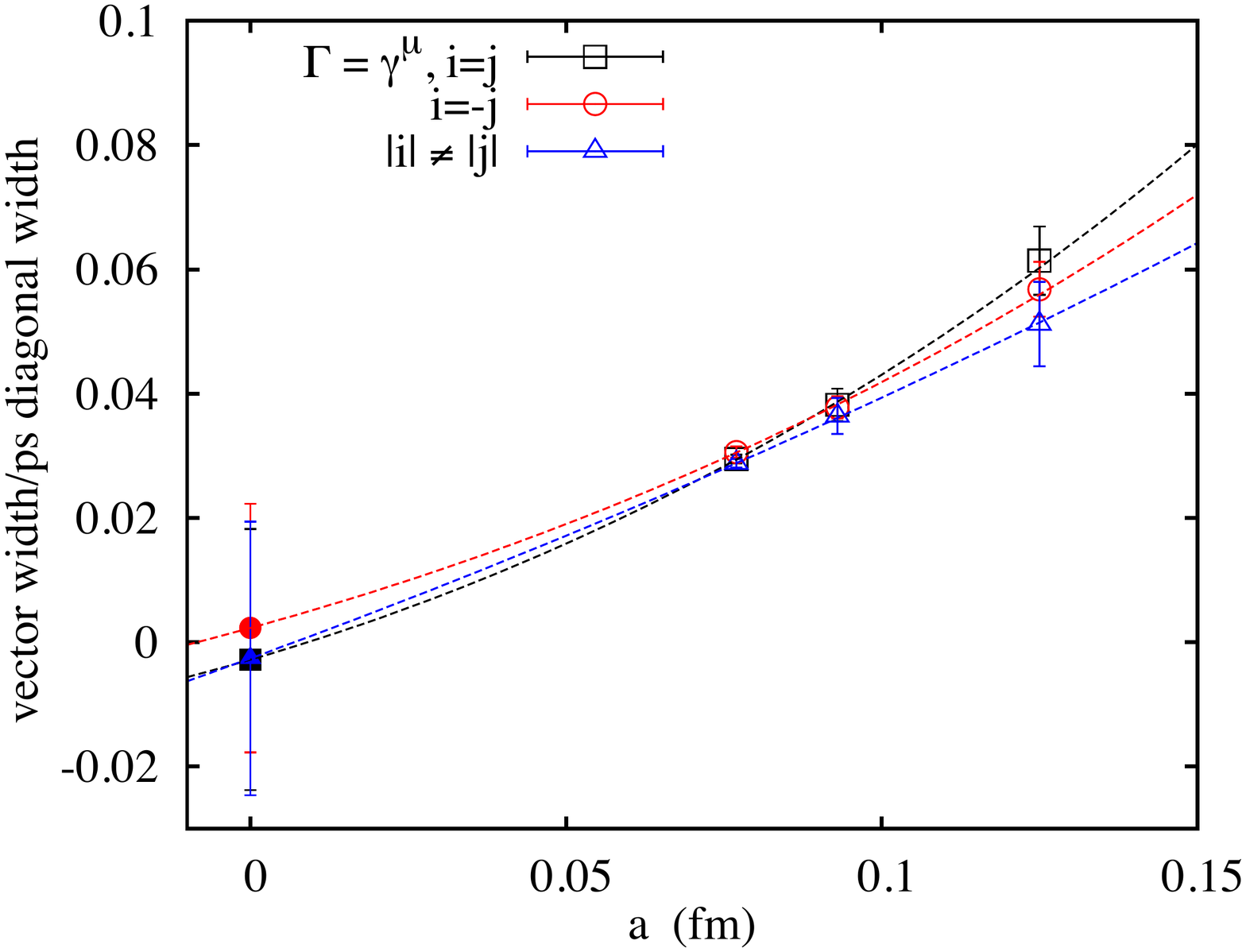} 
    \caption{(color online) 
        The top plot shows the width of $|\zeta^\Gamma_{ij}|$, 
        $j\neq i$, divided by the width of $|\zeta^\Gamma_{ii}|$, 
        for $\Gamma=1,\gamma^5$, $i, j = \pm 1, \pm 2$. 
        The $\Gamma=1$ case is given by open circles (red online), 
        along with a representative polynomial fit in the square of the 
        lattice spacing and the corresponding value in the continuum 
        limit (filled circles).  
        The equivalent results for $\Gamma=\gamma^5$ are given by open 
        and closed squares (blue online). 
        The lower plot shows the width of $|\zeta^{\gamma^\mu}_{ij}|$ 
        divided by the width of $|\zeta^{\gamma^5}_{ii}|$ plotted 
        against the lattice spacing.
        The case $i=j$ is given by (black) open and closed squares, 
        the case $i=-j$ by (red) open and closed circles, and 
        the case $|i| \ne |j|$ by (blue) open and closed triangles.}
    \label{fig:zeta-widths}
\end{figure}

To visualize the lattice-spacing dependence more directly, we plot in 
Fig.~\ref{fig:zeta-widths} the width of the $\zeta^{\Gamma}_{ij}$ 
distributions, appropriately normalized, 
vs~$a^2$. The widths are defined by the central 
66\% of the data in the lower right histogram 
for $|\zeta^{\Gamma}_{ij}|$, but calculating this histogram 
for ten configurations instead of just one or~two. 
The errors are estimated by comparing 
the widths for two subsets of five configurations. 
The values we obtain for the widths, and their errors, 
are given in Table~\ref{tab:widths}. 
Since the eigenvectors are normalized to have modulus 1 at 
each lattice spacing, the widths do not have a physical 
interpretation. 
The best that one can do is to normalize the 
off-diagonal widths against diagonal widths,
as is appropriate for the 
test of Eq.~(\ref{eq:theTest}). 
This ratio of widths is
plotted for the scalar and pseudoscalar in 
Fig.~\ref{fig:zeta-widths}. For the vector, we have no 
diagonal quantity that survives in the continuum 
limit, so we normalize instead against 
the diagonal pseudoscalar width. 
Although it is difficult to be quantitative 
(full ensemble averages of the 
widths are too costly, and the determination of the 
lattice spacing in the quenched approximation is 
ambiguous), the trend in Fig.~\ref{fig:zeta-widths} 
is clear and consistent with what is 
needed according to Eq.~(\ref{eq:theTest}).

Figure~\ref{fig:zeta-widths} shows, with dashed lines, representative 
fits as a polynomial in $a^2$ to our results. The fits include 
a constant plus quadratic, quartic, and sixth powers of $a$. 
The slope of 
the $n$th polynomial term is constrained by a Bayesian prior 
to a size of $(1.0~\mathrm{GeV})^n$ suggested by 
the slope of pion taste splittings~\cite{Bazavov:2010ru}. 
It is very easy to obtain 
good fits with any combination of different polynomials, for example, 
including or not including a linear term, so it is 
not possible to say definitively what the lowest power of $a$ is 
that appears
in the $a$~dependence of the $\zeta_{ij}$. The solid points on the 
plots in Fig.~\ref{fig:zeta-widths} give the $a=0$ value of 
the width ratios, compatible with zero in all cases.   
Thus, our results are consistent with the expectation in 
Eq.~(\ref{eq:theTest}),
although the data are not able to determine $p_{\zeta^\Gamma}$ in a 
definitive way.

We have also investigated the volume dependence of the 
$\zeta^{\Gamma}_{ij}$ for $i, j = \pm 1, \pm 2$ 
at the intermediate lattice spacing 
(i.e., on sets 2, 3, and 4), and these results are also 
included in Table~\ref{tab:widths}. We see that the widths again fall 
as the volume of the lattice increases. Naively this is 
simply a result of the normalization of the eigenvectors 
to 1 over an increasing number of lattice sites. 
Indeed the widths do seem to have simple behavior, inversely 
proportional to $1/V$, at least for the diagonal scalar and 
pseudoscalar widths and the vector widths. 
Note that this is not inconsistent with the fact that, for 
example, the very high values of the pseudoscalar diagonal 
overlaps are localized around the instantons 
that give rise to the near-zero modes. 

\begin{figure}[b]
    \includegraphics[width=80mm]{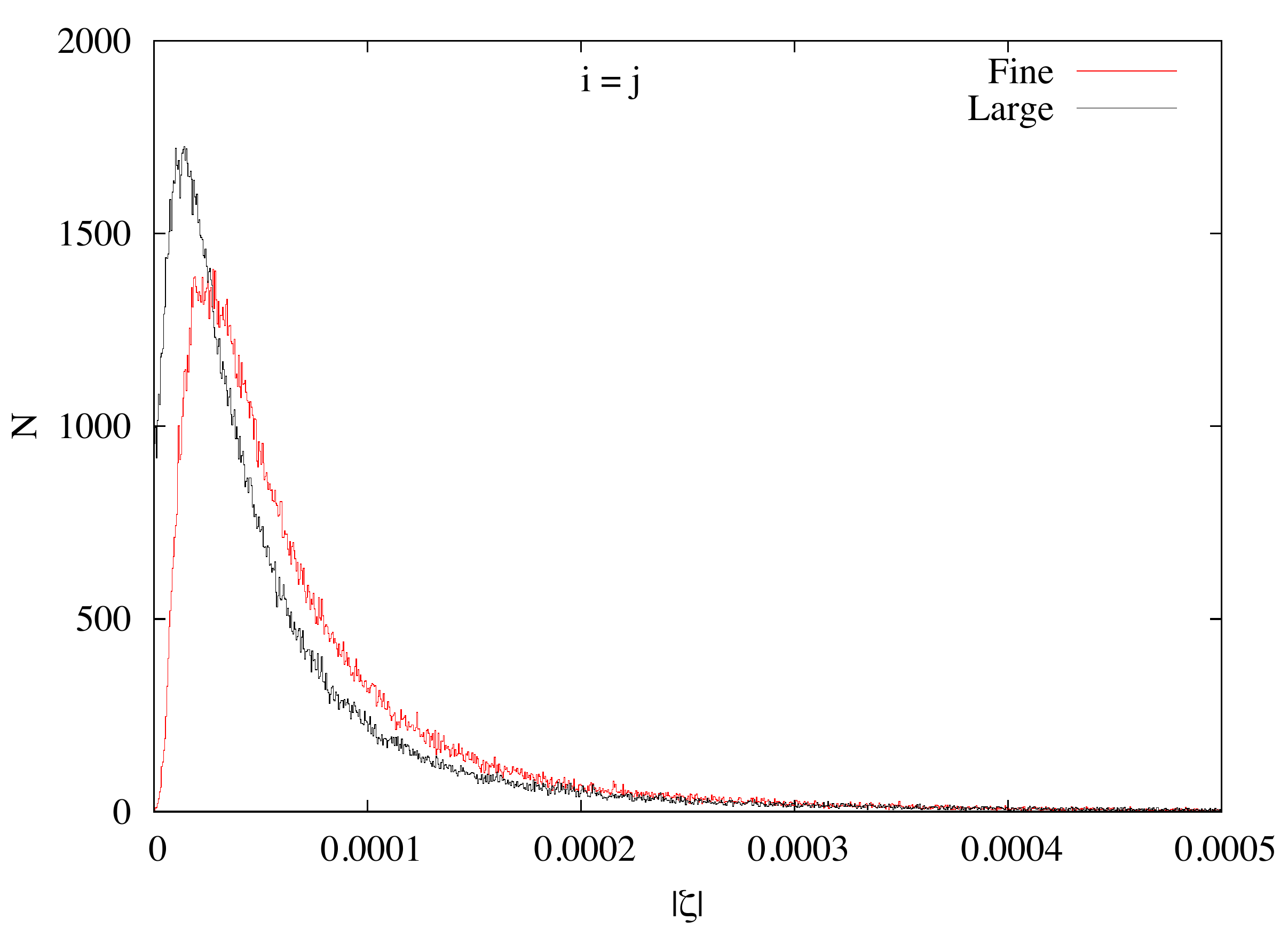} 
    \includegraphics[width=80mm]{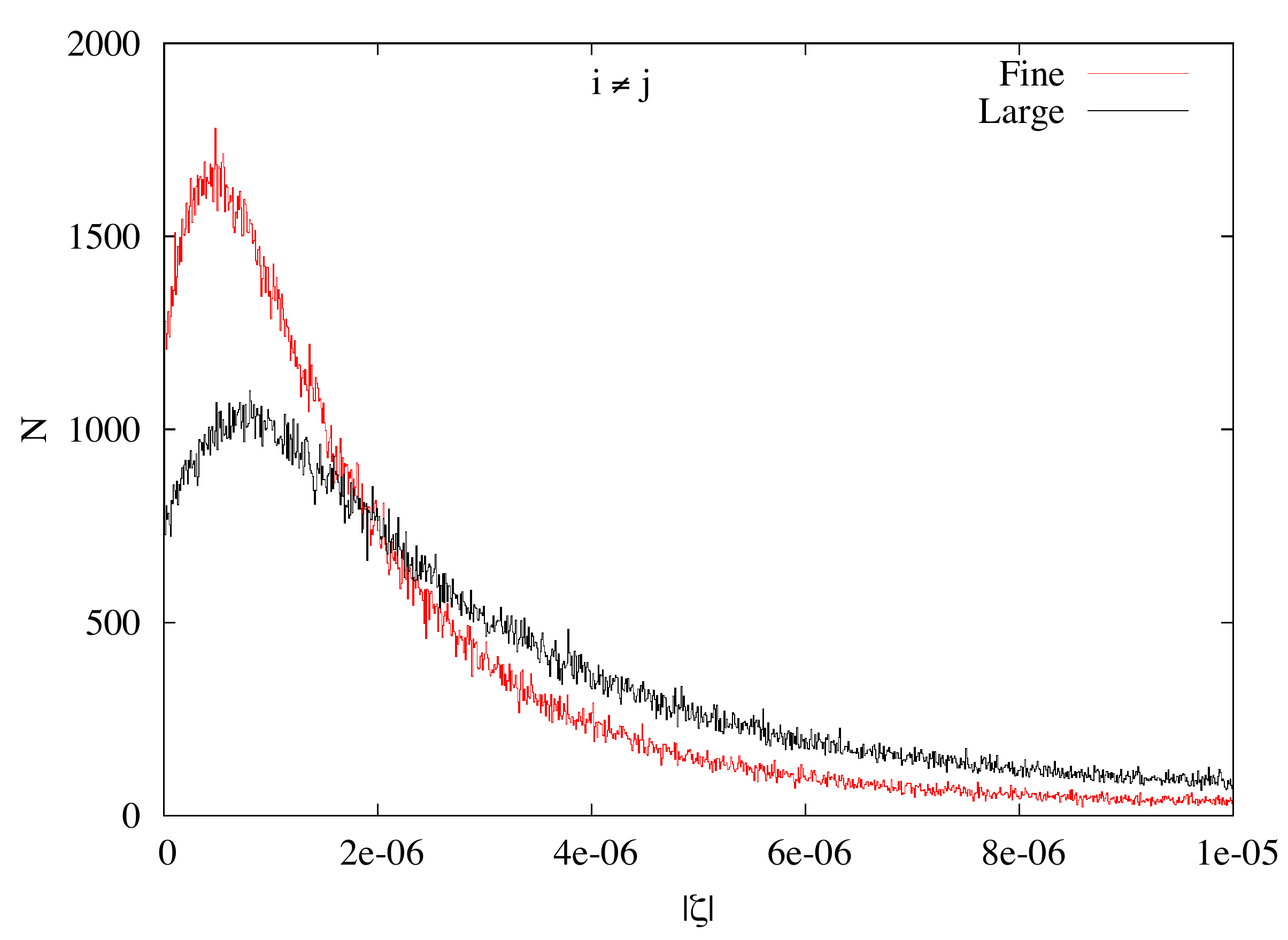} 
    \caption{Histogram of $|\zeta^{\gamma^5}_{ij}|$, for 
        $i=j$ (top) and $i\neq j$ (bottom), comparing results 
        on the fine lattice (set~5---red/gray) and the large intermediate 
        lattice (set~4---black).  
        }
    \label{fig:zeta-vs-L}
\end{figure}

The pseudoscalar and scalar widths behave quite differently 
as a function of lattice spacing than they do as a function 
of volume. We can see this by comparing the histograms 
in Fig.~\ref{fig:zeta-vs-L} for $\zeta^\Gamma_{ij}$ on the 
fine lattices, set~5, and 
the large intermediate volume lattices, set~4. Both of these 
have $20^4$ lattice points. We see that the diagonal 
distribution is broader on the finer lattices and the 
off-diagonal distribution markedly narrower, consistent 
with the fairly rapid fall with lattice spacing of the 
ratio of the widths seen in Fig.~\ref{fig:zeta-widths}.  
For the vector case, as is clear from Table~\ref{tab:widths}, the behavior
of the widths with the lattice spacing is only slightly 
steeper than that with volume. However, this still 
represents a fall to zero with lattice spacing 
when compared to the diagonal scalar and pseudoscalar 
overlaps which survive the continuum limit, as we 
see in Fig.~\ref{fig:zeta-widths}.

We have not shown histograms for the axial vector or 
tensor operators. We have looked at these operators 
in terms of the relevant meson 
correlators (see the following subsections) 
and they give qualitatively identical 
results to the vector and scalar/pseudoscalar cases, 
respectively. 
It therefore seems unlikely that they would 
upset the picture gleaned here.

\subsection{Flavor-singlet meson correlators}
\label{subsec:fs}

Our results in Sec.~\ref{subsec:zeta-plots} show how the taste-singlet 
overlaps, $\zeta_{ij}^{\Gamma}(x)$, of different 
near-zero-mode eigenvectors behave as 
expected to give the correct continuum behavior for 
the 't~Hooft vertex. Here we show explicitly how this translates into 
the correct continuum behavior for the near-zero-mode 
contribution to the flavor-singlet meson correlator. 
We also look at nonzero-mode contributions, as well 
as flavor-nonsinglet correlators, wherever they 
are useful to fill out the picture obtained. 
As discussed in Sec.~\ref{sec:hooft} it is sufficient 
to work in the quenched approximation since 
the structural issue of the behavior of the eigenvector 
overlaps---in a fixed-$|Q|$ sector---is the same whether sea quarks are 
included or not.

To relate results as closely as possible to those of a complete meson 
correlator calculation in lattice QCD, we consider meson correlators 
projected onto zero spatial momentum by summing over spatial sites.
This leads us to consider a modification of the eigenvector overlaps
\begin{equation}
    \overline{\zeta}^{\Gamma}_{rs}(t) =
        \sum_{\bm{x}} f_r^{\dag}(\bm{x},t) \Gamma_I f_s(\bm{x},t),
\end{equation}
summing over a time slice instead of a $2^4$ hypercube.
Then the zero-momentum connected and disconnected contributions can be 
constructed, as in Eqs.~(\ref{eq:Czero}) and~(\ref{eq:Dzero}) from
correlations of time-slice overlaps
\begin{eqnarray}
X^{\Gamma}_{rs}(T) &=& 
\sum_{t} \overline{\zeta}^{\Gamma}_{rs}(t) \overline{\zeta}^{\Gamma}_{sr}(t+T), \\ 
Y^{\Gamma}_{rs}(T) &=& 
\sum_{t} \overline{\zeta}^{\Gamma}_{rr}(t) \overline{\zeta}^{\Gamma}_{ss}(t+T). 
\label{eq:zetaxy}
\end{eqnarray}
Note that $X^{\Gamma}_{rr}(T) = Y^{\Gamma}_{rr}(T)$ by construction, and
we consider values for $r, s$ that correspond to nonzero 
modes as well as near-zero modes. 

The full connected correlator is
$C(T) = \sum_{\bm{x},\bm{y}} C(\bm{x},t;\bm{y},t+T)$, 
with $C(x,y)$ defined in Eq.~(\ref{eq:Cdef}).
Similarly, the disconnected correlator is
$D(T) = \sum_{\bm{x},\bm{y}} D(\bm{x},t;\bm{y},t+T)$, 
following Eq.~(\ref{eq:Ddef}). 
$C(T)$  is then 
made up of $X_{rs}$ correlated overlaps (on the quenched configurations that we 
are studying) as
\begin{eqnarray}
C(T) &=& \left\langle \mathcal{C}(T) \right\rangle_{U} \nonumber \\
&=& \sum_{r,s}\left\langle \frac{X_{rs}(T)}{(i\lambda_r+m)(i\lambda_s+m)} \right\rangle_{\hspace{-0.4em}U}
\label{eq:ct}
\end{eqnarray}
where we have made explicit the dependence on the eigenvalues in the 
denominator. 
For $D(T)$ we have 
\begin{eqnarray}
D(T) &=& \left\langle \mathcal{D}(T) \right\rangle_{U} \nonumber \\
&=& \sum_{r,s}\left \langle \frac{Y_{rs}(T)}{(i\lambda_r+m)(i\lambda_s+m)} \right\rangle_{\hspace{-0.4em}U}.
\label{eq:dt}
\end{eqnarray}
The disconnected correlator factorizes into the product 
of sums over diagonal overlaps $\overline{\zeta}^{\Gamma}_{rr}$,
but the connected correlator contains overlaps between 
different eigenvectors. 

Note that the factor $|m|\mathbb{D}'$ of Eqs. (\ref{eq:Cdef}) 
and (\ref{eq:Ddef}) 
from the $n_f=1$ sea quark determinant is missing. This affects the weighting of 
the particular configurations in the ensemble and therefore the 
quantitative results obtained for $C(T)$ and $D(T)$. However, it 
does not affect qualitatively the properties of 
the $X_{rs}$ factors that we demonstrate here, which 
are evident in a fixed-$|Q|$ sector and even, in some cases,
on a configuration-by-configuration basis in their 
contribution to $\mathcal{C}(T)$ and $\mathcal{D}(T)$.  

As discussed in Sec.~\ref{sec:hooft}, 
we then have to test whether the near-zero 
modes give rise to a divergence in the correlator for 
flavor-singlet meson $H$
as $m\to 0$, when the connected and disconnected contributions are 
combined with their appropriate taste factors of 4 and 16 
[Eq.~(\ref{eq:etacd})]:  
\begin{equation}
M^H(T) = \left \langle \mathcal{M}^H(T) \right \rangle_U = \frac{1}{4} C^H(T) - \frac{1}{16} D^H(T).
\label{eq:mcd}
\end{equation}
To obtain a finite result as $m \to 0$ for $M(T)$ 
we need the near-zero-mode contributions
to cancel between $C(T)$ and $D(T)$. This in turn 
requires the off-diagonal correlated overlaps, $X_{ij}$, $i \ne j$, 
between different near-zero
eigenvectors in the same staggered eigenvalue quartet 
to vanish in the continuum limit. 
Then each quartet behaves as four copies of a single mode and, including 
the factors of 1/4 and 1/16, reproduces within $M(T)$ the 
behavior expected of eigenmodes of the Dirac operator 
in the continuum. We show how this works explicitly 
for the examples of scalar, pseudoscalar, (axial) vector, and tensor 
mesons in the following subsections.  
We do this with the same $|Q|=1$ quenched configurations  
used in the previous subsection. The correlator results 
are, however, averaged over all $|Q|=1$ 
configurations for each ensemble, rather than just 10.  

\subsubsection{Flavor-singlet pseudoscalar mesons}
\label{subsub:etap}

\begin{figure}[t]
\begin{center}
\includegraphics[width=80mm]{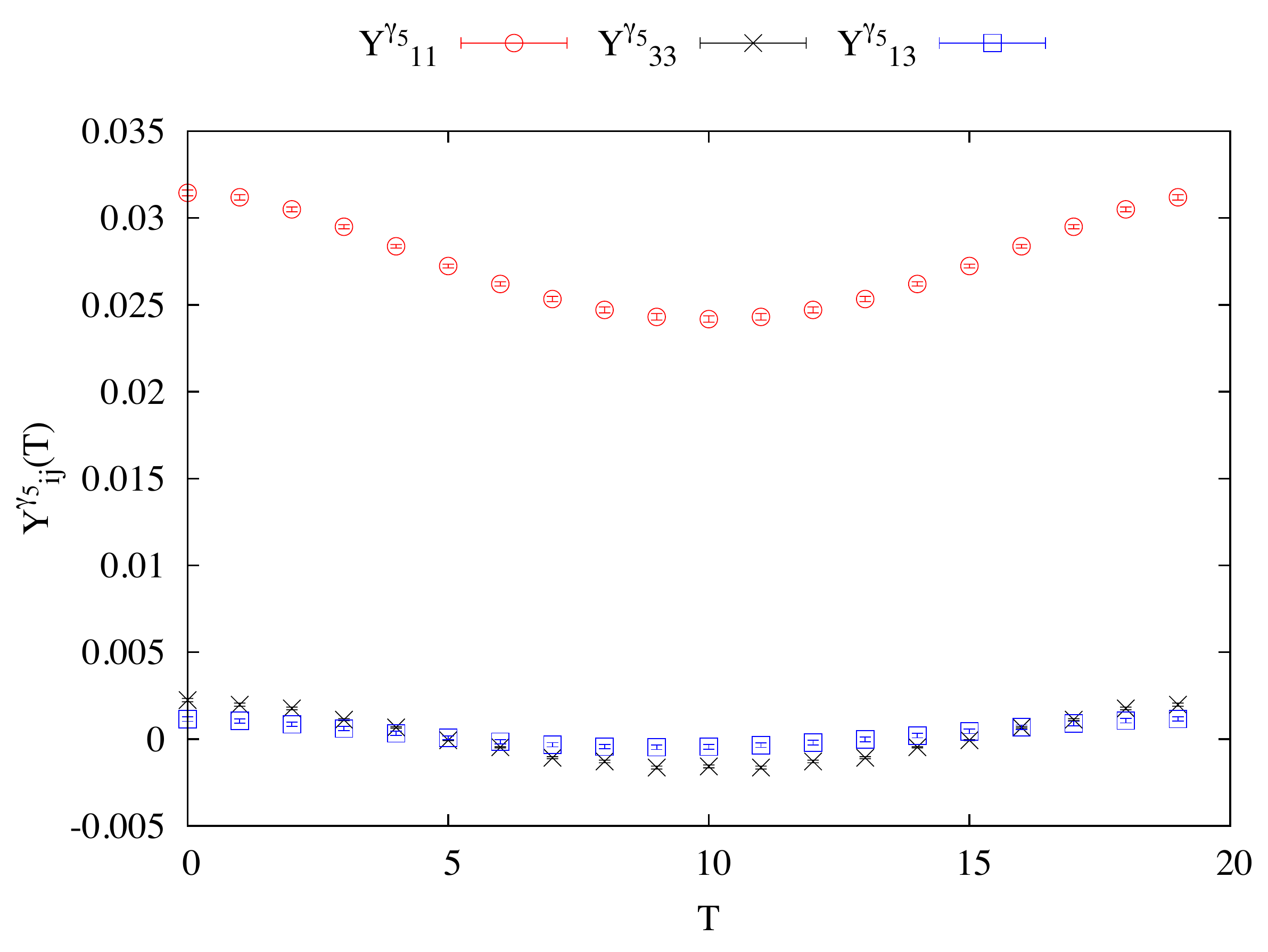}
\includegraphics[width=80mm]{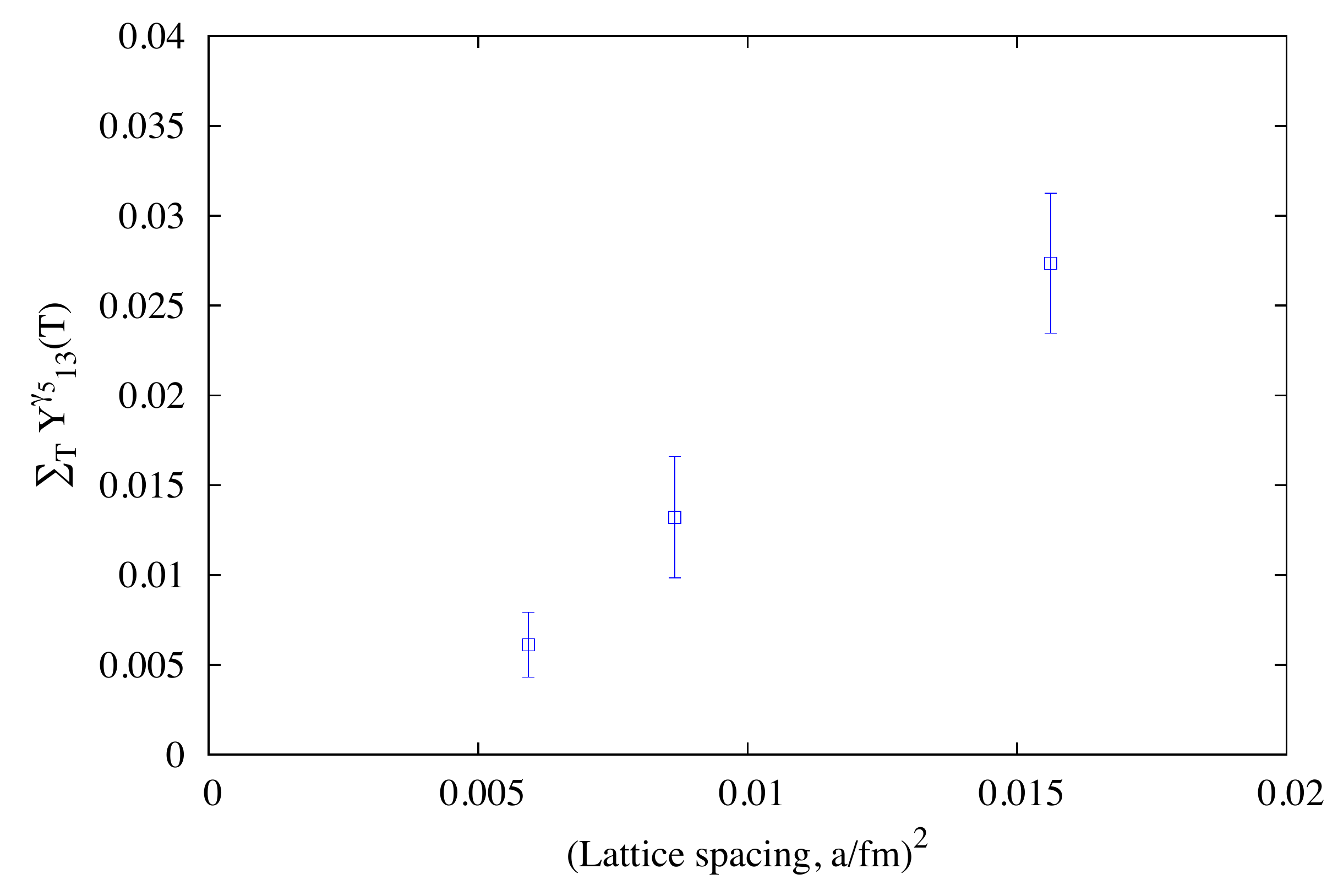}
\end{center}
\caption{Diagonal and off-diagonal correlated overlaps $Y^{\gamma^5}_{ij}(T)$ between 
eigenvectors 1 and 3 that 
contribute to the disconnected piece of the $\eta^{\prime}$ correlator 
at zero spatial momentum. Results are given for the average 
over $|Q|=1$ configurations 
in the fine ensemble, set~5. 
The lower plot shows the off-diagonal 
correlated overlap summed over time separation as a function of the square of the 
lattice spacing. 
}
\label{fig:y13asq}
\end{figure}

\begin{figure}[t]
\begin{center}
\includegraphics[width=80mm]{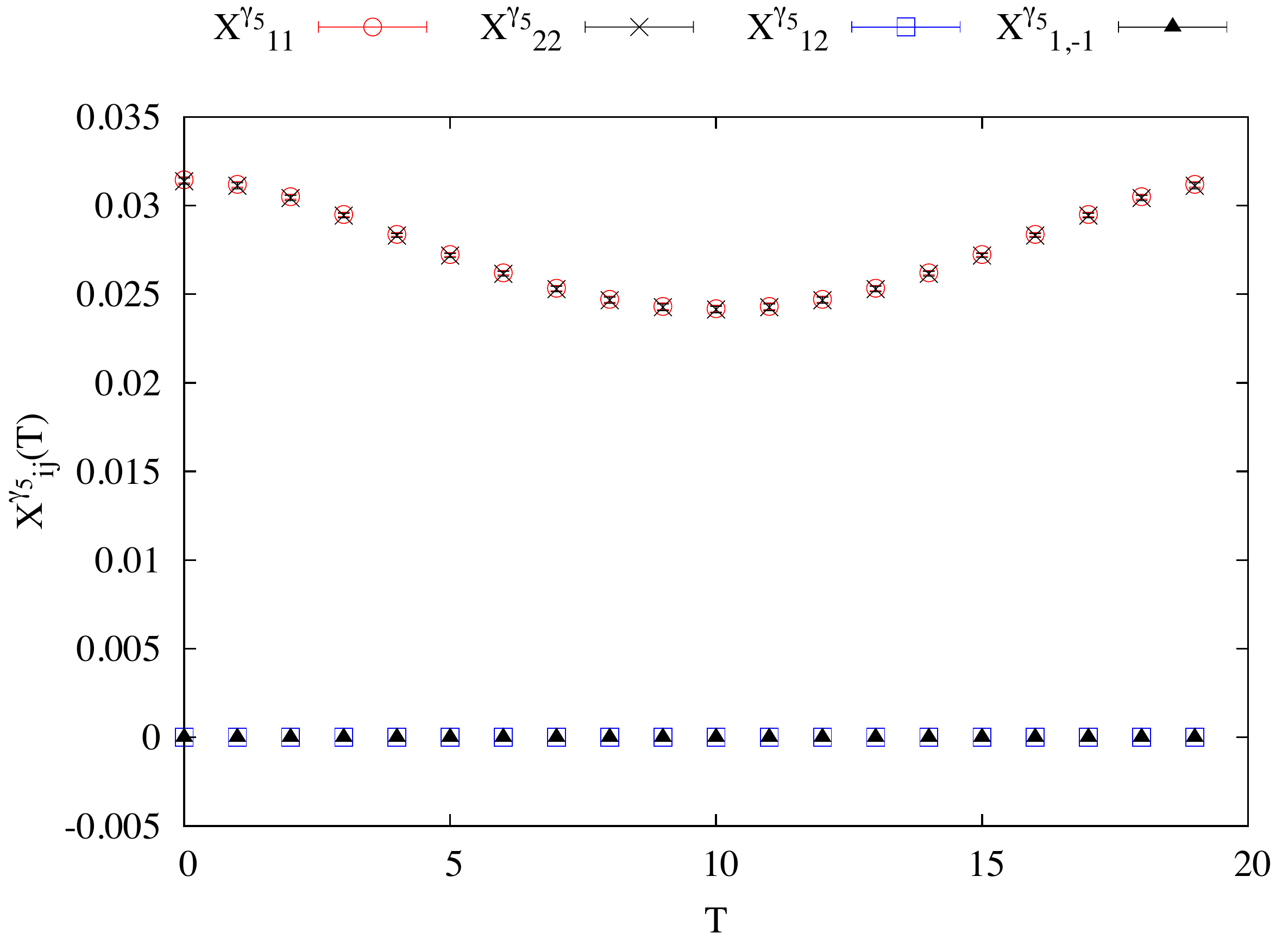}
\includegraphics[width=80mm]{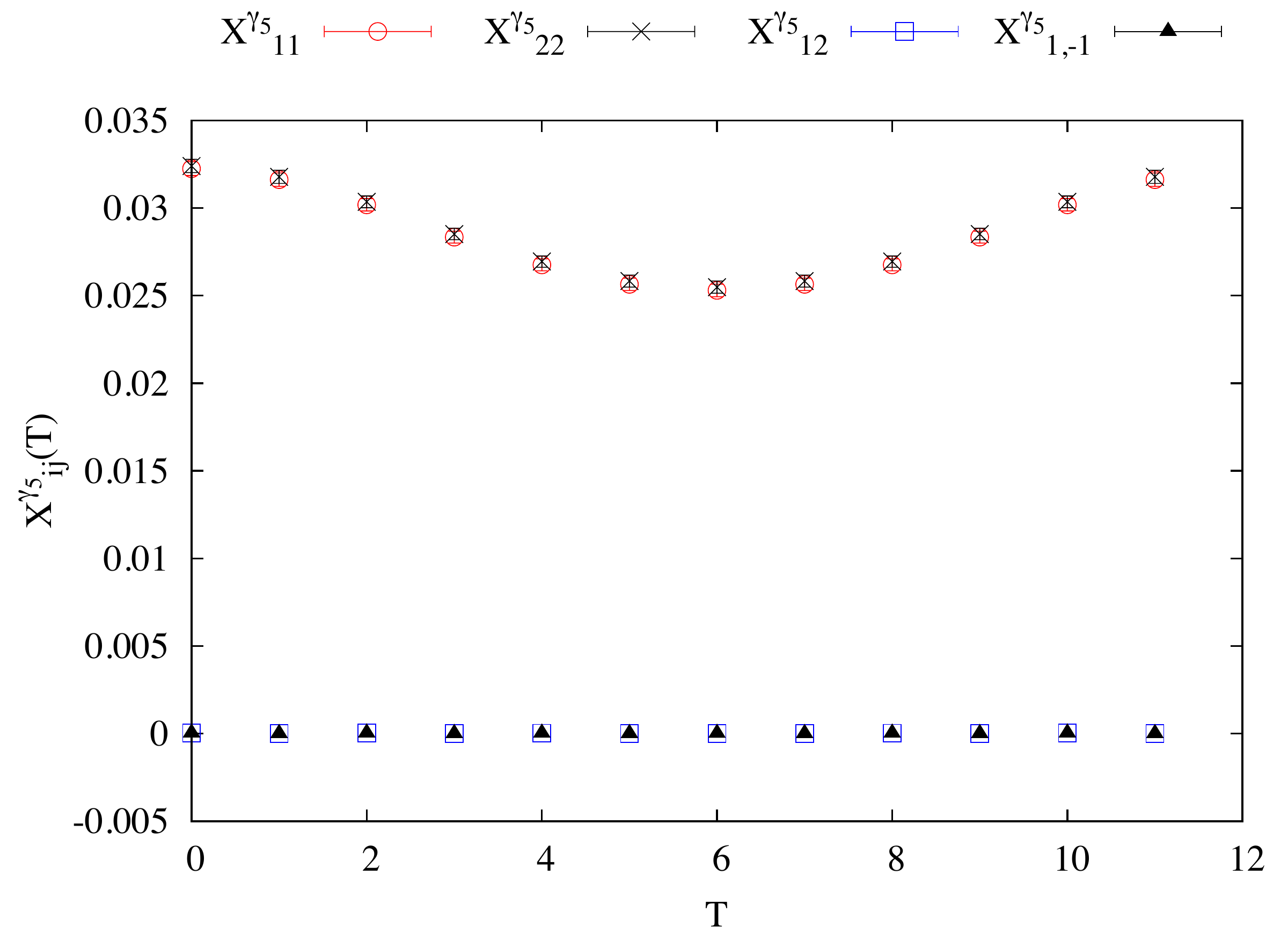}
\end{center}
\caption{The correlated overlaps $X^{\gamma^5}_{ij}(T)$ between 
near-zero modes 1, 2, and $-1$ that 
contribute to the connected piece of the $\eta^{\prime}$ correlator 
at zero spatial momentum. Results are given for the average 
over $|Q|=1$ configurations 
in the fine ensemble, set~5 (top) and the coarse ensemble, set~1 (bottom).  
}
\label{fig:overlaps}
\end{figure}

\begin{figure}[t]
\begin{center}
\includegraphics[width=80mm]{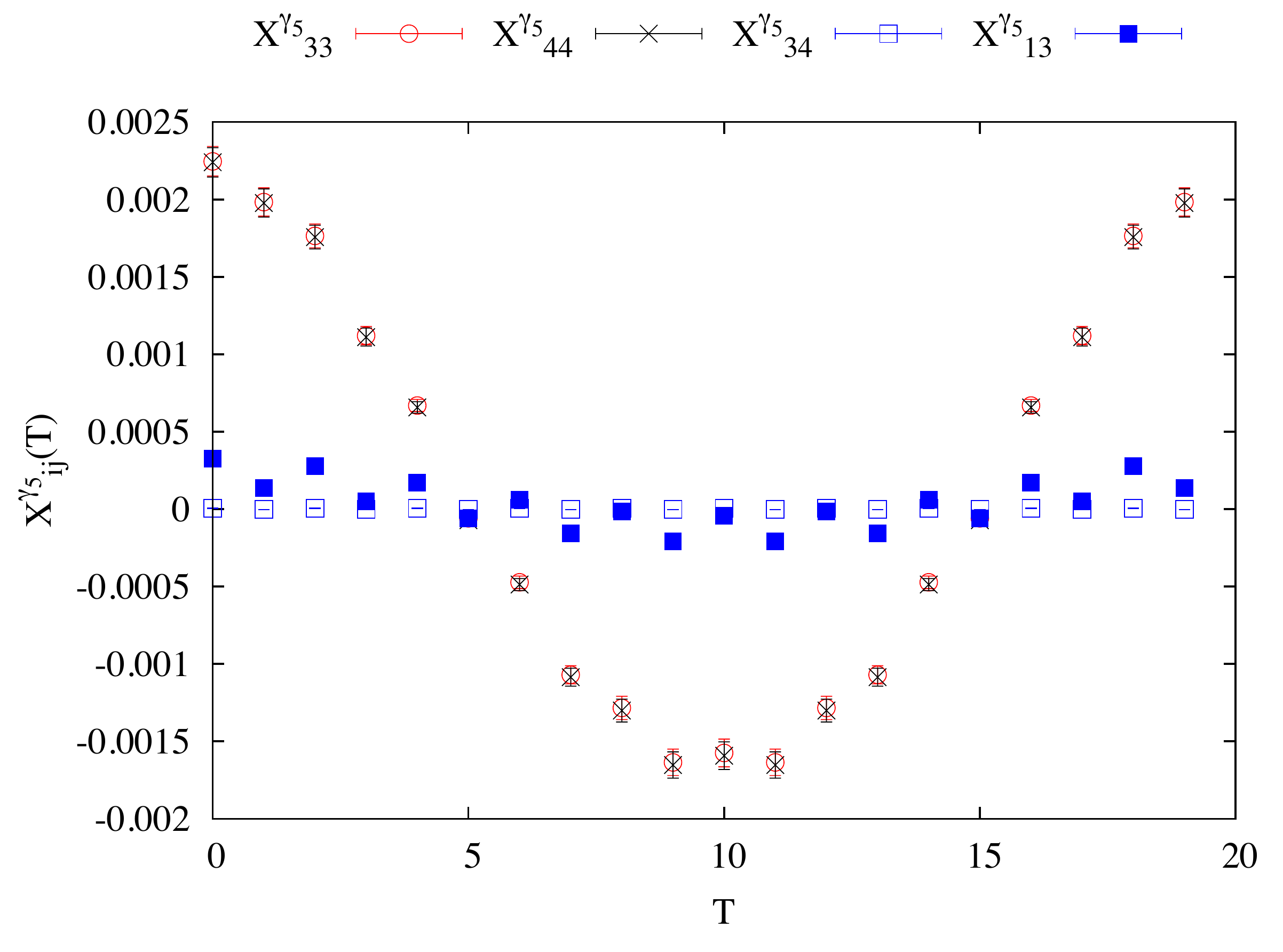}
\end{center}
\caption{The correlated overlaps $X^{\gamma^5}_{rs}(T)$ between 
nonzero modes 3 and 4 that 
contribute to the connected piece of the $\eta^{\prime}$ correlator 
at zero spatial momentum. Results are given for the average 
over $|Q|=1$ configurations 
in the fine ensemble, set~5. 
Results are also shown for the correlated overlap between near-zero
mode~1 and nonzero mode~3. 
}
\label{fig:overlaps34}
\end{figure}

\begin{figure}[t]
\begin{center}
\includegraphics[width=80mm]{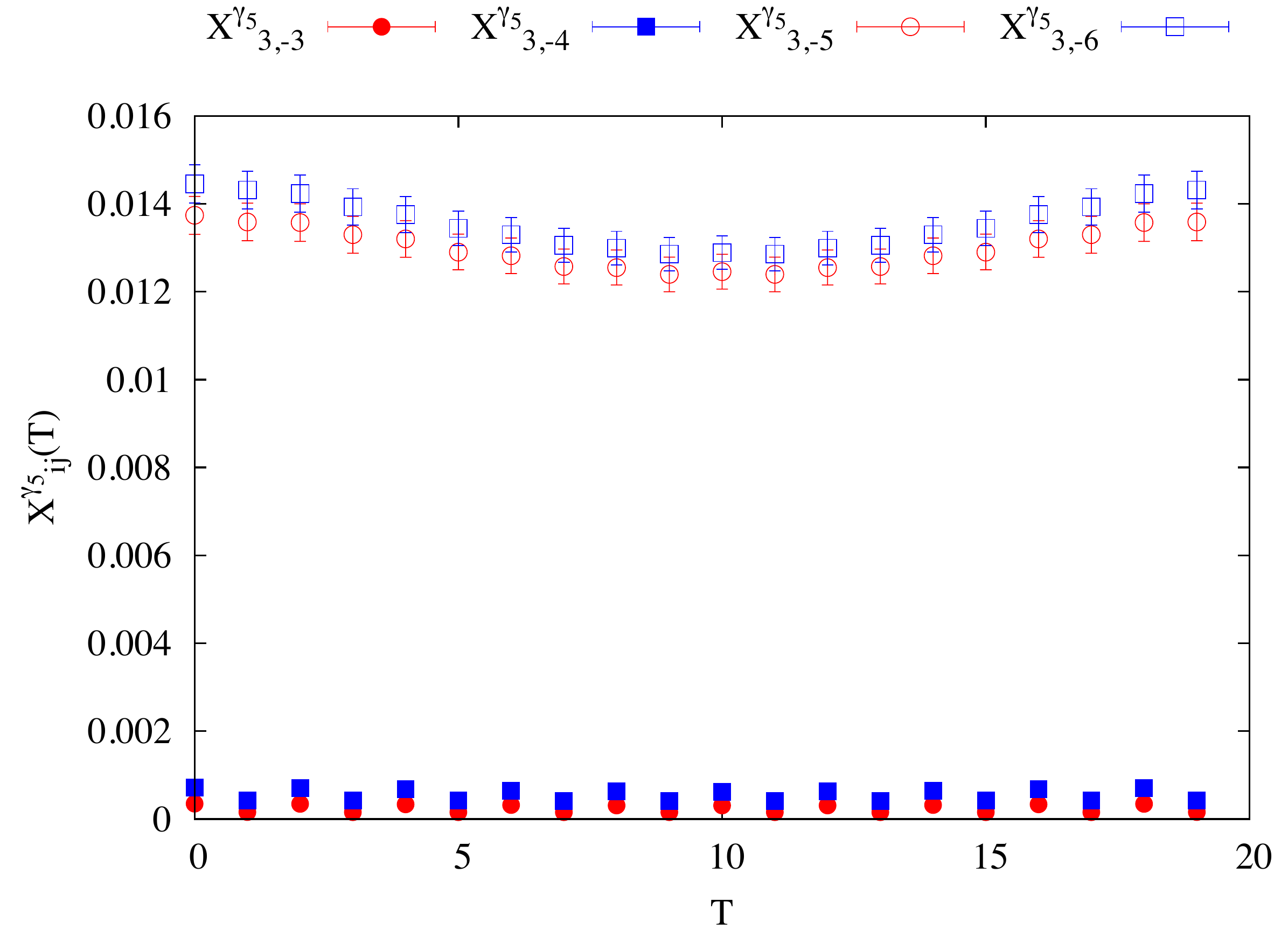}
\end{center}
\caption{The correlated overlaps $X^{\gamma^5}_{rs}(T)$ between 
nonzero mode 3 and modes $-3$, $-4$, $-5$, and $-6$ from 
its $\mathrm{U}_\varepsilon$ mirror quartet. 
Results are averaged 
over $|Q|=1$ configurations for set~5. 
}
\label{fig:overlapminus}
\end{figure}

We first discuss the important case of the 
calculation of the flavor-singlet pseudoscalar 
meson ($\eta'$) correlator, and the associated 
case of the flavor-nonsinglet meson ($\pi$) correlator. 
In continuum QCD with, say, two equal mass light quarks, 
this is readily analyzed in terms of the 
eigenvectors and eigenvalues of the massless Dirac matrix. 
For $|Q|$ = 1 gluon field configurations there 
is one zero mode with a chirality $\phi_0^{\dag} \gamma_5 \phi_0 = \pm 1$. 
The $\pi$ meson 
correlator has no disconnected 
contribution and the connected contribution is 
readily seen to obey, on a given gluon field configuration,
\begin{equation}
\sum_T \mathcal{M}^{\pi}(T) = \sum_{\sigma} \frac{1}{\lambda_{\sigma}^2 + m^2} 
\label{eq:cgold}
\end{equation}
where $m$ is the quark mass and the sum is over all 
eigenmodes of the massless Dirac matrix, 
including the 
zero mode. Each eigenmode contributes a correlated overlap of 1 when 
summed over $T$. For the zero mode, this comes from the square 
of the chirality.  
The nonzero modes have chirality zero but 
still contribute a correlated overlap factor of 1 
because the $\gamma_5$ matrix connects the modes 
$\sigma$ and $-\sigma$ with $\lambda_{-\sigma} = -\lambda_{\sigma}$. Then (continuum)
$\sum_t\overline{\zeta}^{\gamma^5}_{\sigma,-\sigma}(t) = 1$ from eigenfunction 
normalization. 

The $\eta^{\prime}$ meson correlator 
is made from the same eigenvectors but now has a 
disconnected contribution coming from the zero mode 
that exactly cancels the zero-mode contribution to 
the connected correlator. Thus, on a given configuration,  
\begin{equation}
\sum_T \mathcal{M}^{\eta^{\prime}}(T) = \sum_{\sigma \ne 0} \frac{1}{\lambda_{\sigma}^2 + m^2}.
\label{eq:ceta}
\end{equation}
There is now no contribution from the zero mode and the correlator is 
finite as $m \to 0$. This continues to be true on averaging 
over gauge fields and including determinant factors. 

Now let us show how staggered fermions reproduce Eqs.~(\ref{eq:cgold}) 
and~(\ref{eq:ceta}).
For the Goldstone $\pi$ meson correlator, it is straightforward 
and mechanical. Then $\Gamma = \gamma^5_P = \varepsilon(x)$ 
connects eigenvectors $f_s$ and $f_{-s}$,
and the correlated overlap contribution is again 1, when summed over $T$, 
simply from eigenvector normalization. 
The difference 
with the continuum case is that this is \emph{also} true for the 
near-zero modes. Thus, we obtain an  
equation very similar to that in the continuum on a single gluon 
field configuration: 
\begin{eqnarray}
\sum_T \mathcal{M}^{\pi}(T) &=& \frac{1}{4}\sum_{s} \frac{1}{\lambda_s^2 + m^2} \nonumber \\ 
&=& \sum_{q} \frac{1}{\bar{\lambda}_{q}^2 + m^2} + \mathrm{O}(a^2),
\label{eq:cgold2}
\end{eqnarray}
where $s$ is a sum over all modes \emph{including} the near-zero 
modes and the factor of $1/4$ is the same as in 
Eqs. (\ref{eq:etacd}) and (\ref{eq:mcd}).
Since the staggered 
eigenvalues come in quartets that become degenerate in the $a \rightarrow0$ 
limit, the lower equation replaces the 4 eigenvalues in a 
quartet by their mean square and sums all quartets, $q$, including 
the near-zero-mode quartet. 
This then clearly reproduces the continuum Eq.~(\ref{eq:cgold})
as $a \rightarrow 0$.  

The flavor-singlet correlator
is constructed differently and includes both connected 
and disconnected contributions. 
With staggered fermions, we must use the flavor-taste-singlet 
pseudoscalar, $\Gamma=\gamma^5_I$.  
In demonstrating that Eq.~(\ref{eq:ceta})
is reproduced, we also show that the correlated overlaps 
behave so as to give a finite result for the $\eta^{\prime}$ 
correlator. 

From our earlier results on chirality, we can anticipate 
what the disconnected correlated overlaps $Y^{\gamma^5}_{ij}$ look like. 
Because $\sum_t \overline{\zeta}^{\gamma^5}_{ss}(t) = \mathcal{X}_s$ 
we expect the large values of 
$Y^{\gamma^5}_{ij}$ to be those that involve the near-zero modes with 
their large-chirality values. Indeed 
\begin{equation}
    \sum_T Y^{\gamma^5}_{ij}(T) = \mathcal{X}_i\mathcal{X}_j. 
    \label{eq:sumy}
\end{equation}
This expectation is borne out by the numerical results. On averaging over 
$|Q|=1$ gauge fields, $Y^{\gamma^5}_{11}$, $Y^{\gamma^5}_{22}$, 
and $Y^{\gamma^5}_{12}$ are 
all equal, being the ``typical'' product of overlaps for 
two near-zero modes. $\langle Y^{\gamma^5}_{11}\rangle_{|Q|=1}$ is shown as a function 
of $T$ in Fig.~\ref{fig:y13asq}. 
Results for modes $-1$ and $-2$ from the near-zero-mode
quartet match these because, as discussed above, the chirality 
of mode $-1$ is identical to that of 1 and $-2$ to that of~2.
Thus, the sum over all the zero modes, $i,j\in\{\pm 1,\pm 2\}$,
of $Y^{\gamma^5}_{ij}$ gives $4\times 4 = 16$ times the square of the 
chirality for a typical zero mode. This is divided by 16 in 
the contribution to the disconnected correlator, as in 
Eq.~(\ref{eq:mcd}), and so the contribution becomes 
exactly what is required to match that from the one zero 
mode for continuum quarks, up to a renormalization factor 
for the taste-singlet pseudoscalar current. 

The nonzero modes, for example mode 3, have small 
chirality and therefore $Y^{\gamma^5}_{33}$ is small, as also shown 
in Fig.~\ref{fig:y13asq}.  In the continuum this 
would be zero. Here it is not zero for nonzero lattice 
spacing but tends to zero as $a \rightarrow 0$. 
In fact, because we find that $Y_{rr} = Y_{ss} = Y_{rs}$ for 
modes in the nonzero-mode quartet, $r,s\in\{3,4,5,6\}$, 
then the total contribution from the quartet, when 
divided by 16, cancels against the contribution 
from $X^{\gamma^5}_{rr}$ divided by 4 in the total pseudoscalar 
flavor-singlet correlator, as for the near-zero-mode 
quartet. 

It is also worth discussing the cross-term 
$Y^{\gamma^5}_{13}$ between the near-zero-mode quartet and 
the nonzero-mode quartet since this would also 
be identically zero in the continuum. 
Figure~\ref{fig:y13asq} shows the results for $Y^{\gamma^5}_{13}(T)$, which, 
summed over $T$, 
has a value which is the square root of the product of 
the sums over $T$ of $Y^{\gamma^5}_{11}$ and $Y^{\gamma^5}_{33}$. 
The lower plot of Fig.~\ref{fig:y13asq} then shows explicitly 
how $\sum_T Y^{\gamma^5}_{13}(T)$ 
vanishes as $a \rightarrow 0$. Similar behavior is seen for 
other terms that are related to the chirality of nonzero modes. 

For $X^{\gamma^5}$ the results for the 
diagonal case are the same as for 
$Y^{\gamma^5}$.
The results for the off-diagonal $X^{\gamma^5}$ are less clear {\it a priori}. 
In fact, we find in all cases that the off-diagonal correlated overlaps 
within a quartet are zero when averaged 
over gauge fields. Figure~\ref{fig:overlaps} 
illustrates this for modes 1 and 2 in the near-zero-mode quartet.  
$X^{\gamma^5}_{11}$ and $X^{\gamma^5}_{22}$ are large 
(being equal to $Y^{\gamma^5}_{11}$ 
and $Y^{\gamma^5}_{22}$) but $X^{\gamma^5}_{12}$ has an 
average of zero. The same results 
are obtained for the $-1$ and $-2$ modes. 
We also see an average of zero for the correlated overlaps between 
the positive and negative eigenmodes within the quartet. 
This is illustrated for $X^{\gamma^5}_{1,-1}$ in Fig.~\ref{fig:overlaps}. 
$X^{\gamma^5}_{1,-2}$ is very similar. 

The size of correlated overlaps
changes very little with the lattice spacing. 
Figure~\ref{fig:overlaps} also compares correlated 
overlaps $X^{\gamma^5}_{ij}$ on the coarsest lattices, set~1, with 
those for the finest lattices, set~5. 

An average of zero is also 
seen for off-diagonal terms between modes in the first nonzero-mode quartet 
and between modes in the near-zero-mode quartet and modes 
in the first nonzero-mode quartet.  
These points are illustrated in Fig.~\ref{fig:overlaps34}. 

To understand $X^{\gamma^5}_{rs}$ more completely, 
we must also study correlated overlaps between positive and 
negative nonzero eigenmodes. Although this is not relevant 
to the behavior of the 't~Hooft vertex, it shows very clearly 
how the connected contribution to the $\eta^{\prime}$ 
correlator becomes equal to that of the $\pi$ meson 
in the continuum limit, up to a renormalization factor that arises 
because the taste-singlet pseudoscalar current is not
absolutely normalized. 

Figure~\ref{fig:overlapminus} shows the correlated overlaps, $X^{\gamma^5}_{rs}$, 
between the mode $r=3$ and all 
the negative modes that correspond to the first negative 
nonzero quartet (which is the 
$\mathrm{U}_{\varepsilon}$ ``mirror'' of the first positive nonzero
quartet), i.e., $s = -3$, $-4$, $-5$, and $-6$. Interestingly, 
the correlated overlaps that are nonzero here are $X^{\gamma^5}_{3,-5}$ and 
$X^{\gamma^5}_{3,-6}$. These are equal and each about half the size of 
$X^{\gamma^5}_{1,1}$ = $Y^{\gamma^5}_{1,1}$ (compare 
Fig.~\ref{fig:overlaps} and Fig.~\ref{fig:overlapminus}). 
Likewise,
the nonzero correlated overlap for $r=5$ appears with $s=-3$ and $-4$ 
having the same size as $X^{\gamma^5}_{3s}$, $s=-5$, $-6$.  
The correlated overlaps $X^{\gamma^5}_{4s}$ and $X^{\gamma^5}_{6s}$ 
show the corresponding pattern. 
Note the parallel with what happens in 
the case of the near-zero-mode quartet, where $X^{\gamma^5}_{1,-1}$ 
and $X^{\gamma^5}_{1,-2}$ tend to zero
(as shown in Fig.~\ref{fig:overlaps}), 
and $X^{\gamma^5}_{3,-3}$ and $X^{\gamma^5}_{3,-4}$ tend to zero too. 
The difference here is that another pair belongs 
to the mirror quartet, whereas $1$, $2$, $-1$, and $-2$ form a 
single quartet that is its own mirror.
The pattern seen in Fig.~\ref{fig:overlapminus} is repeated for other 
nonzero-mode quartets. For example, $X^{\gamma^5}_{7,-9}$ and 
$X^{\gamma^5}_{7,-10} \approx 0.014$, while $X^{\gamma^5}_{7,-7}$ 
and $X^{\gamma^5}_{7,-8}$ are much much smaller. 

Thus, the large contributions from nonzero modes   
to the connected correlator for the 
taste-singlet pseudoscalar meson
come from correlated overlaps connecting 
members of a quartet and members of its mirror quartet, in fact
members of the opposite pair of the mirror quartet. 
When these correlated overlaps are summed over a quartet
they give a result, per quartet member, approximately equal to 
that of a typical near-zero-mode contribution. 
The near-zero-mode contributions, on the 
other hand, come from diagonal 
terms, as a result of nonzero chirality. 
On adding all modes together, as 
in Eq. (\ref{eq:ct}), and dividing by 4 we 
obtain a result per quartet, similar to that 
in Eqs.~(\ref{eq:cgold}) and (\ref{eq:cgold2}). 
The way in which this is achieved is rather 
different from that for the Goldstone $\pi$ meson, 
and the different mode contributions follow 
more closely that of the continuum. 
A difference with both the continuum and the 
staggered Goldstone $\pi$ is that there 
is a constant of proportionality which is the 
square of the chirality of the zero modes. 
The disconnected terms cancel all diagonal connected 
contributions (having in fact the same 
constant of proportionality), 
and therefore we 
finally obtain, for the $\eta^{\prime}$ correlator, a 
result that tends to Eq.~(\ref{eq:ceta}) in the continuum 
limit, once the taste-singlet pseudoscalar 
current is appropriately normalized. 

\begin{figure}
\begin{center}
\includegraphics[width=80mm]{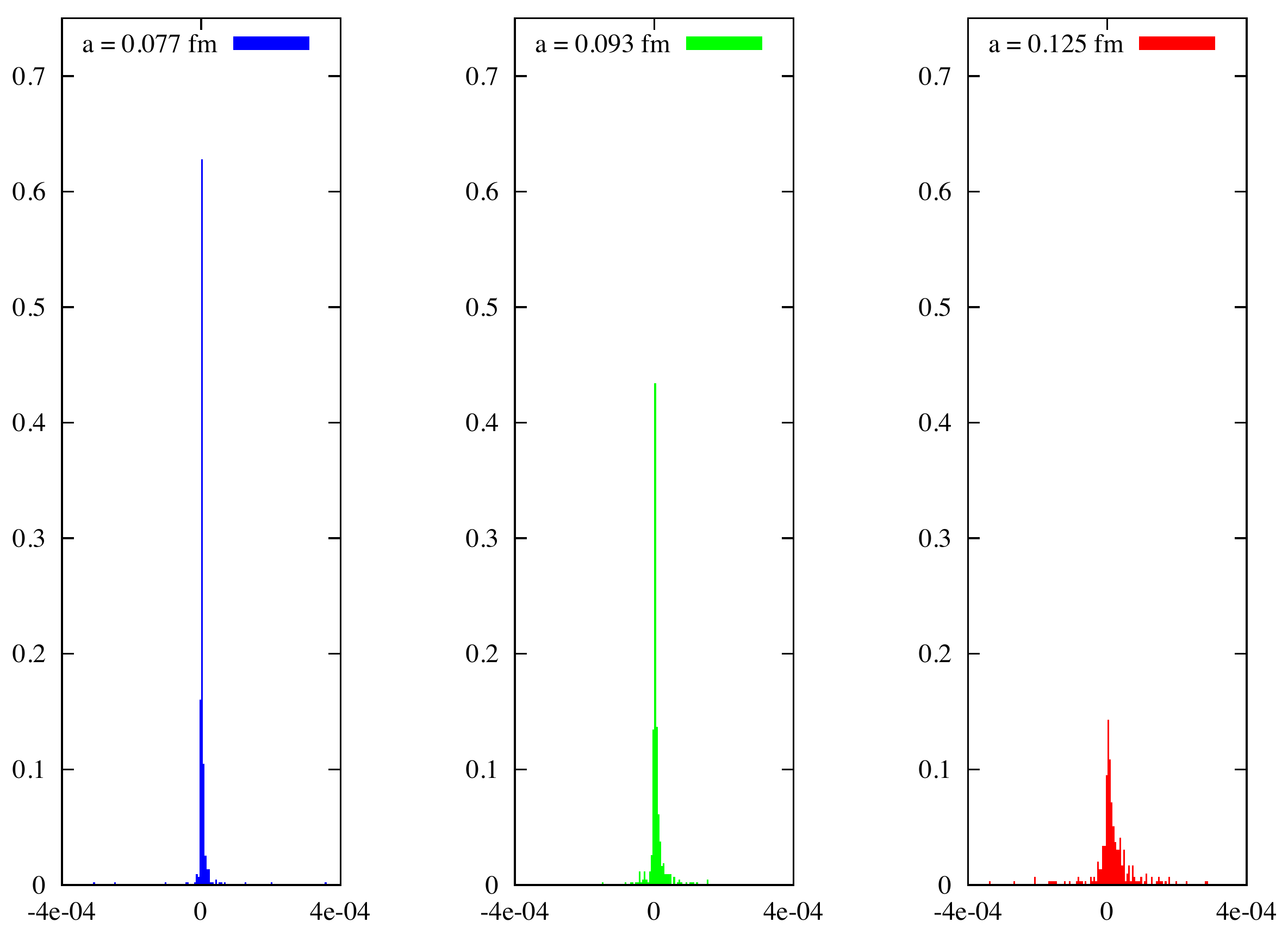}
\end{center}
\caption{ A histogram of values of the combination 
$-X(T_{\mathrm{mid}})/4 + Y(T_{\mathrm{mid}})/16$
calculated from the near-zero modes $i,j = \pm 1, \pm 2$ for
the $|Q|=1$ configurations for sets 1, 3, and 5. 
The results are plotted for time separation, $T_{\mathrm{mid}}$, set to the 
midpoint of the lattice. 
}
\label{fig:midptdist}
\end{figure}

Let us now demonstrate the cancellation 
between the connected and disconnected contributions 
from the near-zero modes more explicitly. 
Figure~\ref{fig:midptdist} shows histograms in the $|Q|=1$ sector for
\begin{equation}
    \sum_{i,j=\pm 1, \pm 2} -\frac{X^{\gamma^5}_{ij}(T)}{4} + 
        \frac{Y^{\gamma^5}_{ij}(T)}{16},
\end{equation}
evaluated at the midpoint of the lattice, $T=T_{\mathrm{mid}}$ for the 
three sets---1, 3, and 5---that have the same physical volume but different 
lattice spacings. Then $T_{\mathrm{mid}}$ corresponds approximately to 
the same physical time separation in each case. 
From Fig.~\ref{fig:midptdist}, it is clear that this combination of $X$ and $Y$, 
which skeptics have worried could be troublesome, is
in fact zero on average at every value of the lattice spacing. 
The histogram 
of values shows that the distribution is somewhat broader on the 
coarser lattices, but there is no other effect from the lattice 
spacing. 

\begin{figure}
    \centering
    \includegraphics[width=80mm]{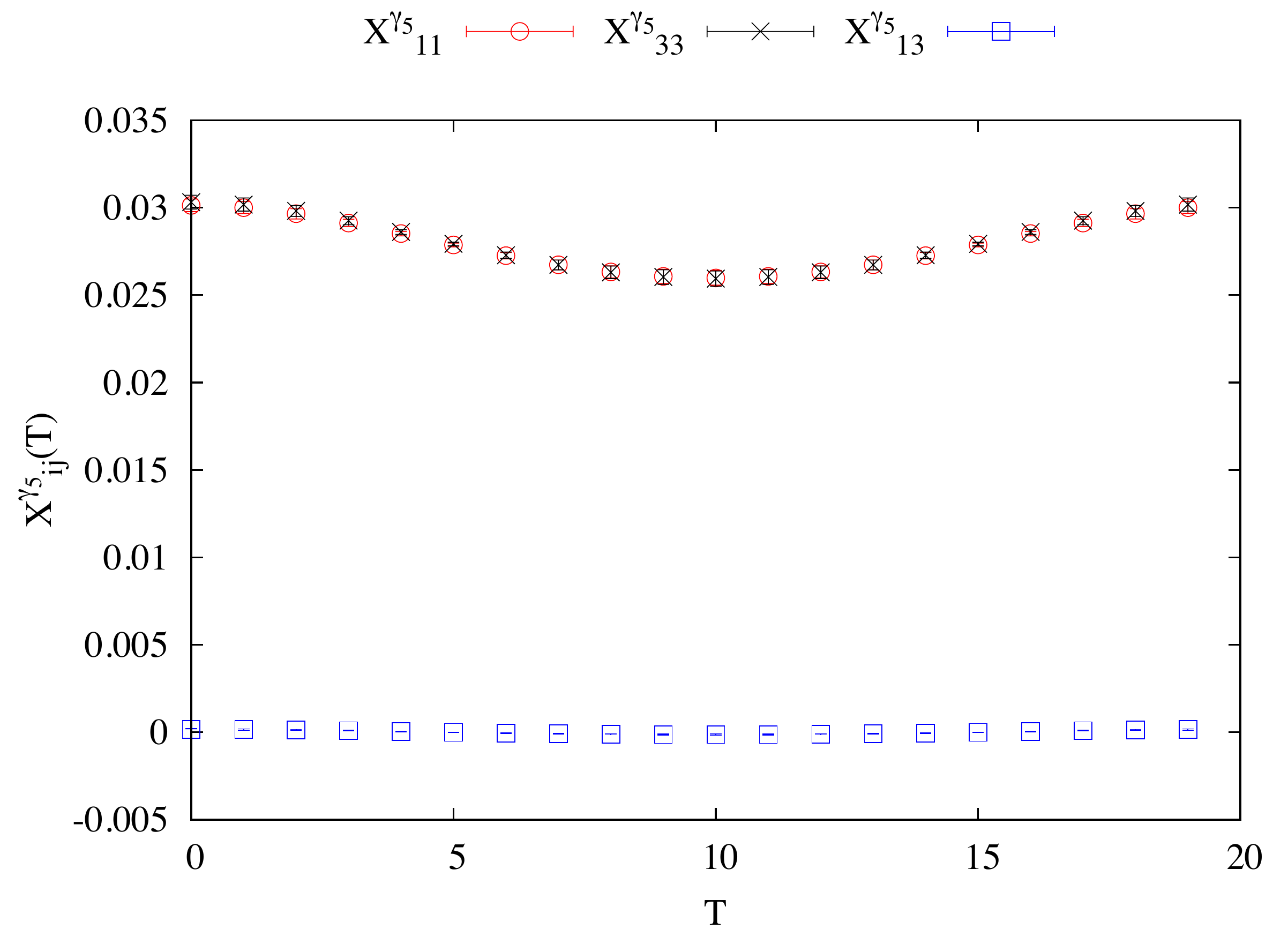}
    \caption{ Overlaps $X^{\gamma^5}_{ij}$ 
    averaged over the 27 lattices from the finest ensemble, set~5, 
    that had topological charge $|Q|=2$. 
    Results are shown for mode 1 from the first near-zero-mode quartet
    and mode 3 from the second near-zero-mode quartet. 
    }
    \label{fig:q2overlap13}
\end{figure}

In the above discussion, we have focused on the $|Q|=1$ case 
because that is the easiest one with which to study near-zero 
and nonzero modes. 
However, results on configurations with other $Q$ values 
also behave exactly as expected from this picture. 
Figure~\ref{fig:q2overlap13} shows results for correlated overlaps 
$X^{\gamma^5}_{ij}$ for 
27 configurations with $|Q|=2$ from the finest, set~5
lattices. The correlated overlaps are between modes 1 and 3 which 
are now members of two separate near-zero-mode quartets.
We see that there is negligible correlated overlap between modes from different 
near-zero quartets, so the counting for each quartet, taken 
care of by the subsequent division by 4 for the connected 
contribution, is exactly 
as for the $|Q|=1$ case. 

\subsubsection{Flavor-singlet scalar mesons}
\label{subsub:scalar}

The flavor-singlet scalar case is easy to analyze 
both in the continuum and for staggered fermions
because of the 
simple form of the taste-singlet scalar, $\Gamma=1_I$. 
The orthogonality and normalization of the eigenvectors give 
$\sum_t \overline{\zeta}^1_{rs}(t) = \delta_{rs}$. Thus, the disconnected 
contribution in the continuum becomes:
\begin{equation}
\sum_T \mathcal{D}^{\sigma}(T) = \sum_{r,s} \frac{1}{(i\lambda_r + m)(i\lambda_s + m)}
\label{eq:cscalardisc}
\end{equation}
where the sum is over all eigenmodes. 
The connected contribution is
\begin{equation}
\sum_T \mathcal{C}^{\sigma}(T) = \sum_{r} \frac{1}{(i\lambda_r + m)^2}
\label{eq:cscalarc}
\end{equation}
and we see that it is canceled by diagonal terms from 
Eq.~(\ref{eq:cscalardisc}). In particular, for $|Q|=1$, 
the single zero-mode contribution to the 
total flavor-singlet correlator cancels between $\mathcal{D}$ and 
$\mathcal{C}$ to give a finite result for $\mathcal{M}^{\sigma}$ as $m\to0$.
For staggered fermions, Eqs.~(\ref{eq:cscalardisc}) 
and~(\ref{eq:cscalarc}) still hold, with a sum over the total number of 
eigenmodes. By taking a suitable average over the eigenvalues 
in a quartet, the cancellation of diagonal terms quartet by quartet 
mimics that of the continuum. In particular, 
neglecting the near-zero $\lambda_i$ relative to $m$ 
and dividing $\mathcal{C}$ by 4 and $\mathcal{D}$ by 16, it 
is clear that exactly the same cancellation 
of the contributions from the near-zero-mode quartet  occurs 
as in the continuum.  

\begin{figure}
\begin{center}
\includegraphics[width=80mm]{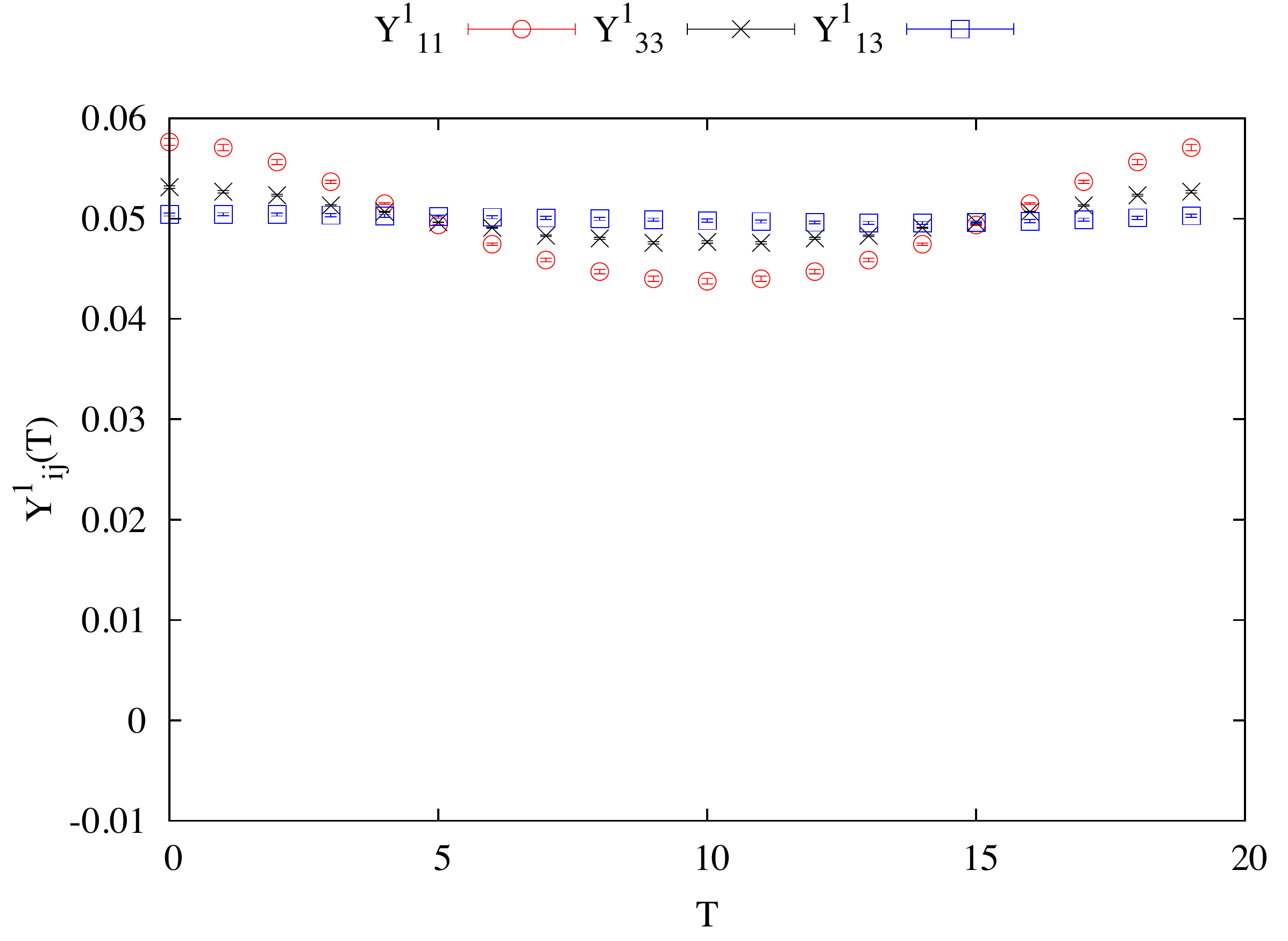}
\end{center}
\caption{Diagonal and off-diagonal correlated overlaps $Y^{1}_{rs}(T)$ between 
eigenvectors 1 and 3 that 
contribute to the disconnected piece of the flavor-singlet scalar
meson correlator 
at zero spatial momentum. Results are given for the average 
over $|Q|=1$ configurations 
in the fine ensemble, set~5. 
}
\label{fig:scalardisc}
\end{figure}

\begin{figure}[]
\begin{center}
\includegraphics[width=80mm]{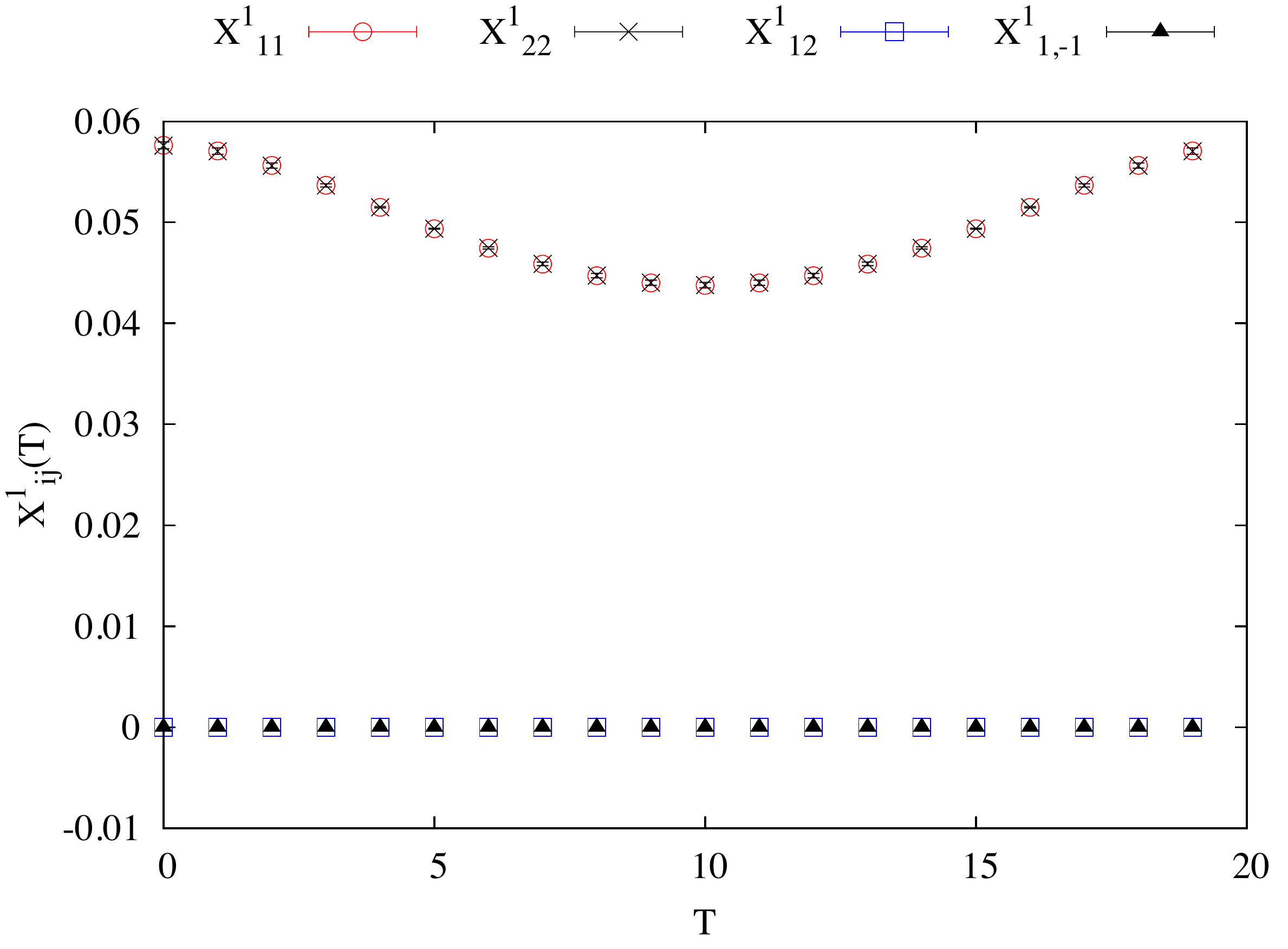}
\includegraphics[width=80mm]{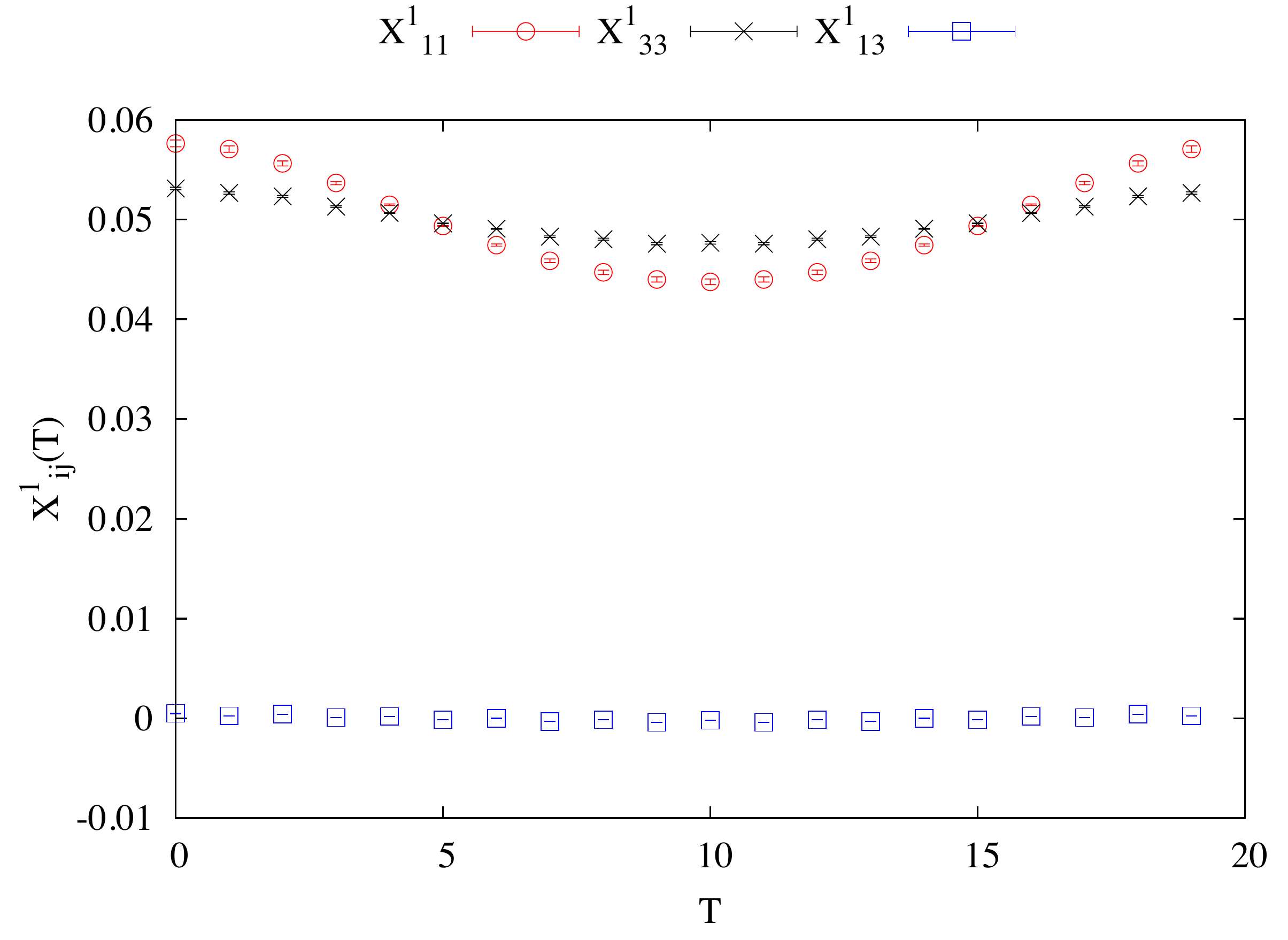}
\end{center}
\caption{The correlated overlaps $X^{1}_{rs}(T)$ between 
near-zero modes 1, 2 and $-1$ (top) and between near-zero 
mode 1 and nonzero mode 3 (bottom) that 
contribute to the connected piece of the flavor-singlet 
scalar meson correlator 
at zero spatial momentum. Results are given for the average 
over $|Q|=1$ configurations 
in the fine ensemble, set~5.  
}
\label{fig:scalaroverlaps}
\end{figure}

Figures~\ref{fig:scalardisc} and~\ref{fig:scalaroverlaps} 
show a representative sample of $Y^1_{rs}$ and $X^1_{rs}$, 
plotted as a function of $T$ for set~5. 
Figure~\ref{fig:scalardisc} shows correlated overlaps $Y^1_{11}$ for 
a near-zero mode and $Y^1_{33}$ for a nonzero mode 
as well as the off-diagonal $Y^1_{13}$. 
Set 5 lattices have a time extent of 20, so 
we expect values around 0.05, such that 
the sum over $T$ yields~1. The results for all 
members of the near-zero-mode quartet agree with those 
of $Y^1_{11}$ and those of the first nonzero-mode quartet 
agree with those of $Y^1_{33}$. Unlike the pseudoscalar 
case, $Y^1_{13}$ is not zero but as large as $Y^1_{11}$ and 
$Y^1_{33}$ since $\sum_T Y^1_{rs}(T) = 1$ for all $r,s$ 
both in the continuum and on the lattice. 

Figure~\ref{fig:scalaroverlaps} shows the correlated overlaps 
$X^1_{rs}(T)$. 
Here $\sum_T X^1_{rs}(T) = \delta_{rs}$, and it is clear 
that the diagonal correlated overlaps are the same as those of the 
appropriate $Y^1_{rs}$ and the off-diagonal correlated overlaps are zero in each case. 

\begin{figure}
\begin{center}
\includegraphics[width=80mm]{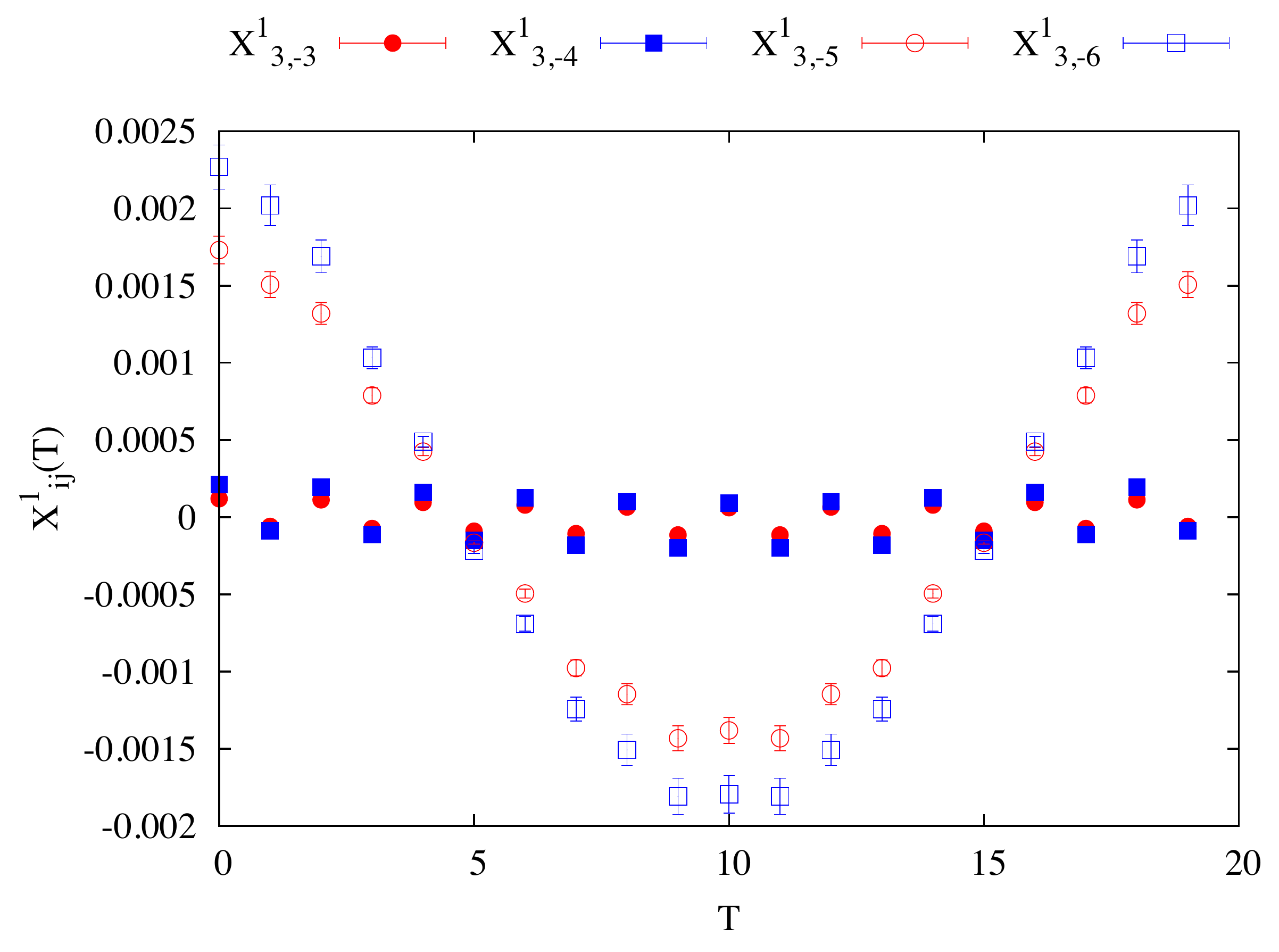}
\end{center}
\caption{The correlated overlaps $X^{1}_{rs}(T)$ between 
nonzero mode 3 and modes $-3$, $-4$, $-5$, and $-6$ from 
its mirror quartet. 
Results are averaged over $|Q|=1$ configurations for set~5.}
\label{fig:scalarminus}
\end{figure}

Further detail is shown in Fig.~\ref{fig:scalarminus}, 
which gives the $X^1_{rs}$ between nonzero modes in mirror 
quartets, and between the positive and negative eigenmodes of the zero 
mode quartet. Some of these correlated overlaps are large in the 
pseudoscalar case. 
None of them is large here and all yield zero 
after summing over $T$. Quite different behavior is 
seen in the different correlated overlaps, however. In particular, we see 
once again in these correlated overlaps the distinction between 
different pairs in the nonzero-mode quartets.  

\begin{figure}
    \includegraphics[width=80mm]{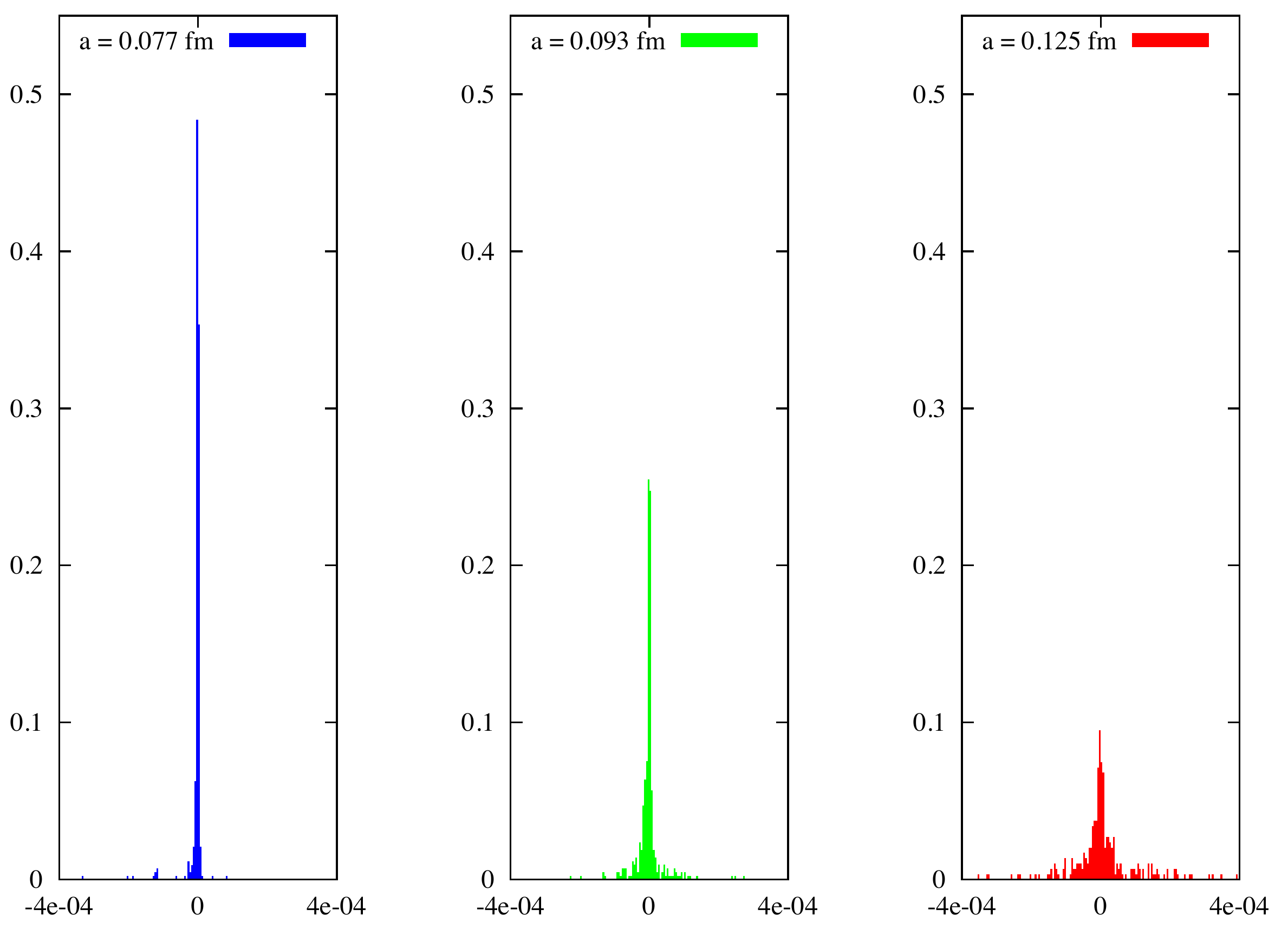}
    \caption{A histogram of values of the combination 
        $-X^1(T_{\mathrm{mid}})/4 + Y^1(T_{\mathrm{mid}})/16$ 
        calculated from the near-zero modes for
        the $|Q|=1$ configurations for sets 1, 3, and 5. 
        The results are plotted for time separation set to the 
        midpoint of the lattice, $T_{\mathrm{mid}}$.}
    \label{fig:scalarmidptdist}
\end{figure}

Figure~\ref{fig:scalarmidptdist} shows histograms in the $|Q|=1$ sector for
\begin{equation}
    \sum_{i,j =\pm 1, \pm 2} -\frac{X^1_{ij}(T)}{4} + 
        \frac{Y^1_{ij}(T)}{16},
\end{equation}
evaluated at $T_{\mathrm{mid}}$ for the three
sets---1, 3, and 5---that have the same physical volume but different 
lattice spacings. 
From Fig.~\ref{fig:scalarmidptdist}, it is clear that, as in the 
pseudoscalar case, this combination of $X$ and $Y$, 
which corresponds to the potentially divergent contribution of 
the near-zero modes to the scalar meson correlator,  again
actually vanishes on average at every value of the lattice spacing. 
The width of the histogram distribution is the quantity which 
changes with lattice spacing, becoming more narrowly peaked 
around zero as the lattice spacing goes to zero. 

\subsubsection{Flavor-singlet vector, axial vector, and tensor mesons}
\label{subsub:others}

\begin{figure}[]
\begin{center}
\includegraphics[width=80mm]{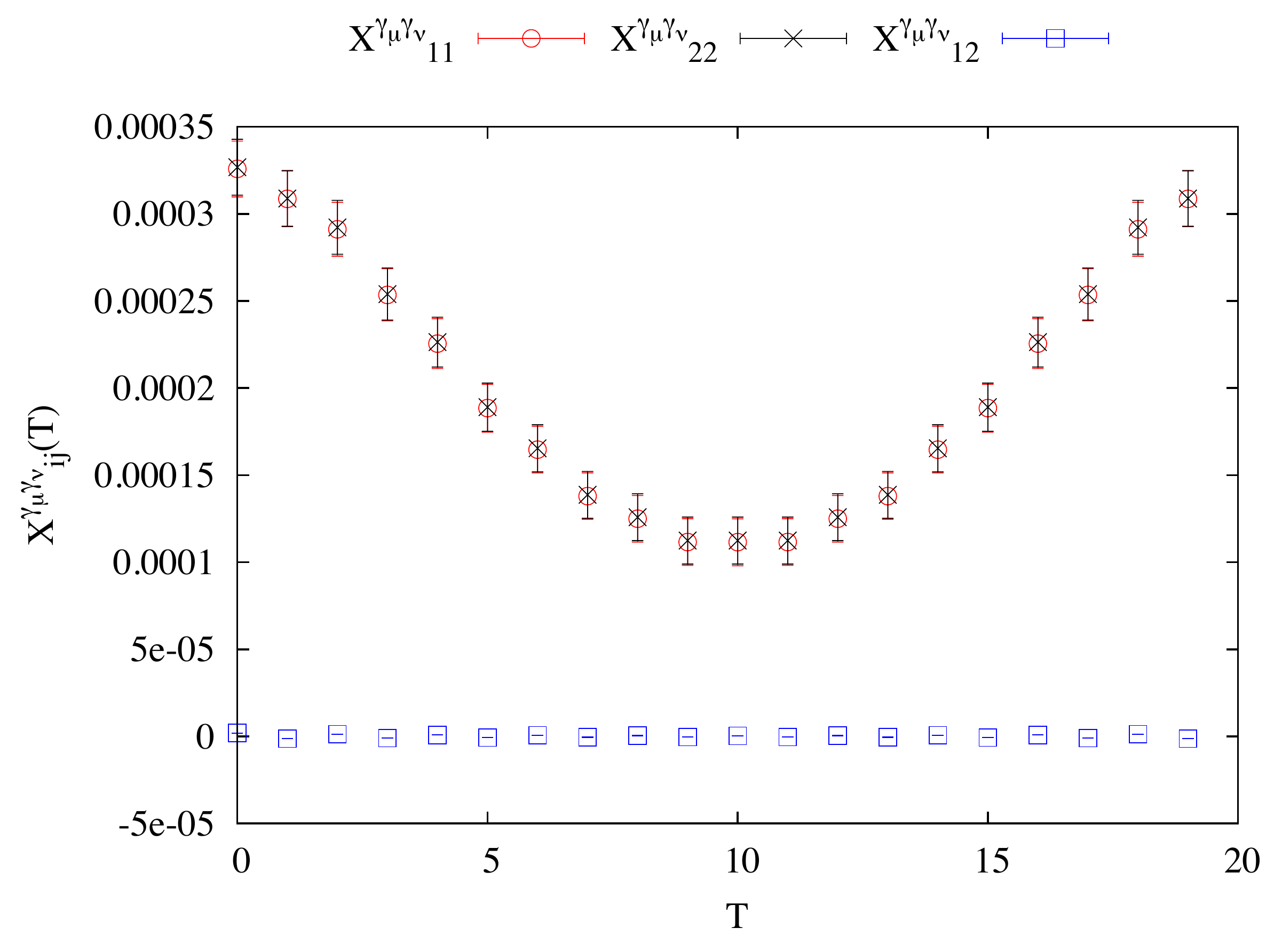}
\end{center}
\caption{The correlated overlaps $X^{\sigma^{xy}}_{ij}(T)$ between 
near-zero modes 1 and 2 that
contribute to the connected piece of the flavor-singlet 
tensor meson correlator 
at zero spatial momentum. Results are given for the average 
over $|Q|=1$ configurations 
in the fine ensemble, set~5.  
}
\label{fig:tensor}
\end{figure}

The correlated overlaps for the flavor-singlet tensor case behave similarly
to the pseudoscalar and scalar. No simple analysis of correlated 
overlaps in terms of the chirality or normalization of the modes 
is possible and, indeed, we find that none of the correlated overlaps is large. 
Figure~\ref{fig:tensor} shows that the key requirement for a sensible 
flavor-singlet correlator holds, i.e., that the off-diagonal 
correlated overlaps between different members of the near-zero-mode quartet 
are consistent with zero. This means, as above, 
that the connected and disconnected near-zero-mode contributions 
cancel rather than giving a potentially divergent piece. 

\begin{figure}[]
\begin{center}
\includegraphics[width=80mm]{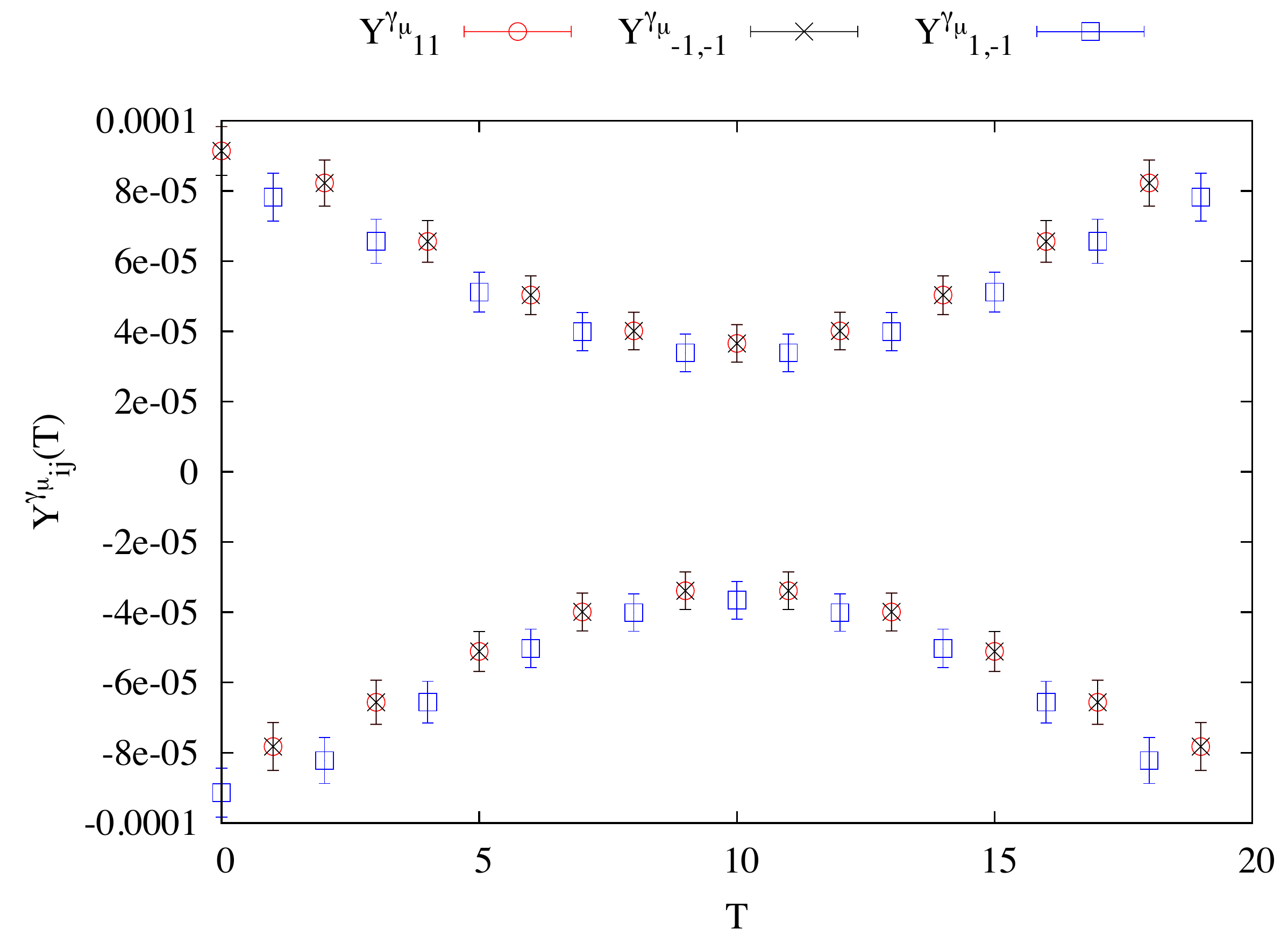}
\end{center}
\caption{The correlated overlaps $Y^{\gamma^{\mu}}_{ij}(T)$ for 
$\mu = x$ between 
near-zero modes 1 and $-1$ that
contribute to the disconnected piece of the flavor-singlet 
vector meson correlator 
at zero spatial momentum. Results are given for the average 
over $|Q|=1$ configurations 
in the fine ensemble, set~5.  
}
\label{fig:vecdisc}
\end{figure}

The flavor-singlet vector and axial-vector cases behave somewhat 
differently, which can be traced back to the fact that the 
taste-singlet versions of these operators couple even and odd 
lattice sites together rather than even-to-even or odd-to-odd 
as with the other examples. The axial vector and 
vector behave in the same way, so we only show results here for the 
vector case. As discussed in Sec.~\ref{sec:hooft}, in the 
continuum there is no zero-mode 
contribution to the disconnected piece of the flavor-singlet vector 
meson correlator because
$\gamma^\mu$ 
and $\gamma^5$ anticommute. We show in Fig.~\ref{fig:vecdisc} 
how this works for staggered fermions. 
Because $\overline{\zeta}^{\gamma^\mu}_{ii}$ couples 
odd and even sites, and $f_{-1}$ has the opposite sign 
on odd sites to $f_{1}$, then $\overline{\zeta}^{\gamma^{\mu}}_{11}$ and 
$\overline{\zeta}^{\gamma^{\mu}}_{-1-1}$ have opposite sign. This means 
that the near-zero-mode contribution to the disconnected 
correlator from $Y^{\gamma^{\mu}}_{11}$ has opposite sign 
to that from $Y^{\gamma^{\mu}}_{1-1}$. This is seen clearly 
for $\mu = x$ in Fig.~\ref{fig:vecdisc}. Summing over $i,j\in\{\pm 1,\pm 2\}$ 
then clearly gives a total disconnected contribution to 
the flavor-singlet vector meson correlator of zero.  

\begin{figure}[]
\begin{center}
\includegraphics[width=80mm]{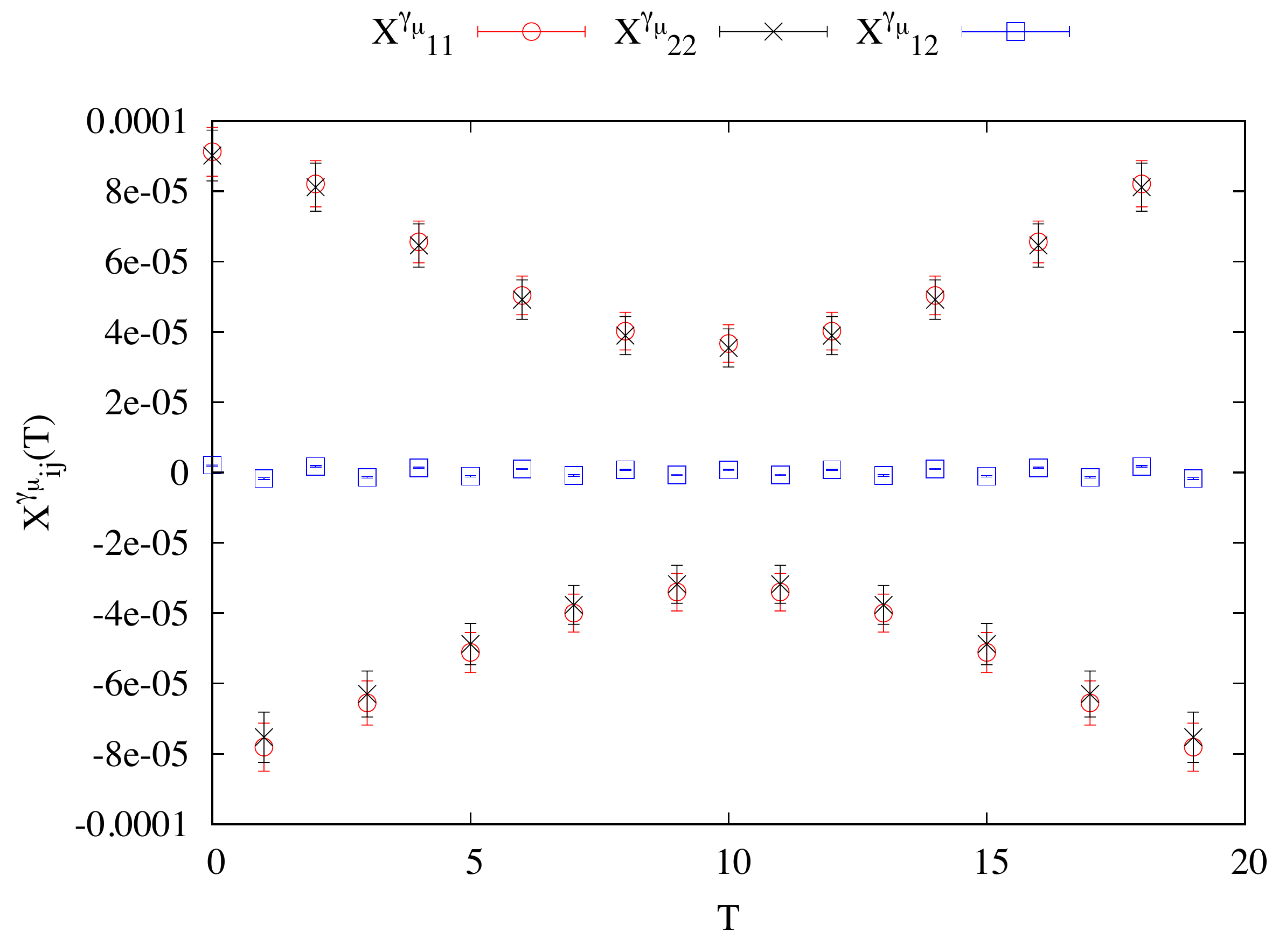}
\end{center}
\caption{The correlated overlaps $X^{\gamma^{\mu}}_{ij}(T)$ for 
$\mu = x$ between 
near-zero modes 1 and 2 that
contribute to the disconnected piece of the flavor-singlet 
vector meson correlator 
at zero spatial momentum. Results are given for the average 
over $|Q|=1$ configurations 
in the fine ensemble, set~5.  
}
\label{fig:vector}
\end{figure}

The connected contributions are not zero time slice by time slice, 
as we see from Fig.~\ref{fig:vector}. 
Correlated overlaps $X^{\gamma^\mu}_{ii}$ (equal to their counterparts 
$Y^{\gamma^\mu}_{ii}$ in Fig.~\ref{fig:vecdisc})
become pure oscillations, $(-1)^T$, that cancel upon summing over $T$. 
Oscillatory terms are a 
feature of staggered meson correlators, stemming from opposite-parity 
contributions to the correlator and, ultimately, the remaining 
time doubling.
They do not 
then affect the properties of the ground state meson, in this 
case the flavor-singlet vector meson. 
The off-diagonal $X^{\gamma^\mu}_{ij}$ are close to zero and 
also oscillatory.  Thus, once again there is no significant net 
contribution from near-zero modes to the flavor-singlet vector 
meson. 

\subsubsection{Summary}
\label{subsub:summary}

We conclude that the behavior of the staggered flavor-taste 
singlet meson correlators 
in every case follows that expected in continuum QCD. 
In particular, no chiral-limit divergence results from 
the near-zero modes. 
In all cases, we find that the $Y^{\Gamma}_{ij}$ take the 
same average value for 
$i, j$ both in a given quartet. 
Thus, the disconnected contribution from a quartet
of degenerate eigenvalues
is $4 \times 4 = 16$ times that of a single mode. 
The off-diagonal correlated overlaps $X_{ij}$ are zero for $i,j$ within 
a quartet in every case. This means that the connected 
contributions give instead $4$ times 
that of a single mode. 
Including the factors of 4 and 16 in 
Eq.~(\ref{eq:mcd}) means that the correlator 
is effectively made of single-species contributions,
as in continuum QCD, 
and, in particular, the contribution 
from near-zero modes cancels as it does there. 

The pseudoscalar and scalar cases are particularly simple 
to analyze, both for zero and nonzero modes, 
and to see the clear correspondence with continuum 
behavior. The correlated overlaps for the taste-singlet 
pseudoscalar between mirror quartets are a striking 
demonstration of how staggered fermions conspire to 
give the ``right'' answer, but sometimes in a rather nontrivial way. 
The match demonstrated between the taste-singlet 
and the Goldstone pseudoscalar 
also leads to a practical 
suggestion that may improve the determination of 
the $\eta^{\prime}$ mass using staggered fermions. 
The calculation of 
the taste-singlet connected and disconnected contributions 
is particularly statistically noisy because of the 
point-split nature of the taste-singlet operator. 
It may be preferable, although numerically challenging, to 
determine instead the near-zero-mode eigenvectors and then subtract 
their contribution from the Goldstone pseudoscalar correlator. 
This must agree with the $\eta^{\prime}$ correlator in the 
continuum limit and yet is constructed of local operators and 
so has significantly less gauge noise. 

\section{Conclusions}
\label{sec:conclusions}

This paper adds weight to the growing evidence that shows that 
staggered fermions behave in the correct way 
to reproduce QCD in the continuum limit, even with the rooted 
determinant. Here we have focused on 
the eigenvectors of the staggered-fermion Dirac operator and 
the way in which the 't~Hooft vertex and flavor-singlet meson 
correlators are built  
from the overlaps between different eigenvectors, using
the appropriate taste singlets.
The important overlaps are those between eigenvectors 
within a near-zero quartet since these could have
generated dangerous singular terms as
$m\to 0$. 
From our theoretical results we determine a condition 
for the local overlaps that needs to hold and 
then test this numerically and demonstrate that it does. 
Indeed we see that the near-zero-mode quartet in all cases
behaves functionally in such a way to reproduce the required behavior 
of four copies of a single mode that mimics the expected behavior 
in the continuum. 

Most of our results are not surprising, but in providing 
a clear link between the theoretical requirements 
and the numerical results for the eigenvector overlaps, we 
add further confidence to the soundness of the framework 
for the accurate phenomenology that is being done with staggered fermions.  
We demonstrate most directly that a calculation 
of flavor-singlet meson 
masses, notably that of the $\eta^{\prime}$ meson, should 
give the correct QCD result. This is not a substitute for doing the 
full calculation and
 this is underway~\cite{Gregory:2007ev,Gregory:2011lat}. 

\acknowledgments

We thank Alistair Hart for generating the 
configurations~\cite{oureigsshort,oureigslong}, and
we thank Junko Shigemitsu for alerting us to 
Ref.~\cite{gplshortpaper:2007}. 
This work was funded by STFC, the Scottish Universities Physics 
Alliance, MICINN (under Grants No.~FPA2009-09638 and No.~FPA2008-10732), 
DGIID-DGA (Grant No.~2007-E24/2), and by the EU under 
ITN-STRONGnet (PITN-GA-2009-239353). EF is supported by the MICINN Ram\'on
y Cajal program. 
Fermilab is operated by Fermi Research Alliance, LLC, under Contract
No.~DE-AC02-07CH11359 with the United States Department of Energy.
The calculations were performed at 
the High Performance Computing Centre in Cambridge as part of 
the DiRAC facility, jointly funded by STFC, 
the Large Facilities Capital Fund of BIS, 
and the Universities of Cambridge and Glasgow.

\appendix

\begin{widetext}
\section{Detailed Formulas}
\label{app:formulas}

For Sec.~\ref{subsec:zeta-plots} it is convenient to spread 
staggered-fermion bilinears over a hypercube, inserting the lattice 
gauge field to preserve gauge invariance.
An explicit construction is 
\begin{eqnarray}
    S_I(x) & = & \frac{1}{16} \sum_b \bar{\chi}(x+b)\chi(x+b), 
    \label{eq:scalar} \\
	V^\mu_I(x) & = & \frac{i}{16} \sum_b \eta^\mu(x+b)\,
        \bar{\chi}(x+\bar{b}^{(\mu)})U(x+\bar{b}^{(\mu)},x+b)\chi(x+b) ,
	\label{eq:vector} \\
	T^{\mu\nu}_I(x) & = & -\frac{1}{16} \sum_b
        \eta^\mu(x+\bar{b}^{(\mu\nu)})\eta^\nu(x+b) \,
        \bar{\chi}(x+\bar{b}^{(\mu\nu)})\bar{U}(x+\bar{b}^{(\mu\nu)},x+b) 
        \chi(x+b) , \nu\neq\mu,
	\label{eq:tensor} \\
	A^\mu_I(x) & = & \frac{i}{16} \sum_b
        \eta^\mu(x+d-\bar{b}^{(\mu)})\,
        \eta_1(x+b)\eta_2(x+b)\eta_3(x+b)\eta_4(x+b) \times
        \nonumber \\ & & \hspace{8em}
        \bar{\chi}(x+d-\bar{b}^{(\mu)})\bar{U}(x+d-\bar{b}^{(\mu)},x+b) \chi(x+b),
    \label{eq:axial} \\
	P_I(x) & = & \frac{1}{16} \sum_b 
		\eta_1(x+b)\eta_2(x+b)\eta_3(x+b)\eta_4(x+b)\,
		\bar{\chi}(x+d-b)\bar{U}(x+d-b,x+b) \chi(x+b),
	\label{eq:pseudoscalar}
\end{eqnarray}
\end{widetext}
\noindent
where $b$ runs over the $2^4$-site hypercube with origin~$x$;
$\bar{b}_\mu^{(\mu)}=a-b_\mu$ but $\bar{b}_\rho^{(\mu)}=b_\rho$, 
$\rho\neq\mu$;
$\bar{b}_\lambda^{(\mu\nu)}=a-b_\lambda$, $\lambda=\mu,\nu$, 
but $\bar{b}_\rho^{(\mu\nu)}=b_\rho$, $\rho\neq\mu,\nu$;
and $d=(\hat{1}+\hat{2}+\hat{3}+\hat{4})a$.
Gauge invariance is ensured via averages of parallel transport over 
paths from $x$ to $x'$, $\bar{U}(x,x')$.%
\footnote{There is no bar on $U(x+\bar{b}^{(\mu)},x+b)$ because only the 
one-link path enters.}
Under shift symmetry these are all taste singlets.
The vector current and scalar density satisfy the Ward identity 
corresponding to quark-number conservation for all~$a$, and the 
axial-vector and pseudoscalar density satisfy the anomalous 
Ward identity as $a\to0$.
In practice, we use in place of~$U$ the HISQ-smeared gauge field $W$, 
defined in Eq.~(\ref{eq:W4HISQ}) below.

For brevity and clarity, it is then helpful to write
\begin{eqnarray}
    S_I(x)          & = & \bar{\chi}1_I               \chi, \\
    V^\mu_I(x)      & = & i\bar{\chi}\gamma^\mu_I     \chi, \\
    T^{\mu\nu}_I(x) & = & \bar{\chi}i\sigma^{\mu\nu}_I\chi, \\
    A^\mu_I(x)      & = & i\bar{\chi}\gamma^{\mu5}_I  \chi, \\
    P_I(x)          & = & \bar{\chi}\gamma^5_I        \chi, 
\end{eqnarray}
which with Eqs.~(\ref{eq:scalar})--(\ref{eq:pseudoscalar})
\emph{define} $1_I$, $\gamma^\mu_I$, $i\sigma^{\mu\nu}_I$, 
$\gamma^{\mu5}_I$, and $\gamma^5_I$, when acting on $\chi$, 
$\bar{\chi}$, and the eigenvectors of $\Dstag$ for the analysis 
of $\zeta^{\Gamma}_{ij}$ in Sec.~\ref{subsec:zeta-plots}.

When constructing the correlators for the $\eta'$ and other flavor-taste-singlet 
correlators, it is more customary to restrict the operators to 
one time slice.
In Sec.~\ref{subsub:etap}, therefore, we average over spatial cubes 
only [and then time slices, cf.\ Eqs.~(\ref{eq:zetaxy})].
$V^4_I$ and $\bm{A}_I$ remain as in Eqs.~(\ref{eq:vector}) 
and~(\ref{eq:axial}), because they naturally 
extend over a timelike link or three-dimensional cube.
For $P_I$, the operator is defined as attached to point 
on a time slice by averaging over all hypercubes that 
have a corner at that point, i.e., extending both 
forwards and backwards in time. Then averaging over a time slice 
is straightforward.   

\section{Improved Staggered Actions}
\label{app:actions}

To introduce improved staggered-fermion actions, it is convenient to 
proceed in steps, introducing notation along the way.
The first step is ``Fat7'' smearing~\cite{Orginos:1999cr},
\begin{equation}
	\mathcal{F}_\mu U_\mu =
		\prod_{\rho\neq\mu}^{\rm sym} \left[1 +
		\case{1}{4}\left(T_\rho + T_{-\rho} - 2 \right) \right] U_\mu,
\end{equation}
which yields paths of length 3, 5, and~7.
Here $T_{\pm\rho}U_\mu(x)=U_{\pm\rho}(x)U_\mu(x\pm\hat{\rho}a)%
U_{\mp\rho}(x\pm\hat{\rho}a)$, 
$U_{-\rho}(x)=U_\rho^\dagger(x-\hat{\rho}a)$.
It is easy to check that the smearing introduces a form factor that 
reduces the coupling to taste-changing gluons \cite{Lepage:1996jw}.

As is often the case with smearing algorithms,
Fat7 smearing introduces additional discretization errors.
These can be removed by introducing an order-$a^2$ 
improvement~\cite{gplasqtad}
\begin{equation}
	V_\mu = \left(\mathcal{F}_\mu - \case{1}{4}\mathcal{L}_\mu\right)U_\mu,
    \label{eq:V4Fat7}
\end{equation}
where
\begin{equation}
    \mathcal{L}_\mu U_\mu = \sum_{\rho\neq\mu}(T_\rho-T_{-\rho})^2U_\mu,
\end{equation}
introduces the five-link Lepage term.
The discretization error of the simple difference operator in 
Eq.~(\ref{eq:Sstag}) can be removed with the three-link Naik 
term~\cite{Naik:1986bn}, 
\begin{equation}
    S_{\mathrm{Naik}} = -\case{1}{12} a^3 \sum_{x,\mu} 
        \eta_\mu(x)\bar{\chi}(x)\left(
		T_\mu - T_{-\mu}\right)^3\chi(x) ,
\end{equation}
where now $T_{\pm\mu}\chi(x)=U_{\pm\mu}(x)\chi(x\pm\hat{\mu}a)$.

For the HISQ action, Fat7 smearing is applied twice, with the Lepage 
correction taken at the second step
\begin{equation}
    W_\mu = \left(\mathcal{F}_\mu-\case{1}{2}\mathcal{L}_\mu\right)
        \mathcal{U}\mathcal{F}_\mu U_\mu,
    \label{eq:W4HISQ}
\end{equation}
where $\mathcal{U}$ denotes a reunitarization and projection to SU(3).
[The SU(3) projection makes little difference in practice.]
The HISQ action is then 
\begin{equation}
    S_{\mathrm{HISQ}} = \Sstag(W_\mu) + 
        S_{\mathrm{Naik}}(\mathcal{U}\mathcal{F}_\mu U_\mu),
    \label{eq:SHISQ}
\end{equation}
substituting for the original gauge field $U_\mu$ as shown.

For completeness we write the Fat7$\times$Asqtad~\cite{oureigslong} and 
Asqtad~\cite{gplasqtad} actions in this notation:
\begin{eqnarray}
    S_{\mathrm{Fat7}\times\mathrm{Asqtad}} & = & \Sstag(\check{W}_\mu) + 
        S_{\mathrm{Naik}}(\mathcal{U}\mathcal{F}_\mu U_\mu), 
    \label{eq:SFat7Asqtad} \\
    \check{W}_\mu & = & 
        \left(\mathcal{F}_\mu-\case{1}{4}\mathcal{L}_\mu\right)
        \mathcal{U}\mathcal{F}_\mu U_\mu, 
    \label{eq:W4Fat7Asqtad} \\
    S_{\mathrm{Asqtad}} & = & \Sstag(V_\mu) + S_{\mathrm{Naik}}(U_\mu), 
    \label{eq:SAsqtad}
\end{eqnarray}
Unfortunately, Ref.~\cite{oureigsshort} referred to Fat7$\times$Asqtad 
as ``HISQ.''
The Asqtad action defines the rooted determinant in the MILC 
ensembles~\cite{milcdecay04,milcreview}, which have been used by the 
zero-temperature results cited in the Introduction.
For this action there is an additional tadpole-improvement 
step in which one replaces $T_{\pm\rho}\chi$ and
$T_{\pm\rho}U_\mu$ by $u_0^{-1}T_{\pm\rho}\chi$ and 
$u_0^{-2}T_{\pm\rho}U_\mu$, respectively, 
where $u_0$ is a measure of the mean link. 
In MILC's simulations of the Asqtad action~\cite{milcreview}, $u_0$ is 
set by the fourth root of the $1\times1$ Wilson loop (the plaquette).
(The reunitarization in HISQ makes tadpole improvement unnecessary.)

\section{Further remarks on Refs.~\cite{Creutz:2007rk,Creutz:2008nk}}
\label{app:creutz}

Creutz~\cite{Creutz:2007rk,Creutz:2008nk} makes several remarks that 
sound simple, and thus seem to be accepted by nonexperts, 
but they do not withstand careful scrutiny.
One, explained elsewhere~\cite{asklat07}, is that the different tastes 
have different chirality.
As discussed above, all near-zero modes within a common quartet possess 
(identically for mirrors; empirically otherwise) the same taste-singlet 
chirality, Eqs.~(\ref{eq:Adams}) or (\ref{eq:chirality}).
The nonzero modes all have (nearly) zero taste-singlet chirality.
Finally, all modes have no net Goldstone chirality, 
$\sum_x f_s(x)\gamma^5_Pf_s(x)=
 \sum_x \varepsilon(x)f_s(x)f_s(x)=
 \sum_x f_{-s}(x)f_s(x)=0$.

Another incorrect statement~\cite{Creutz:2008nk} concerns the $\theta$ 
angle of the strong $CP$ problem, which can appear via a modified mass 
term
\begin{equation}
    m\,\bar{\psi}\psi \mapsto m\cos(\theta)\,\bar{\psi}\psi + 
        im\sin(\theta)\,\bar{\psi}\gamma_5\psi.
    \label{eq:mass}
\end{equation}
Creutz states correctly that $\theta$ obtains a physical meaning via 
the anomaly and, hence, the ultraviolet regulator.
He also states, incorrectly, that staggered fermions cannot possess 
this property, owing to the exact $\mathrm{U}_\varepsilon$ symmetry.

With this symmetry, the following two mass terms are, of course, 
equivalent:
\begin{equation}
    m\,\bar{\chi}\chi(x) \leftrightarrow 
         m\cos(\varphi)\,\bar{\chi}\chi(x) + 
        im\sin(\varphi)\,\varepsilon(x)\bar{\chi}\chi(x).
    \label{eq:blass}
\end{equation}
In the continuum limit, however, this corresponds to
\begin{equation}
    m\,\bar{q}q(x) \leftrightarrow
         m\cos(\varphi)\,\bar{q}q(x) + 
        im\sin(\varphi)\,\bar{q}\gamma_5\xi_5q(x),
\end{equation}
namely is a taste \emph{nonsinglet}.
It is superficially the kind of transformation used to set up 
twisted-mass Wilson fermions~\cite{Frezzotti:2004pc}, but we have seen 
no argument that proves it is the same.
In particular, unlike the unsubstantiated mass terms posited in 
Refs.~\cite{Creutz:2009zq,Creutz:2011hy}, 
$\varepsilon(x)=\sum_sf_s(x)f^{\dagger}_{-s}(x)$ is \emph{off-diagonal} 
in any basis where Eq.~(\ref{eq:yigal}) makes sense.

The correct analog of Eq.~(\ref{eq:mass}) is
\begin{equation}
    m\bar{\chi}\chi(x) \mapsto m\bar{\chi}\left[\cos(\theta) +
        i\sin(\theta)\gamma^5_I\right]\chi(x).
    \label{eq:mUndertastesinglet}
\end{equation}
The taste singlet $\gamma^5_I$ extends across a hypercube and depends 
on the lattice gauge field.
It thus relies on the regulator for its definition, as it must.
To simulate the $\theta$ vacuum via the fermion mass, 
one needs to implement Eq.~(\ref{eq:mUndertastesinglet}), not 
Eq.~(\ref{eq:blass}) \cite{Vink:1988ss,Durr:2006ze}.

%

\end{document}